\documentclass[%
 reprint,
nofootinbib,
 amsmath,amssymb,
 aps,
]{revtex4-2}

\usepackage{graphicx}
\usepackage{dcolumn}
\usepackage{bm}
\usepackage[pdftex,pagebackref=false]{hyperref}
\usepackage{amsmath,amssymb,amstext}
\usepackage{caption}
\usepackage{subcaption}
\usepackage{wrapfig}
\usepackage{cleveref} 
\usepackage{enumitem}
\usepackage{float}
\usepackage{mathtools}
\usepackage[export]{adjustbox}
\usepackage{xcolor}

\begin{document}

\preprint{APS/123-QED}

\title{Slowly Rotating Black Holes\\in 4D Einstein Gauss-Bonnet Gravity}

\author{Michael Gammon}
\email{mgammon@uwaterloo.ca}
\author{Robert B. Mann}%
\email{rbmann@uwaterloo.ca}
\affiliation{%
 University of Waterloo\\
 Department of Physics
}%

\date{\today}

\begin{abstract}
Since the recent derivation of a well-defined $D\rightarrow 4$ limit for regularized 4D Einstein Gauss-Bonnet (4DEGB) gravity, there has been considerable interest in testing it as an alternative to Einstein's general theory of relativity. In this paper we construct slowly rotating black hole solutions for 4DEGB gravity in asymptotically flat, de Sitter, and anti-de Sitter spacetimes. At leading order in the rotation parameter, exact solutions of the metric functions are derived and studied for all three of these cases. We compare how physical properties (innermost stable circular orbits, photon rings, black hole shadow, etc.) of the solutions are modified by varying coupling strengths of the 4DEGB theory relative to standard Einstein gravity results. We find that a vanishing or negative cosmological constant in 4DEGB gravity enforces a minimum mass on the black hole solutions, whereas a positive cosmological constant enforces both a minimum \textit{and} maximum mass with a horizon root structure directly analogous to the Reissner-Nordström de Sitter spacetime. Besides this, many of the physical properties are qualitatively similar to general relativity, with the greatest deviations typically being found in the low (near-minimal) mass regime.
\end{abstract}

\maketitle


\section{Introduction}

Despite the empirical success and predictive power of Einstein's general theory of relativity (GR), its modifications continue  attract attention. Early attempts can be seen in the writings of Weyl \cite{weyl1921} or Eddington \cite{eddington1921}, and continue through to present day, motivated by the singularity problem (ie. geodesic incompleteness), the need for reconciling quantum physics and gravity, and the need for phenomenological competitors so as to test GR in the most stringent manner possible. 

A common class of modifications to GR are higher curvature theories (or HCTs), in which it is assumed that a sum of powers of the curvature tensor is proportional to stress-energy, extending the assumed linear relationship between spacetime curvature (the Einstein tensor) 
and stress-energy in GR. Logically this linear relationship is not required, and it is conceivable that the empirical success of Einstein's theory could be improved upon by modifying the left-hand-side of the Einstein equations with a sum of powers of the curvature tensor.  

These HCTs play important roles in many different areas of physics  - they appear in a number of proposals for quantum gravity \cite{ahmed2017}, and may be necessary to account for observational evidence of dark matter, early-time inflation, or late-time acceleration \cite{bueno2016}. Lovelock theories \cite{lovelock1971} are the best-known examples of HCTs, and have the distinct feature that their differential equations are of 2nd order.
However, until recently, their equations only had non-trivial solutions in spacetime dimensions larger than four ($D>4$) \cite{bueno2019}, so their physical significance had been unclear. Recently, a new class of HCTs have been proposed that do allow for higher-order gravity in four dimensions and satisfy reasonable physical requirements such as positive energy excitations on constant curvature backgrounds. Such higher curvature theories are referred to as “generalized quasi-topological gravities”, or GQTGs \cite{bueno2019}. The original HCT was cubic in curvature \cite{lovelock1971,poshteh2019}; it was soon followed by a class quartic in curvature \cite{ahmed2017}. Most recently a procedure was found for constructing physically reasonable HCTs to any desired power of curvature \cite{bueno2019}.  

Amidst the plethora of higher curvature gravity theories, the quadratic Lovelock theory, or so-called ``Einstein Gauss-Bonnet" gravity has been of special interest. In addition to having 2nd order equations of motion, it is the simplest HCT.  For a long time it was thought that the Gauss-Bonnet (GB) action term 
\begin{equation}\label{eq:dgbaction}
\begin{aligned}
S_{D}^{G B} &= \alpha \int d^{D} x \sqrt{-g} \left[R^{\mu\nu\rho\tau}R_{\mu\nu\rho\tau} - 4 R^{\mu\nu}R_{\mu\nu} + R^2 \right] \\
&\equiv \alpha \int d^{D} x \sqrt{-g} \mathcal{G}
\end{aligned}
\end{equation}
could not contribute to a system's gravitational dynamics in $D \leq 4$ (since it becomes a total derivative in such cases), and hence the GB contribution \eqref{eq:dgbaction} is often referred to as a topological term of no relevance. Furthermore, a new 3+1-dimensional gravity theory which possesses diffeomorphism invariance, metricity, and second order equations of motion would be a violation of the Lovelock theorem \cite{lovelock1971} and thus should only be possible by introducing an additional field into the theory besides the metric tensor.  

In 2020, Glavan and Lin \cite{glavan2020} claimed to have bypassed the Lovelock theorem via the following rescaling of the Gauss-Bonnet coupling constant:
\begin{equation}\label{eq:alpharescale}
	(D-4) \alpha \rightarrow \alpha,
\end{equation}
and then taking the $D \rightarrow 4$ limit of exact solutions to Einstein-Gauss-Bonnet gravity. This interesting approach allowed for non-vanishing contributions from the GB action term in $D = 4$. In doing so, a number of 4-dimensional metrics can be obtained (for spherical black holes \cite{glavan2020,kumar2022,fernandes2020,kumar2020,kumar2022_2}, cosmological solutions \cite{glavan2020,li2020,kobayashi2020}, star-like solutions \cite{doneva2021,charmousis2022}, radiating solutions \cite{ghosh2020}, collapsing solutions \cite{malafarina2020}, etc.) carrying imprints of higher curvature corrections inherited from  their $D > 4$ counterparts. 

Unfortunately the existence of  limiting solutions  does not actually imply the existence of a 4D theory, and a number of objections in this vein quickly appeared \cite{gurses2020,ai2020,shu2020}. However, the conclusion that there ultimately was no four-dimensional Gauss-Bonnet theory of gravity proved to be premature when it was shown that a $D\to 4$ limit of the action \eqref{eq:alpharescale} could be taken \cite{hennigar2020,Fernandes:2020nbq}, generalizing a previous procedure for taking the $D\to 2$ limit of GR \cite{Mann:1992ar}. It is also possible to employ a Kaluza-Klein-like procedure \cite{Lu:2020iav}, compactifying D-dimensional Gauss-Bonnet gravity on a $(D-4)$-dimensional maximally symmetric space and then re-scaling the coupling constant according to equation \eqref{eq:alpharescale}. The resultant 4D scalar-tensor theory is a special case of  Horndeski theory \cite{horndeski1974}, which surprisingly has spherical black hole solutions whose metric functions match those from the naïve $D \rightarrow 4$ limiting solutions derived by Glavan \& Lin \cite{glavan2020}. Such solutions can be obtained without ever referencing a higher dimensional spacetime \cite{hennigar2020}.  

This theory, referred to as 4D Einstein Gauss-Bonnet gravity, provides an interesting phenomenological competitor to GR \cite{Clifton:2020xhc}. Conformally transforming the metric  $g_{\mu \nu} \to e^{-2\phi} g_{\mu \nu}$ in \eqref{eq:dgbaction} and subtracting it from the original GB action \eqref{eq:dgbaction} yields the 4DEGB action after trivial field redefinitions:
 \cite{hennigar2020}
\begin{equation}\label{eq:4DEGBaction}
\begin{aligned}
S_{4}^{G}=\alpha \int d^{4} x \sqrt{-g}&\Big[  \phi \mathcal{G}+4 G_{\mu \nu} \nabla^\mu \phi \nabla^\nu \phi-4(\nabla \phi)^2 \square \phi\\
&+2(\nabla \phi)^4\Big]
\end{aligned}
\end{equation}
using \eqref{eq:alpharescale}, 
where $\phi$ is an additional scalar field.
No further assumptions about particular solutions to higher dimensional theories or background spacetimes are required.  

Adding to this the Einstein-Hilbert action with a cosmological term, 
\begin{equation}
   S =  \int d^{4} x \sqrt{-g} \left[R - 2\Lambda \right] + S_{4}^{G}
\end{equation}
the equations of motion follow from the standard variational principle, with that for the scalar being given by
\begin{equation}\label{eq:eomscalar}
\begin{aligned}
\mathcal{E}_{\phi}&=-\mathcal{G}+8 G^{\mu \nu} \nabla_{\nu} \nabla_{\mu} \phi+8 R^{\mu \nu} \nabla_{\mu} \phi \nabla_{\nu} \phi-8(\square \phi)^{2}\\
&+8(\nabla \phi)^{2} \square \phi+16 \nabla^{a} \phi \nabla^{\nu} \phi \nabla_{\nu} \nabla_{\mu} \phi +8\nabla_{\nu} \nabla_{\mu} \phi \nabla^{\nu} \nabla^{\mu} \phi\\
&= \; 0
\end{aligned}
\end{equation}
and the variation with respect to the metric yields
\begin{equation}\label{eq:eommetric}
\begin{aligned}
&\mathcal{E}_{\mu \nu} =\Lambda g_{\mu \nu}+G_{\mu \nu}+\alpha\Bigg[\phi H_{\mu \nu}-2 R\left[\left(\nabla_{\mu} \phi\right)\left(\nabla_{\nu} \phi\right)+\nabla_{\nu} \nabla_{\mu} \phi\right]\\
&+8 R_{(\mu}^{\sigma} \nabla_{\nu)} \nabla_{\sigma} \phi+8 R_{(\mu}^{\sigma}\left(\nabla_{\nu)}\phi\right)\left(\nabla_{\sigma} \phi\right) -2 G_{\mu \nu}\left[(\nabla \phi)^{2}+2 \square \phi\right]\\
&-4\left[\left(\nabla_{\mu} \phi\right)\left(\nabla_{\nu} \phi\right)+\nabla_{\nu} \nabla_{\mu} \phi\right] \square \phi+8\left(\nabla_{(\mu} \phi\right)\left(\nabla_{\nu)} \nabla_{\sigma} \phi\right) \nabla^{\sigma} \phi\\
&-\left[g_{\mu \nu}(\nabla \phi)^{2}-4\left(\nabla_{\mu} \phi\right)\left(\nabla_{\nu} \phi\right)\right](\nabla \phi)^{2}+4\left(\nabla_{\sigma} \nabla_{\nu} \phi\right)\left(\nabla^{\sigma} \nabla_{\mu} \phi\right)\\
&-4 g_{\mu \nu} R^{\sigma \rho}\Big[\nabla_{\sigma} \nabla_{\rho} \phi+\left(\nabla_{\sigma} \phi\right)\left(\nabla_{\rho} \phi\right)\Big]-2 g_{\mu \nu}\left(\nabla_{\sigma} \nabla_{\rho} \phi\right)\left(\nabla^{\sigma} \nabla^{\rho} \phi\right)\\
&-4 g_{\mu \nu}\left(\nabla^{\sigma} \phi\right)\left(\nabla^{\rho} \phi\right)\left(\nabla_{\sigma} \nabla_{\rho} \phi\right)+4 R_{\mu \nu \sigma \rho}\left[\left(\nabla^{\sigma} \phi\right)\left(\nabla^{\rho} \phi\right)+\nabla^{\rho} \nabla^{\sigma} \phi\right] \\
&+2 g_{\mu \nu}(\square \phi)^{2}\Bigg]=\;0
\end{aligned}
\end{equation}
where $H$ is  the Gauss-Bonnet tensor:
\begin{equation}\label{eq:gbtensor}
    \begin{aligned}
    H_{\mu \nu}&=2\Big[R R_{\mu \nu}-2 R_{\mu \alpha \nu \beta} R^{\alpha \beta}+R_{\mu \alpha \beta \sigma} R_{\nu}^{\alpha \beta \sigma}-2 R_{\mu \alpha} R_{\nu}^{\alpha}&\\
    &-\frac{1}{4} g_{\mu \nu}\left(R_{\alpha \beta \rho \sigma} R^{\alpha \beta \rho \sigma}-4 R_{\alpha \beta} R^{\alpha \beta}+R^{2}\right)\Big].
    \end{aligned}
\end{equation}

These field equations satisfy the following relationship
\begin{equation}\label{eq:fieldeqntrace}
0=g^{\mu \nu} \mathcal{E}_{\mu \nu}+\frac{\alpha}{2} \mathcal{E}_{\phi}=4 \Lambda-R-\frac{\alpha}{2} \mathcal{G}
\end{equation}
which can act as a useful consistency check to see whether prior solutions generated via the Glavin/Lin method are even possible solutions to the theory. 
For example, using \eqref{eq:fieldeqntrace}
it is easy to verify that the rotating metrics generated from a Newman-Janis algorithm \cite{kumar2020,wei2020} are not solutions to the field equations of the 4DEGB theory. 

Despite much exploration of the theory \cite{Fernandes:2022zrq}, there has been relatively little work investigating rotating black hole solutions. Attempts have been made using a naïve rescaling of the 4DEGB coupling constant \cite{papnoi2021}, or by implementing the Newman-Janis approach \cite{kumar2020,heydari2021,wei2020}, neither of which produce valid solutions to this theory in general (the latter case not satisfying the positive energy condition). 
A notable exception is a recently obtained class of asymptotically flat slowly rotating black hole solutions \cite{charmousis2022} that were obtained for the  4DEGB theory. However, a detailed study of the geodesics of particles surrounding such black holes (particularly in de Sitter and anti-de Sitter spacetimes) was not done.

In this paper we address this issue, obtaining slowly rotating black hole solutions to the field equations of the 4DEGB theory in asymptotically flat/(A)dS space. We then analyze their  physical properties (innermost stable circular orbits, photon rings, black hole shadow, etc.) to see how they differ from  standard results in Einstein gravity. We find for $\Lambda \leq 0$ that 4DEGB gravity enforces a minimum mass on the black hole solutions, whereas for $\Lambda > 0$ both a minimum \textit{and} maximum mass occur, whose horizon structure is directly analogous to that of Reissner-Nordström de Sitter spacetime. Besides this, the results are similar in form to general relativity. The incredible empirical success of GR necessitates this similarity of solutions for a correct theory, but makes differentiation via measurement difficult. We find that the greatest deviations from GR are typically in the low (near-minimal) mass regime, motivating a search for the smallest observable astrophysical black holes.

\section{Solutions}

\subsection{Metric Functions}

To construct slowly rotating solutions for the new 4DEGB theory, we begin with the following metric ansatz:
\begin{equation}\label{sloro}
\begin{aligned}
d s^{2}&=-f(r) d t^{2}+\frac{d r^{2}}{h(r)}+2 a r^{2} p(r) \sin ^{2} \theta d t d \phi\\
&+r^{2}\left[d \theta^{2}+\sin^{2} \theta d \phi^{2}\right]
\end{aligned}
\end{equation}
where $a$ is a small parameter governing the rate of rotation. Of particular interest are Schwarzschild-like solutions where $h(r) = f(r)$. With this, and the substitution $x = \cos\theta$, our line element can be written
\begin{equation}\label{eq:metric}
\begin{aligned}
	ds^2 &= -f(r) dt^2 + \frac{dr^2}{f(r)}+2ar^2p(r)(1-x^2)dtd\phi\\
 &+r^2\Big[\frac{dx^2}{1-x^2}+(1-x^2)d\phi^2\Big].
 \end{aligned}
\end{equation}

Inserting this into the equations of motion \cref{eq:eomscalar,eq:eommetric} and considering the combination $\mathcal{E}^0_0 - \mathcal{E}^1_1$, we derive the following equation for the scalar field
\begin{equation}\label{phisol}
	(\phi'^2 + \phi'')(1-(r\phi' - 1)^2 f)=0
\end{equation}
which admits either the solution $\phi = \ln(\frac{r-r_0}{l})$ (with $r_0, l$ integration constants), or the solutions
\begin{equation}\label{eq:phi}
	\phi_\pm = \int \frac{\sqrt{f} \pm 1}{\sqrt{f}r}dr,
\end{equation}
where  the latter solution with a minus sign  reproduces previous results 
\cite{glavan2020} in the spherically symmetric case, and falls off as $1/r$ when $\Lambda = 0$. We shall choose this solution henceforth.

Using \eqref{phisol} we can solve for the metric function $f(r)$ from the geometric expression (\ref{eq:fieldeqntrace}). It can easily be shown that
\begin{equation}
	r^2 (1-f(r))-\frac{\Lambda  r^4}{3}+\alpha  f(r)^2-2 \alpha  f(r)-\alpha  C_2 r + C_1 = 0.
\end{equation}
As we wish to recover the Schwarzschild-AdS solution when $\alpha=0$, we set $C_1=\alpha$ and $C_2=\frac{2 M}{\alpha}$ yielding
\begin{equation}\label{eq:f}
	f_\pm = 1 + \frac{r^2}{2 \alpha} \Big( 1 \pm \sqrt{1 + \frac{8 \alpha M}{r^3} + \frac{4}{3} \alpha \Lambda } \Big)
\end{equation}
where the $f_-$ (or Einstein) branch is the one yielding the Schwarzschild AdS solution in the limit $\alpha \rightarrow 0$. From here it is straightforward to show that $\phi_-$ falls off as $1/r$ when $\Lambda=0$, whereas all other solutions for $\phi$ diverge logarithmically at large $r$. \newline

Remarkably the solution \eqref{eq:f} is still valid to leading order in $a$.  The only remaining independent equation from \eqref{eq:eommetric} to this order is given by $\mathcal{E}_{03}$:
\begin{equation}
	r p'' \left(24 \alpha  M+r^3 (4 \alpha  \Lambda +3)\right)+4 p' \left(15 \alpha  M+r^3 (4 \alpha  \Lambda +3)\right)=0
\end{equation}
and admits the exact solution
\begin{equation}\label{psol}
	p=C_2-C_1\frac{ \sqrt{1+\ \frac{8 \alpha  M}{r^3}+\frac{4}{3} \alpha  \Lambda}}{12 \alpha  M }.
\end{equation}

We require that the metric  match the slowly rotating Kerr-(A)dS metric function $p$  in the large $r$ limit. An expansion of the function $f_-(r)$ in this limit indicates that  
\begin{equation}\label{Leff}
\Lambda_\mathrm{eff} =  \frac{3}{2 \alpha }\left[  \sqrt{1+ \frac{4 \alpha  \Lambda}{3}}
-1\right]
= \frac{2\Lambda}{1+\sqrt{1+ \frac{4 \alpha  \Lambda}{3}}}
\end{equation}
is the effective cosmological constant, yielding
\begin{equation}\label{c1c2}
C_1 = 2  M \sqrt{12 \alpha  \Lambda +9}
\qquad
C_2=\frac{3+ 4\alpha  \Lambda -2\alpha \Lambda_\mathrm{eff} }{6 \alpha }
\end{equation}
This leaves us with the final expression
\begin{align}\label{eq:p}
	p(r) &=  {\frac{2\Lambda}{3}} 
 \left( \frac{\sqrt{1+ \frac{4 \alpha  \Lambda}{3}}}{1+\sqrt{1+ \frac{4 \alpha  \Lambda}{3}}}\right) \\
&\quad + \frac{1}{2 \alpha }\left[ 1 -  \sqrt{1+ \frac{4 \alpha  \Lambda}{3}} \sqrt{1 + \frac{4 \alpha}{3}  \left(\Lambda +\frac{6 M}{r^3}\right)} \right]
\nonumber
\end{align}
which, in the $\Lambda=0$ limit, matches\footnote{up to an errant factor of the rotation parameter included in the  metric function in \cite{charmousis2022}} the result  \cite{charmousis2022} for asymptotically flat spacetime. Additionally we note that we can rewrite the metric functions in terms of 
$\Lambda_\mathrm{eff}$ as
\begin{align}\label{eq:feff}
	f_\pm &= 1 + \frac{r^2}{2 \alpha} \left( 1 \pm \sqrt{ \left(\frac{2}{3} \alpha \Lambda_\mathrm{eff} +1\right)^2 + \frac{8 \alpha M}{r^3} } \right) \\
\label{eq:peff}
	p(r) &=  {\frac{\Lambda_\mathrm{eff}}{3}}
 \left(\frac{2}{3} \alpha \Lambda_\mathrm{eff} +1\right)
 \\
&+ \frac{1}{2 \alpha }\left[ 1 -   \left(\frac{2}{3} \alpha \Lambda_\mathrm{eff} +1\right)
\sqrt{ \left(\frac{2}{3} \alpha \Lambda_\mathrm{eff} +1\right)^2 + \frac{8 \alpha M}{r^3} } \right].
\nonumber
\end{align}
 
It is straightforward to show that the metric \eqref{sloro} is asymptotically of constant curvature, with 
\begin{equation}
    R_{\mu\nu} \to \Lambda_\mathrm{eff} g_{\mu\nu}
\end{equation}
as $r\to \infty$.
 
We pause to comment on the range of validity of the metric
\eqref{sloro} (with
$f_-(r)$ and $p(r)$ given by respectively by \eqref{eq:f} and
\eqref{eq:p}), which  
  is within the  class of generalized Lense-Thirring metrics recently discussed in \cite{Gray:2021roq}. Unlike Einstein gravity, the solution \eqref{sloro} is characterized by two  metric functions, as was shown for this class 
\cite{Gray:2021roq}.  It
generalizes the asymptotically flat case previously obtained \cite{charmousis2022} 
and reduces to the slowly rotating Kerr-(A)dS metric if
$\alpha=0$. 
 We first note that if $M=0$ the metric 
is that of a  spacetime
of constant curvature $\Lambda_\mathrm{eff}$
in rotating coordinates, and that if $a=0$ the metric is an exact solution to 
 the field equations \eqref{eq:eomscalar}, \eqref{eq:eommetric}, with  event horizons 
given by the  real solutions of
\begin{equation}\label{horsol1}
    \Lambda r^4 - 3 r^2 + 6Mr - 3\alpha = 0
\end{equation}
where $f_-(r) = 0$. 
If $\Lambda>0$ there are, in decreasing order of magnitude, three solutions  $r_c$, $r_+$, and $r_{-}$ to \eqref{horsol1}; if 
$\Lambda<0$ only the latter two are present. In both cases $r_+$ is the outer horizon of the black hole.
Note that the values of these  solutions to
\eqref{horsol1} are independent of $a$, as is the case for the Kerr solution to leading order in $a$,
though the latter has only a single horizon in this approximation. 
Using analytic continuation via a Kruskal-type extension to continue
to values of $r<r_+$, our solution
will be valid for all values of $r>r_{-}$ provided $a << r_{-}$.    In analyzing the structure of the metric, we shall make this assumption\footnote{It is possible to consider values of the scalar field for $r<r_+$
if the solution \eqref{eq:phi} is extended to the time-dependent solution 
$\phi  = qt + \int \frac{f-\sqrt{f+q^2 r^2} }{f r}dr$
where $q$ is a constant of integration. Remarkably this
does not affect the solution for the metric \cite{charmousis2022}.}. However phenomenologically we need only consider physics in regions where $r>r_+$, in which case it is sufficient to require 
$a << r_+$, ensuring that the $g_{t\phi}$ component of the metric remains small in this region. For $\alpha=0=\Lambda$ this criterion becomes $a < 2m$.

\subsection{Analytic Properties}

The slowly rotating 4DEGB metric 
\eqref{eq:metric}, with
$p$ given by \eqref{eq:p}
and $f=h$ given by
\eqref{eq:f}, 
is singular near $r=0$, which can be seen by computing the Ricci scalar to lowest order in $r$. In doing so we find that
\begin{equation}
	R(r)_{r\rightarrow 0} \sim \frac{15}{4}\sqrt{\frac{2M}{\alpha r^3}}
\end{equation}
regardless of the value of $\Lambda$. We  consider here only solutions where this singularity is  behind the horizon. Computed explicitly to leading order in the rotation parameter, the Kretschmann scalar is
\begin{equation}
    K = f''(r)^2+\frac{4 f'(r)^2}{r^2}+\frac{4 (f(r)-1)^2}{r^4},
\end{equation} 
and the only other condition leading to a curvature singularity is
\begin{equation}
	4 \alpha  \Lambda +\frac{24 \alpha  M}{r^3}+3 = 0
\end{equation}
which always occurs inside the horizon and is never true  since
the quantity $4 \alpha  \Lambda  +3 \geq 0$ in the allowed region of $(\alpha, \Lambda)$ parameter space (see section \ref{sec:propsoln}).
Therefore the  4DEGB metric is regular everywhere but the black hole singularity. 

The metric \eqref{sloro} has two Killing vectors $\xi_{(t)} = \partial/\partial t$ and $\xi_{(\phi)} = \partial/\partial \phi$. 
It can be shown easily that  
parameter $M= M_K \equiv {\cal Q}(\xi_{(t)})$ 
is the Komar mass, 
where 

\begin{equation}
{\cal Q}(\xi) = \frac{1}{8\pi}\int_{\partial\Sigma} \star K  
= \frac{c}{8\pi}\int_{\partial\Sigma}\star \left[ d\xi + \frac{3\nabla^2 d\xi}{2\Lambda_\mathrm{eff}}\right]
\end{equation}
is the 
Komar charge \cite{peng_2021_komar}
 associated with the Killing form
 $\xi = \xi_\mu dx^\mu$
 (with $\star$ the Hodge dual);  the normalization constant $c = \frac{1}{2} \sqrt{1+\frac{4}{3}\alpha\Lambda}$, and  $\partial\Sigma$ is the $2$-surface of constant $t$ at spatial infinity.  Likewise, it is straightforward to show that $J=2 M a c$ is the angular momentum associated with the Komar charge 
${\cal Q}(\xi_{(\phi)})$.


For black hole solutions, the scalar field \eqref{eq:phi} is ill-defined inside the horizon where $f(r) < 0$. Such solutions can be made regular across the horizon \cite{charmousis2022}. 
As we are primarily interested in the phenomenological aspects of the slowly rotating solutions,
we shall  focus on exterior solutions in the sequel.  

The structure of the field equations is such that all quantities can be rescaled into a unitless form, relative to some length scale. Writing 
\begin{equation}
	\Lambda = \pm \frac{3}{L^2}
\end{equation}
where the positive and negative branches correspond to de Sitter and anti-de Sitter space respectively, and $L$ is the Hubble length,  we shall perform the rescalings:
\begin{align}
    &\alpha \rightarrow \bar{\alpha} L^2 \qquad
	M \rightarrow \bar{M} L 
	\qquad
	r \rightarrow \bar{r} L  \qquad
	a \rightarrow \bar{a} L  
\end{align}
in units of $L$.  The 4DEGB metric functions then become
 	\begin{equation}
	\bar{f}(\bar{r}) = 1 + \frac{\bar{r}^2}{2 \bar{\alpha}} \left( 1 - \sqrt{1 \pm 4 \bar{\alpha} + \frac{8 \bar{\alpha} \bar{M}}{\bar{r}^3}  } \right)
\end{equation}
and $r^2 p(r) \to \bar{r}^2\bar{p}(\bar{r})$, where
\begin{equation}
	\bar{p}(\bar{r}) = \pm \frac{2\sqrt{1 \pm 4 \bar{\alpha}}}{1+\sqrt{1 \pm 4 \bar{\alpha}}} + \frac{1}{2 \bar{\alpha} } \left( 1 - \sqrt{1\pm 4 \bar{\alpha} } \sqrt{1 \pm 4 \bar{\alpha} + \frac{8 \bar{\alpha}  \bar{M}}{\bar{r}^3}} \right).
\end{equation}
 
If instead we consider an asymptotically flat spacetime, we can directly set $\Lambda = 0$ in \cref{eq:f,eq:p}  and 
can rescale all quantities in units of some fiducial mass. 
In what follows we shall take this to be a solar mass.

\section{Properties of the Solution}\label{sec:propsoln}

In this section we study physical properties of the solutions derived above. We discuss the location and angular velocity of the black hole horizons, the equatorial geodesics -- including the innermost stable circular orbit, photon rings (and associated Lyapunov exponents) -- as well as the black hole shadow. In each case we compare the properties of the 4DEGB solution for multiple values of the coupling constant to the analogous GR result, and discuss how the rotational corrections affect these properties.

\subsection{Location and Angular Velocity of the Black Hole Horizons}

The angular velocity of the black hole horizon is defined as 
\begin{equation}
	\Omega_h = -\frac{g_{t\phi}}{g_{\phi \phi}}|_{r=r_h} = -a p(r_h) 
\end{equation}
where $r_h$ is the radius at which $g_{rr}$ diverges (ie. $f(r_h)=0$).  

To determine the locations of the horizons, it is convenient to rewrite the metric function $f_{-}(r)$ as 
\begin{equation}\label{fFdef}
	f_-(r) = \frac{r^2}{\alpha} \frac{F(r)}{f_+(r)}
\end{equation}
where 
$$
F(r) = 1 - \frac{2 M}{r} + \frac{\alpha}{r^2} - \frac{1}{3} \Lambda r^2 \; .
$$  
The denominator $f_+$ does not vanish, and the horizons are given by the roots of the numerator, which obey the equation
\begin{equation}\label{horsol}
   \Lambda r^4_h - 3 r^2_h + 6 M   r_h - 3 \alpha  = 0
\end{equation}
which has exact solutions for all values of $\Lambda$.

\subsubsection{$\mathbf{\Lambda = 0}$}

In asymptotically flat space, \eqref{horsol} admits the following simple solutions:
\begin{equation}
	r_h = M \pm \sqrt{M^2 - \alpha}
\end{equation}
with $r_+$ (the outer horizon) recovering the Schwarzschild value as $\alpha \rightarrow 0$. Assuming $\alpha > 0$, this equation sets a minimum value for black hole mass in the theory, namely 
\begin{equation}\label{eq:mminl0}
M_{min} = \sqrt{\alpha}
\end{equation}
when $\Lambda=0$. For smaller masses, the metric function $f(r)$ does not vanish anywhere and thus no horizon exists. These solutions have naked singularities.  
 
\begin{figure}[H]
	\includegraphics[width=8cm]{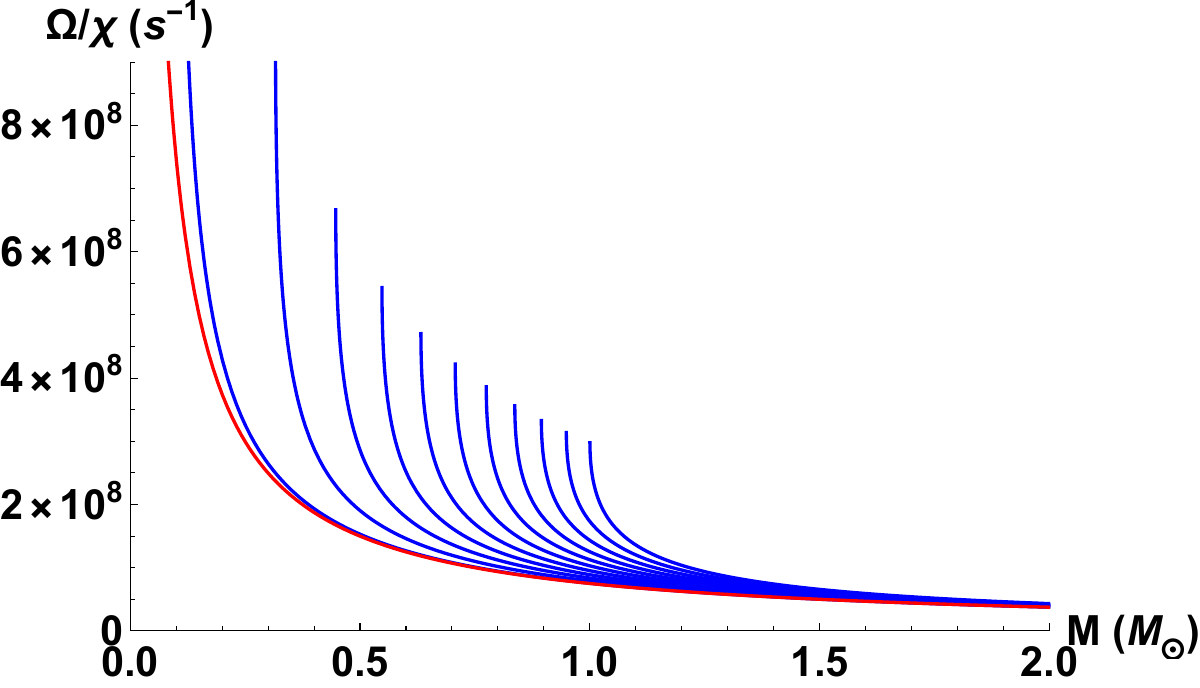}
	\centering
	
	\caption[Angular velocity of the black hole horizon when $\Lambda = 0$.]{Angular velocity of the black hole horizon as a function of mass when $\Lambda = 0$ for $\alpha/M_\odot^2$ = 0.01, 0.1, 0.2, 0.3, 0.4, 0.5, 0.6, 0.7, 0.8, 0.9, 1 (in blue from left to right) plotted against the Einstein ($\alpha=0$) solution $\Omega = \chi/4M$ (in red).
	}
	
	\label{fig:hav l0}
\end{figure}

Since we have an exact solution for $p(r)$ from  \eqref{eq:p} and a simple analytic form for $r_h$, the angular velocity of the   horizon 
\begin{equation}
	\Omega_h = a    \frac{1 - \sqrt{1 + \frac{8\alpha M}{\left(M + \sqrt{M^2 - \alpha}\right)^3}}}{2\alpha}
\end{equation}
is straightforward to compute.
In figure \ref{fig:hav l0} we plot $\frac{\Omega}{\chi}$ (in units of
inverse seconds) as a function
of $M$ (in units of solar mass) 
 for a variety of values of $\alpha$ alongside the GR solution for comparison (where $\chi = a/M$). The main new feature introduced by the 4DEGB theory is the existence of a maximal angular velocity, indicated by the termination points of the blue curves at any given $\alpha$, due to the presence of a minimum mass. As $\alpha$ increases, this maximal value decreases.  

For a fixed $\alpha$ we observe that the 4DEGB theory predicts a significantly larger angular velocity at a given (small) mass than does GR, but quickly converges to the GR result when the mass is large.

\subsubsection{AdS ($\mathbf{\Lambda < 0}$)}\label{subsec:adshorizons}

\begin{figure}

\subcaptionbox[Location of the black hole horizon in AdS.]{The locations of the anti-de Sitter black hole horizon plotted as a function of mass. The red line represents the solution from GR (ie. when $\alpha=0$), and the blue lines represent the 4DEGB solutions for $\frac{\alpha}{\alpha_C} = 0.01$, $0.1$, $0.4$, $0.7$, $0.85$, $1$ from left to right. We see that as soon as a nonzero coupling constant is introduced, the horizon structure includes an inner horizon and a minimum mass point. \label{fig:horizons ads}}
{\includegraphics[width=8cm]{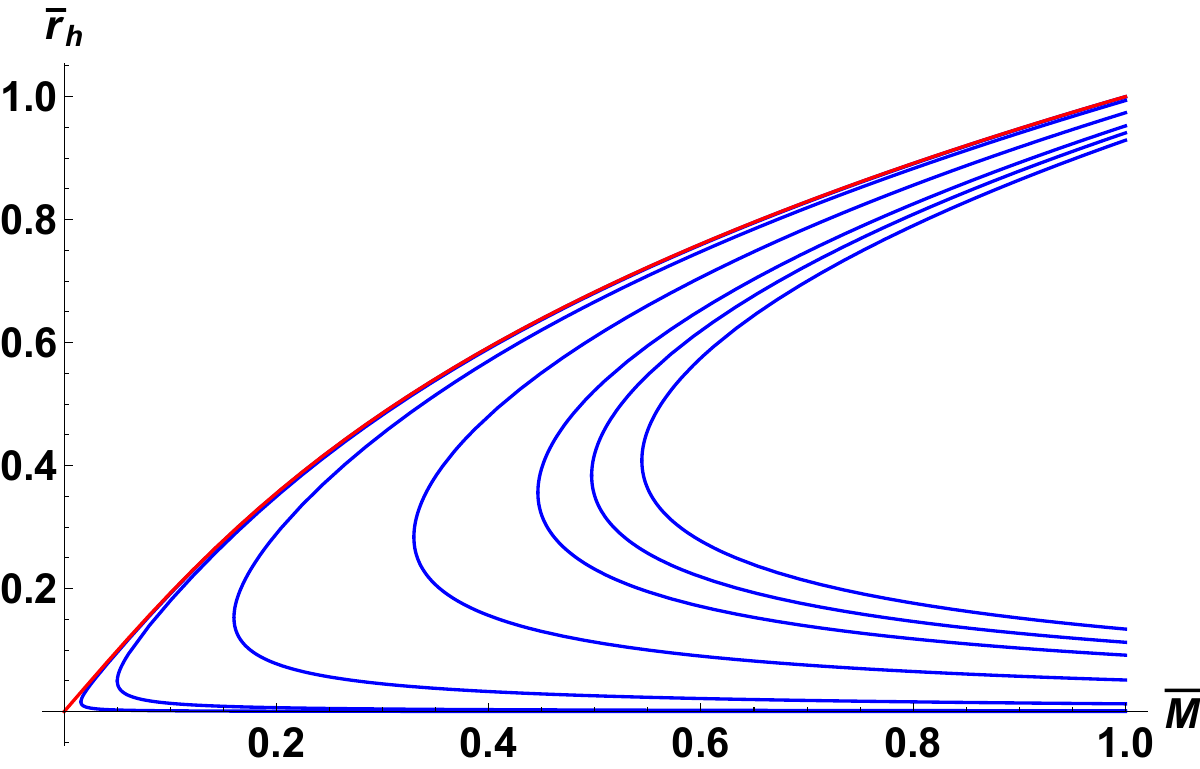}}
\hspace{0.05\textwidth} 
\subcaptionbox[Angular velocity of the black hole horizon in AdS.]{Angular velocity of the black hole horizon in the AdS case as a function of mass for $\frac{\alpha}{\alpha_C}$ = 0.01, 0.1, 0.2, 0.3, 0.4, 0.5, 0.6, 0.7, 0.8, 0.9, 0.95, 0.99, 1 in blue from left to right, plotted against the Einstein ($\alpha=0$) solution $\Omega = \chi/4M$ (in red). \label{fig:hav ads}}
{\includegraphics[width=8cm]%
    {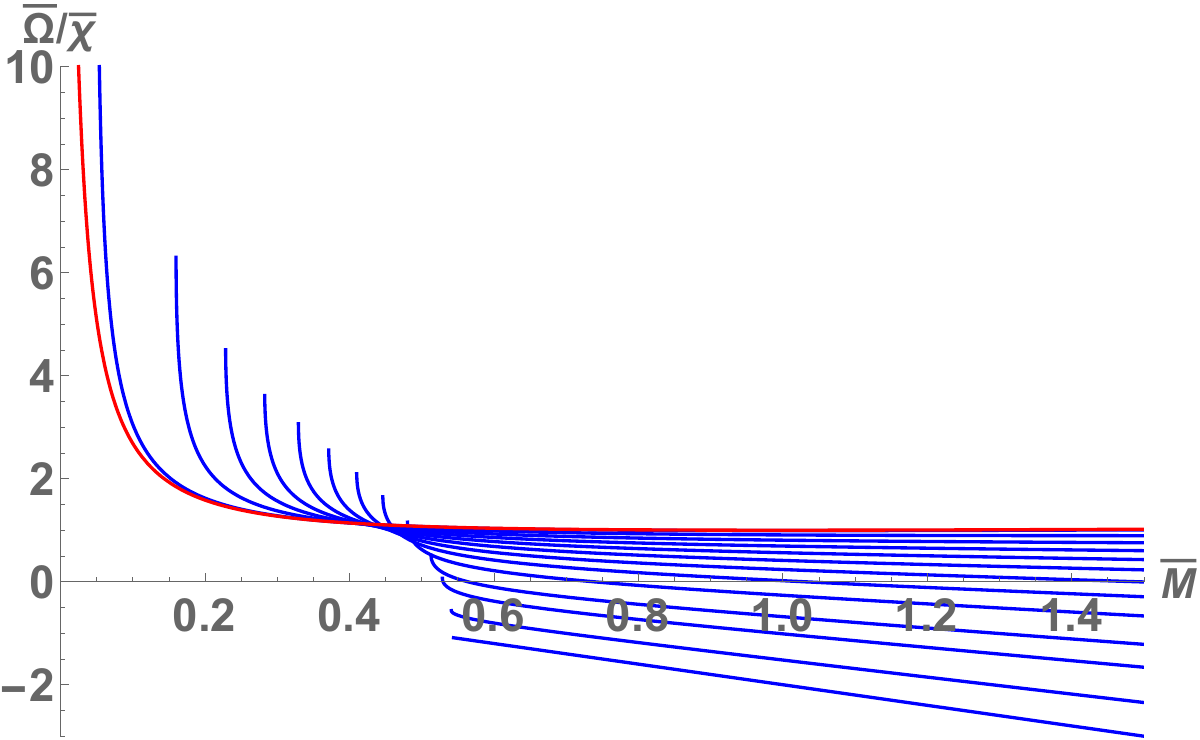}%
}
\caption{Black hole horizons and angular velocities in AdS.}
\end{figure}

Solutions to \eqref{horsol} yield  rather cumbersome expressions for the two different positive values of $r_h$.  For this reason, we solve numerically for the horizon as a function of black hole mass,  illustrating the results in figure \ref{fig:horizons ads}. As in the asymptotically flat case, a minimum mass black hole exists at which the inner and outer horizons merge, given by
\begin{eqnarray}\label{eq:mminneg}
	{M_{min}}&=&\frac{\sqrt{1+ 12\alpha\Lambda - \sqrt{( 1-4\alpha\Lambda)^3}}}{3\sqrt{2\Lambda}}
\end{eqnarray}
which yields \eqref{eq:mminl0} in the limit  $\Lambda \rightarrow 0$.  

There are also upper and lower bounds on the Gauss-Bonnet coupling constant $\alpha$ for
any fixed $\Lambda$. These are given by the two conditions $4 \alpha  \Lambda +3>0$ (so that $p(r)$ is real) and $1 - 4 \alpha \Lambda > 0$ (so that $M_{min}$ is real). This corresponds to
\begin{equation}
	-\frac{3}{4} \leq \alpha \Lambda \leq \frac{1}{4}.
\end{equation}
Since we restrict ourselves to positive values of the 4DEGB coupling constant, only one of these inequalities is relevant when $\Lambda$ is non-zero, depending on its sign. In asymptotically anti-de Sitter space, $\Lambda
=-|\Lambda|$ and these inequalities reverse. We can then define a critical value for the coupling constant at its upper limit:
\begin{eqnarray}\label{eq:adsalphacrit}
	\alpha_C = \frac{3}{4 |\Lambda|}.
\end{eqnarray}

The angular velocity at the horizon is
\begin{equation}\label{OmGB}
   \Omega_h =  -\frac{4 a \Lambda^2\alpha}{3(2\alpha\Lambda + \sqrt{12\Lambda\alpha + 9} + 3)} + \frac{a\sqrt{12\Lambda\alpha + 9}}{3 r_h^2}
\end{equation}
where the horizon radius $r_h$ is obtained from setting $f_-(r_h) = 0$. 
In figure \ref{fig:hav ads} we plot   $\Omega_h$ for the AdS case for various allowed values of the coupling constant $\alpha$. We again observe the existence of a maximal angular velocity, indicted by the termination points of the blue curves at any given $\alpha$ due to the presence of a minimum mass.  Interestingly, for large enough $\alpha$ the angular velocity of the horizon can become negative due to frame dragging effects. This behaviour can also be noted in the behaviour of equation \eqref{OmGB}, where in the large $\alpha/M$ regime the negative constant term starts to dominate.





Another interesting feature is the existence of a crossover mass ($\bar{M}_{\rm{x}}$) at which the small $\alpha$ 4DEGB solutions become less than the GR solutions. We can find this crossover point analytically for solutions where $\bar{\alpha} << 1$. We begin by fixing a black hole mass, and describe it using both the GR black hole horizon radius ($\bar{r}_{h(GR)}$), and the 4DEGB horizon radius ($\bar{r}_{h(GB)}$):
\begin{equation}\label{eq:ads fixed mass}
2\bar{M} = \bar{r}_{h \rm{(GR)}} + \bar{r}_{h \rm{(GR)}}^3 = \bar{r}_{h \rm{(GB)}} + \bar{r}_{h \rm{(GB)}}^3 +  {\frac{\bar{\alpha}}{ \bar{r}_{h \rm{(GB)}}}}.
\end{equation}

The horizon angular velocity of an Einsteinian black hole can then be written as
\begin{equation}
\frac{\bar{\Omega}_{GR}}{\bar{\chi}} = \frac{\bar{M} \bar{r}_{h \rm{(GR)}}}{2 \bar{M}-\bar{r}_{h \rm{(GR)}}}.
\end{equation}
 Similarly, from \eqref{OmGB} this quantity for a 4DEGB black hole can be expanded in a power series in $\bar{\alpha}$:
\begin{equation}\label{eq:ads hav gb series}
\begin{aligned}
\frac{\bar{\Omega}_{GB}}{\bar{\chi}} &= 
\frac{\bar{\Omega}_{GR}}{\bar{\chi}}
+ \frac{1-2 \left(\bar{r}_{h \rm{(GB)}}^3+\bar{r}_{h \rm{(GB)}}\right)^2}{2 \bar{r}_{h \rm{(GB)}}^3}\bar{\alpha}\\
&+ \mathcal{O} \left(\bar{\alpha} ^2\right)
\end{aligned}
\end{equation}
after using equation \eqref{eq:ads fixed mass} to replace $M(\bar{r}_{h \rm{(GB)}},\alpha)$ in the higher order terms.  
The crossover  occurs when 
 the subleading contribution in \eqref{eq:ads hav gb series} vanishes,  yielding 
\begin{equation}
\bar{r}_{h \rm{(GB)}} = 0.545121
\end{equation}
which corresponds to a crossover mass of
\begin{equation}
\bar{M}^{\bar{\Omega}}_{\rm{x}} \approx 0.353553 + 0.917228\bar{\alpha}
\end{equation}
to leading order in $\bar{\alpha}$.

\subsubsection{dS ($\mathbf{\Lambda > 0}$)}

If $\Lambda > 0$, the 4DEGB theory returns three positive solutions to \eqref{horsol}: $r_-$ (inner),  $r_+$ (outer), and  $r_c$ (cosmological). The horizon structure is  identical to that of the so-called ``charged Nariai" solutions in the Reissner-Nordstrom de Sitter metric \cite{mann1995, hawking1995,bousso97}, with  the 4DEGB coupling constant $\alpha$ playing a role analogous to that of $Q^2$, where $Q$ is the total charge. Indeed, upon replacing $\alpha$ with $Q^2$, the function $F(r)$ in \eqref{fFdef} is 
equivalent to
the charged Nariai metric, and asymptotes to \cite{bousso97}

\begin{equation}
	F(r) \rightarrow 1 - \frac{1}{3} \Lambda r^2
\end{equation}

as $r\rightarrow\infty$.

\begin{figure}[h]

\subcaptionbox[Location of the black hole horizon in dS.]{The locations of the de Sitter black hole horizons plotted as a function of mass. In all cases the red line represents the solution from GR (ie. when $\alpha = 0$), and the blue lines represent the 4DEGB solutions for $\frac{\alpha}{\alpha_C}$ = 0.01, 0.1, 0.4, 0.7, 1 from left to right. We see that as soon as a nonzero coupling constant is introduced the horizon structure includes an inner horizon and a minimum mass point. Once $\alpha$ passes criticality, no physical black hole solutions can exist. \label{fig:horizonsds}}
{\includegraphics[width=8cm]{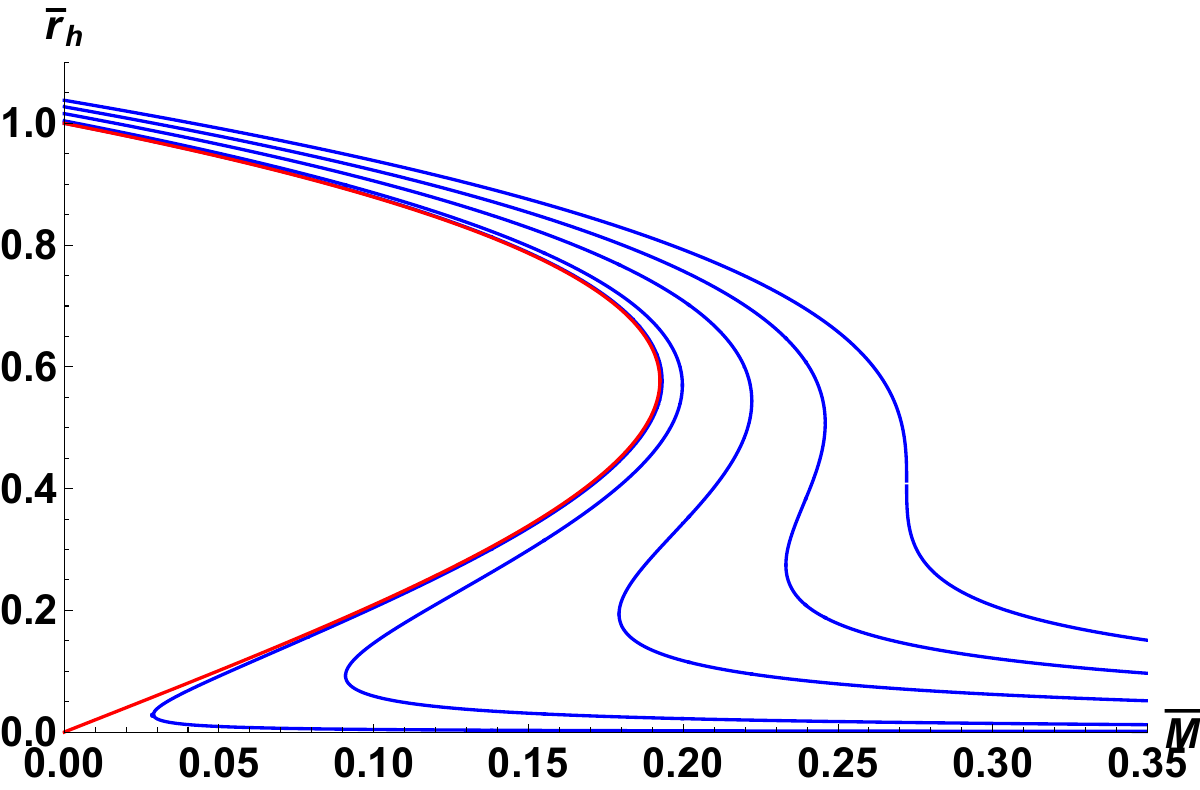}}
\hspace{0.05\textwidth} 
\subcaptionbox[Angular velocity of the black hole horizon in dS.]{Angular velocity of the black hole horizon as a function of mass in an asymptotically de Sitter spacetime for $\frac{\alpha}{\alpha_C}$ = 0.1, 0.2, 0.3, 0.4, 0.5, 0.6, 0.7, 0.8, 0.9, 0.95 in blue from left to right, plotted against the Einstein (${\alpha}=0$) solution $\bar{\Omega} = \bar{\chi}/4\bar{M}$ (in red). \label{fig:hav ds}}
{\includegraphics[width=8cm]{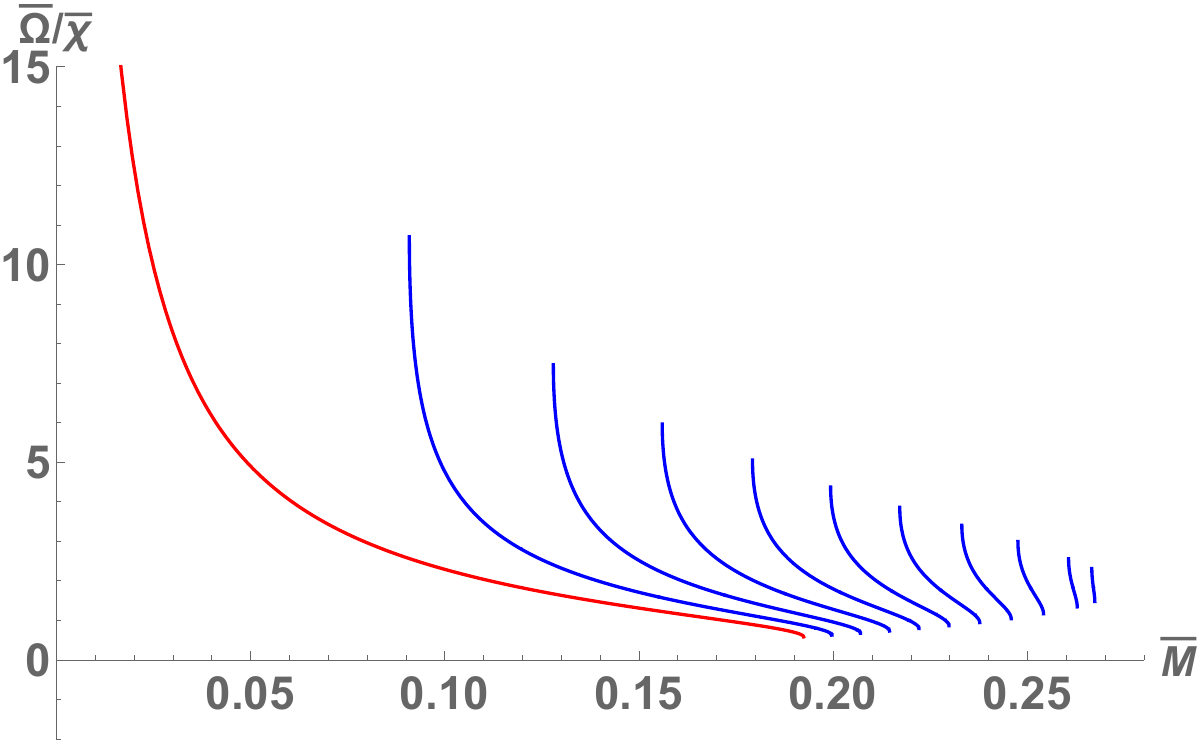}}
\caption{Black hole horizons and angular velocities in dS.}
\end{figure}

Unlike the asymptotically flat and AdS cases, the de Sitter 4DEGB theory enforces both a minimum mass black hole (below which only a naked singularity surrounded by a cosmological horizon exists), and a maximum mass black hole, corresponding to the merging of the outer and cosmological horizons, which is the Nariai solution. We find the location of these extremal mass points to be
\begin{eqnarray}\label{eq:mmaxminpos}
	{M_{max/min}} &=& \frac{\sqrt{1 + 12\alpha\Lambda \pm \sqrt{(1 - 4\alpha\Lambda)^3}}}{3\sqrt{2\Lambda}}
\end{eqnarray}	
which for $M_{min}$ (the minus value) is identical in form to that of the limiting mass for the Reissner–Nordström-de Sitter black hole \cite{bousso97}.  \newline

From   \eqref{eq:mmaxminpos} we can again set an upper and lower bound on the $\alpha$ parameter for a given $\Lambda$:
\begin{equation}\label{eq:alphalimitsposalph}
	-\frac{3}{4} < \alpha \Lambda < \frac{1}{4},
\end{equation} 
using the same reality criteria as in the AdS case. When $\alpha$ is outside this allowed range we find solution sets that lie in a region of parameter space with no black hole event horizon. Since we restrict ourselves to $\alpha>0$, we can then define a critical value for the coupling constant in de Sitter from the upper bound

\begin{eqnarray}
	\alpha_C = \frac{1}{4 |\Lambda|}.
\end{eqnarray}

Armed with an understanding of the de Sitter horizon structure and the allowed regions of parameter space, we solve numerically for the black hole horizon radius and angular velocity, the results of which are plotted in figures \ref{fig:horizonsds} and \ref{fig:hav ds} respectively. Once again we observe that the presence of a minimum mass yields a maximal angular velocity, indicated by the leftmost termination points of the blue curves for any given $\alpha$. In all cases the angular velocity reaches its global minimum at the maximal mass.  As $\alpha$ increases the allowed range of mass/angular velocity increasingly diminishes, vanishing when $\alpha = \alpha_C$.

\subsection{Geodesics in the Equatorial Plane}

We begin the treatment of equatorial geodesics by setting $x = 0$ ($\theta = \pi/2$) in the metric ansatz \eqref{eq:metric} and multiplying both sides of the line element by $\frac{1}{2}f(r)$, yielding
\begin{equation}
\begin{aligned}
	-\xi\frac{ f(r)}{2} &= -\frac{f(r)^2}{2} \dot{t}^2 + \frac{1}{2} \dot{r}^2 + \frac{1}{2} r^2 f(r)\dot{\phi}^2 \\
 &+  a f(r) P(r) \dot{t} \dot{\phi}
 \end{aligned}
\end{equation}
where the overdot refers to a derivative with respect to the affine parameter $s$, $P(r) = r^2 p(r)$, and $\xi$ is a constant which takes the value 0 or 1, for null and timelike geodesics respectively. Using the static and rotational Killing fields, we know that $E = f(r)\dot{t} - a P(r) \dot{\phi}$ and $j = r^2 \dot{\phi} + a P(r) \dot{t}$ are constants of motion along the geodesic, and thus
\begin{equation}\label{eq:geodesicpt2}
-\frac{1}{2} \xi  f(r) = -\frac{1}{2}E^2+ \frac{1}{2} \dot{r}^2 +\frac{j^2 f(r)}{2 r^2} - a \frac{P(r) j E}{r^2}
\end{equation}
to leading order in $a$.

Asymptotically, $j^2$ represents the total orbital angular momentum of the body following the geodesic. Since we are confined to the equatorial plane, clearly $j^2 = \ell_z^2$ for non-spinning particles. If we rewrite equation \eqref{eq:geodesicpt2} in the following form
\begin{equation}\label{photogeo}
	 \frac{1}{2} \dot{r}^2 + V_{\text{eff}} = 0
\end{equation}
it becomes clear that
\begin{equation}\label{eq:effpot}
    V_{\text{eff}} = \frac{f(r)}{2}\Bigg( \frac{\ell_z^2}{r^2} + \xi \Bigg) - \frac{1}{2}E^2 - a \frac{P(r) \ell_z E}{r^2}
\end{equation}
to leading order in the rotation parameter. Note that the rescaled version of this equation is unchanged besides swapping parameters with their barred counterparts. In the following sections we further specialize this analysis by considering the innermost stable circular orbit (or ISCO) for timelike geodesics, and the photon ring for null geodesics.

\subsection{Timelike Geodesics: Innermost Stable Circular Orbits}

For timelike geodesics we set $\xi = 1$ in \eqref{eq:effpot}. The  condition for the existence of circular geodesics  is
\begin{eqnarray}\label{ISCOcond}
	V'_{\text{eff}}(r)=0 
\end{eqnarray}
where 
the sign of $V''_{\rm{eff}}(r)$ dictates the stability of such orbits. Stable orbits are described by $V''_{\rm{eff}}(r)>0$, while $V''_{\rm{eff}}(r)<0$ indicates instability. 
Of particular interest is the \textit{innermost} stable circular orbit (or ISCO), for which $V''_{\text{eff}}(r)=0$. It is always possible to choose $E$ so that $V(r)=0$ for a circular orbit.

Next, we solve these three equations to leading order in the rotation parameter by making the following perturbative expansions
\begin{eqnarray}
	r_{\text{ISCO}} &=& r_{\text{ISCO}}^{(0)} + a r_{\text{ISCO}}^{(1)} \nonumber\\
	j_{\text{ISCO}} &=& j_{\text{ISCO}}^{(0)} + a j_{\text{ISCO}}^{(1)} \label{iscos}\\
	E_{\text{ISCO}} &=& E_{\text{ISCO}}^{(0)} + a E_{\text{ISCO}}^{(1)}\; . \nonumber
\end{eqnarray}
Substituting these into \eqref{ISCOcond} and expanding to leading order in $a$ yields a system of six equations for the six unknowns on the right-hand side of \eqref{iscos}, which can be solved for numerically. 

\subsubsection{$\mathbf{\Lambda = 0}$}

\begin{figure*}
	\begin{subfigure}{0.5\textwidth}
		\includegraphics[width=8.6cm]{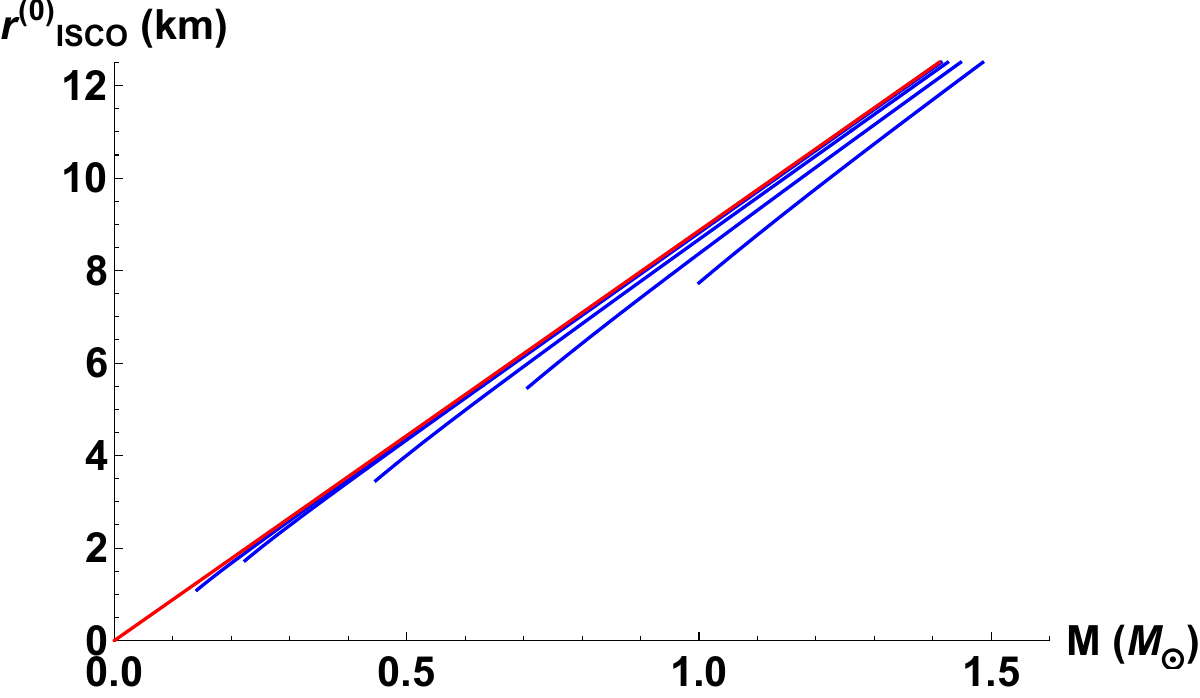}
		\label{fig:}
	\end{subfigure}\hfill%
	\begin{subfigure}{0.5\textwidth}
		\includegraphics[width=8.6cm]{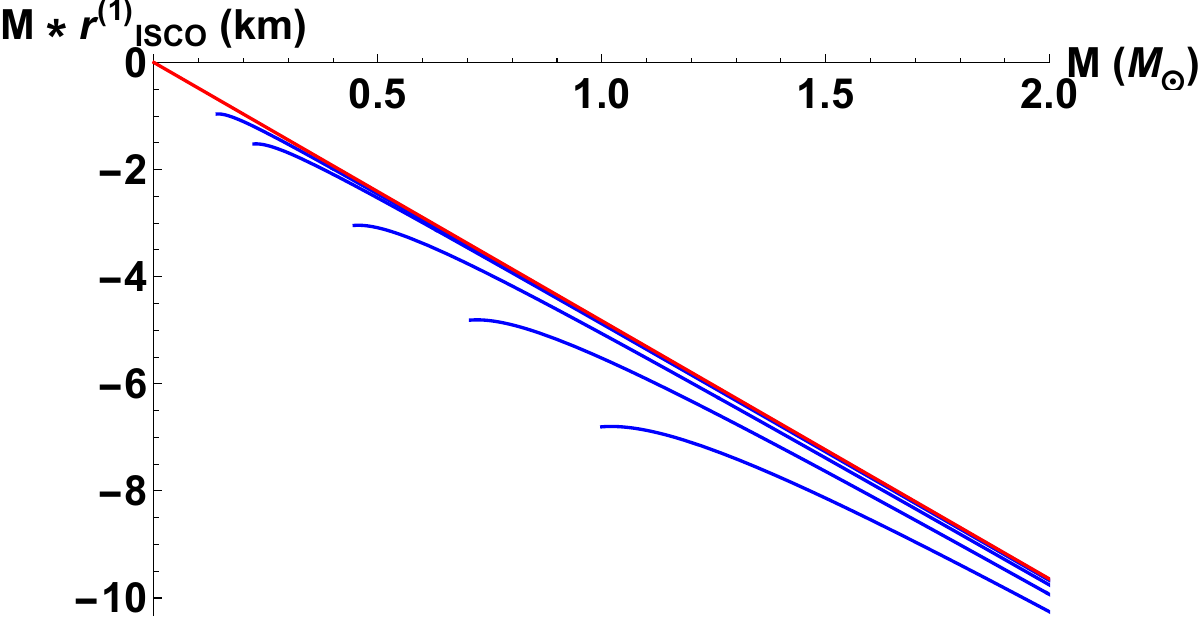}
		\label{fig:}
	\end{subfigure}\hfill%
 \vspace{10mm}
	\begin{subfigure}{0.5\textwidth}
		\includegraphics[width=8.6cm]{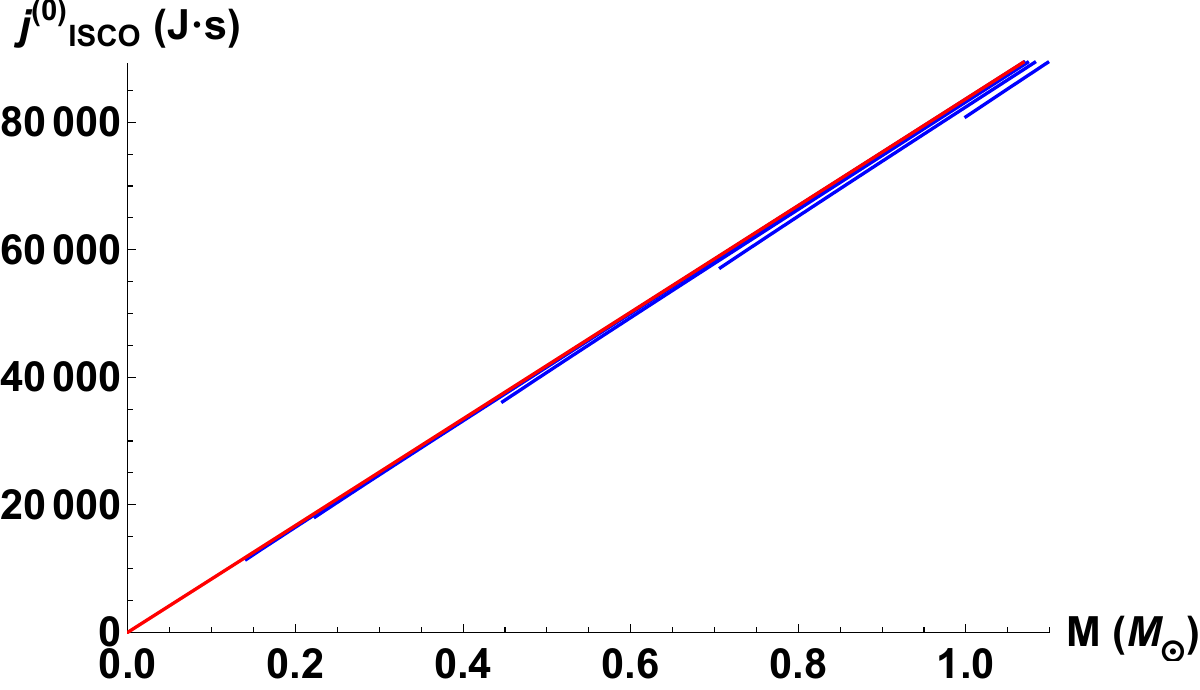}
		\label{fig:}
	\end{subfigure}\hfill%
	\begin{subfigure}{0.5\textwidth}
		\includegraphics[width=8.6cm]{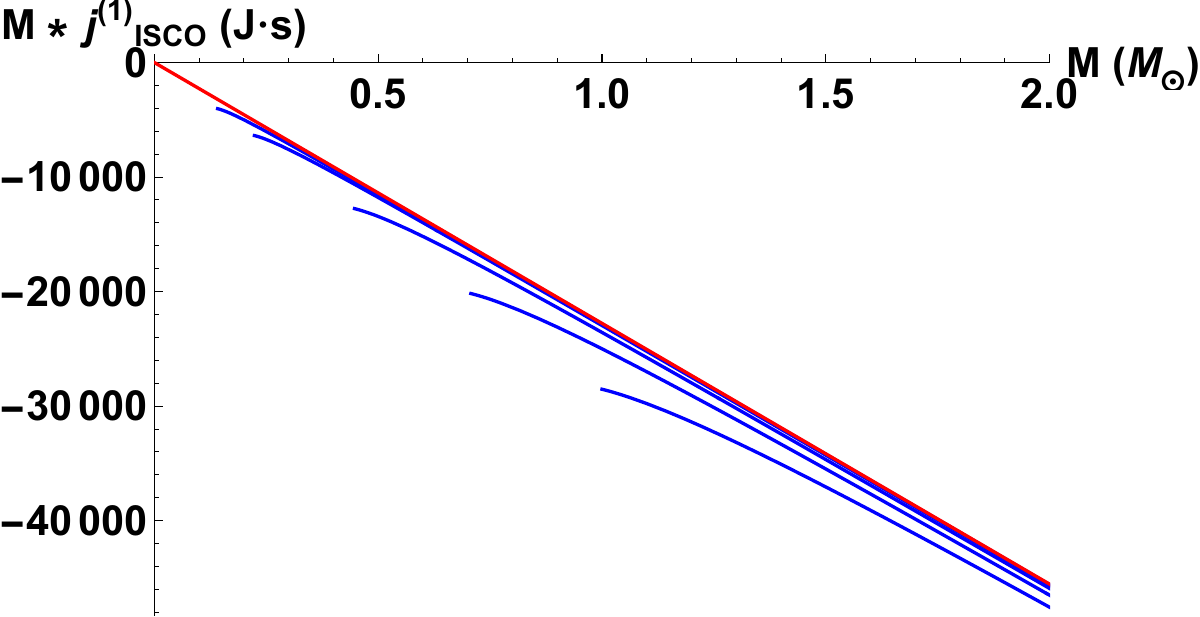}
		\label{fig:}
	\end{subfigure}\hfill%
 \vspace{10mm}
	\begin{subfigure}{0.5\textwidth}
		\includegraphics[width=8.6cm]{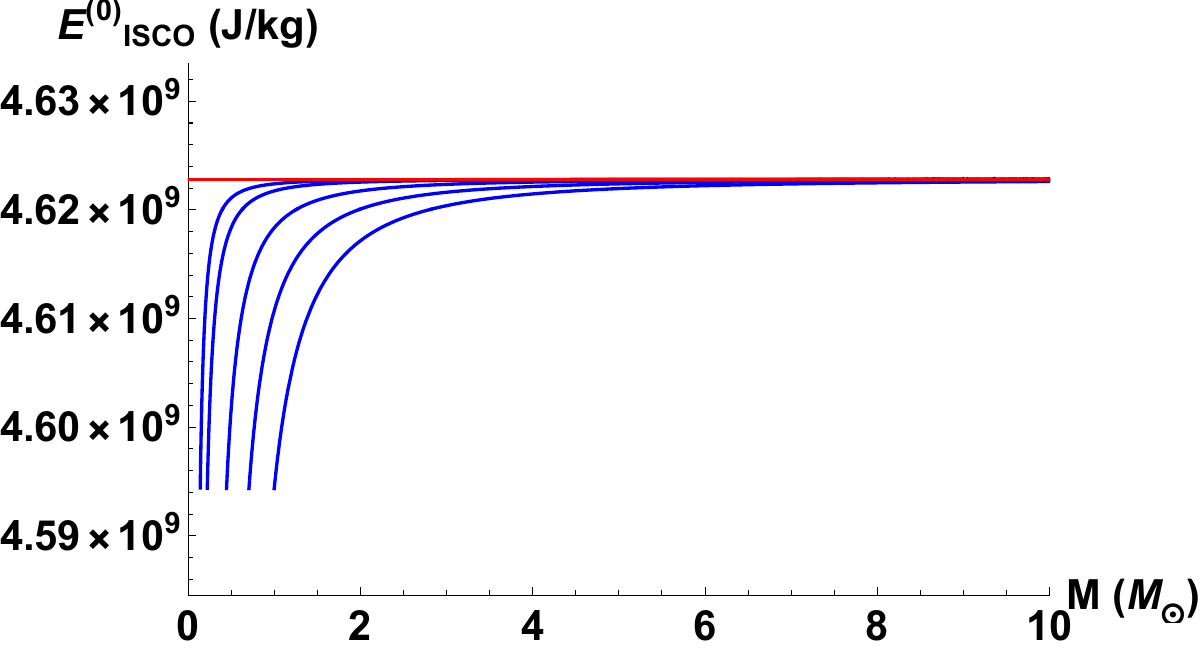}
		\label{fig:}
	\end{subfigure}\hfill%
	\begin{subfigure}{0.5\textwidth}
		\includegraphics[width=8.6cm]{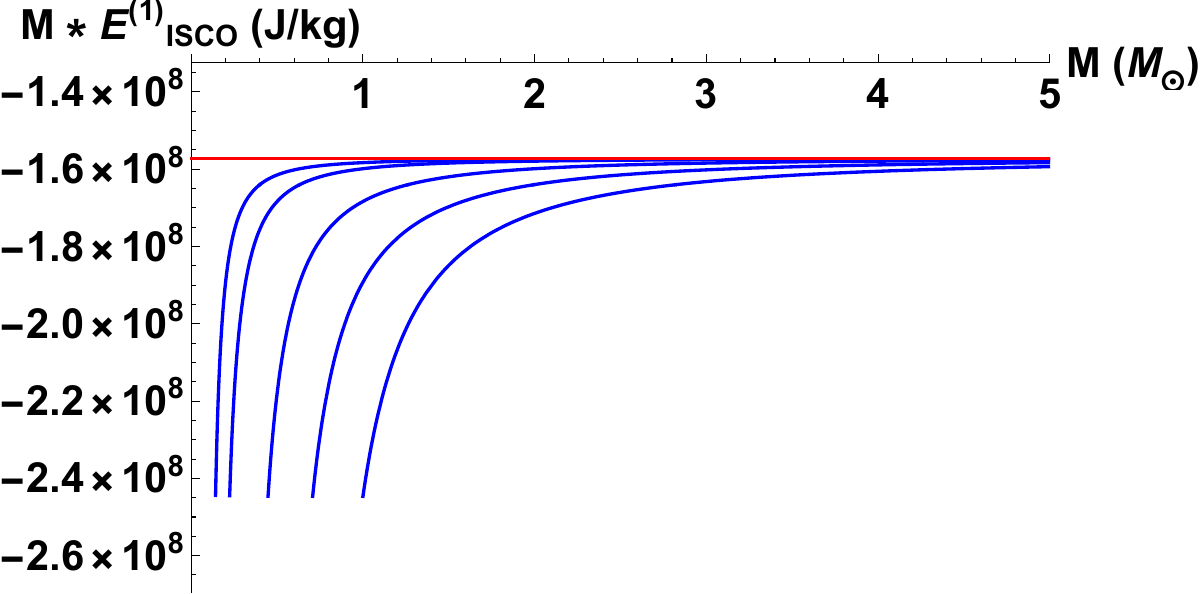}
		\label{fig:}
	\end{subfigure}
\caption[ISCO parameters for $\Lambda = 0$.]{Plots of the ISCO parameters when $\Lambda = 0$. The leftmost column contains the 0th order (static) terms, whereas the rightmost column contains the leading order corrections due to rotational effects. In all cases the red line represents the solution from GR (ie. when $\alpha=0$), and the blue lines represent the 4DEGB solutions for $\alpha/M_\odot^2$ = 0.02, 0.05, 0.02, 0.50, 1.0 from left to right.}
\label{fig:isco}
\end{figure*}

In this section the numerical results of the aforementioned six equations are plotted as a function of mass for five different values of the coupling constant $\alpha$, and are compared to the GR solution where $\alpha=0$ (see figure \ref{fig:isco}). Note that these results correspond to the prograde solutions (namely $j^{(0)}_\mathrm{ISCO}>0$), and that an analogous retrograde solution set exists. In all cases we find results similar in form to those from GR, with the greatest deviation from occurring as $M \rightarrow M_{min}$ for any given $\alpha$. As with the horizons, the 4DEGB theory induces a minimum mass black hole for a non-zero coupling constant. We find that a static black hole described by the 4DEGB theory should a have slightly smaller innermost orbital radius, corresponding to a particle with slightly lower angular momentum and energy. The effect of prograde slow rotation in the Einstein case is to subtract slightly from these otherwise positive parameters. The 4DEGB rotation corrections act similarly to the GR corrections, but with slightly larger magnitudes (with this difference again being more pronounced near minimum mass).

\subsubsection{AdS ($\mathbf{\Lambda < 0}$)}

\begin{figure*}
	\begin{subfigure}{.5\textwidth}
		\includegraphics[width=8.6cm]{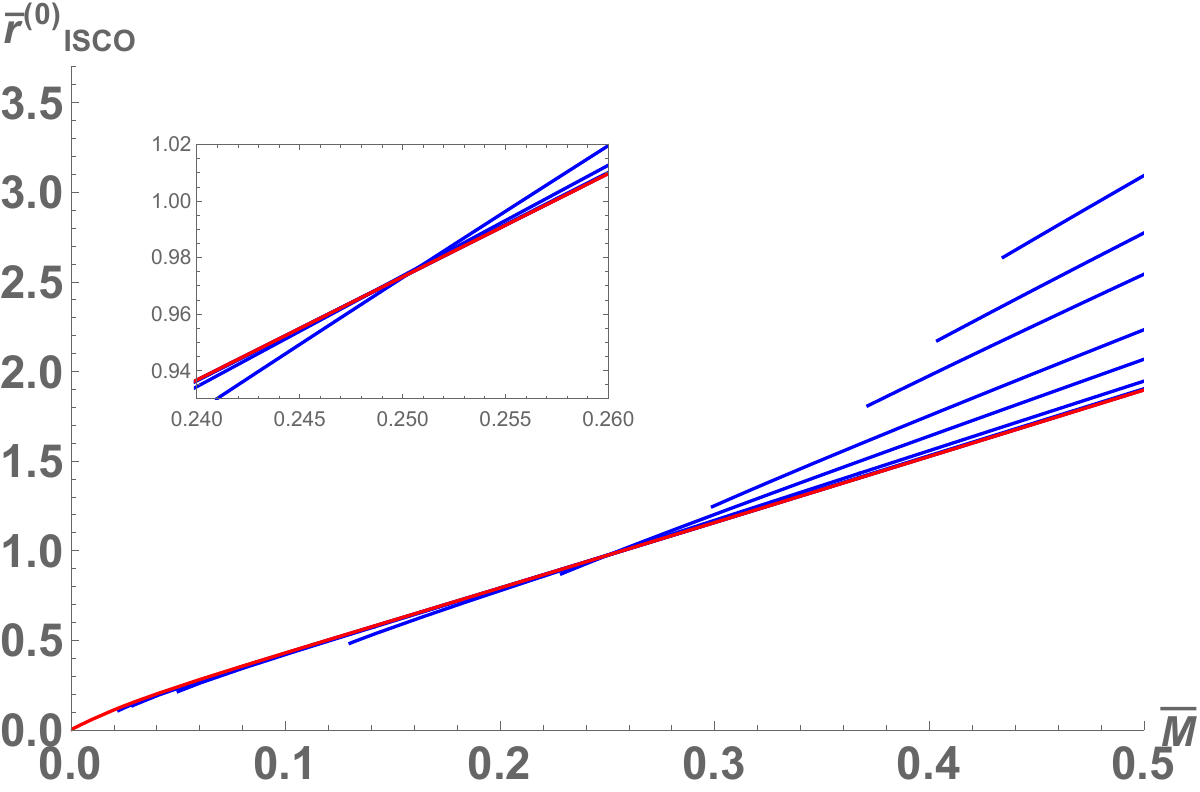}
		
		\label{fig:}
	\end{subfigure}\hfill%
	\begin{subfigure}{.5\textwidth}
		\includegraphics[width=8.6cm]{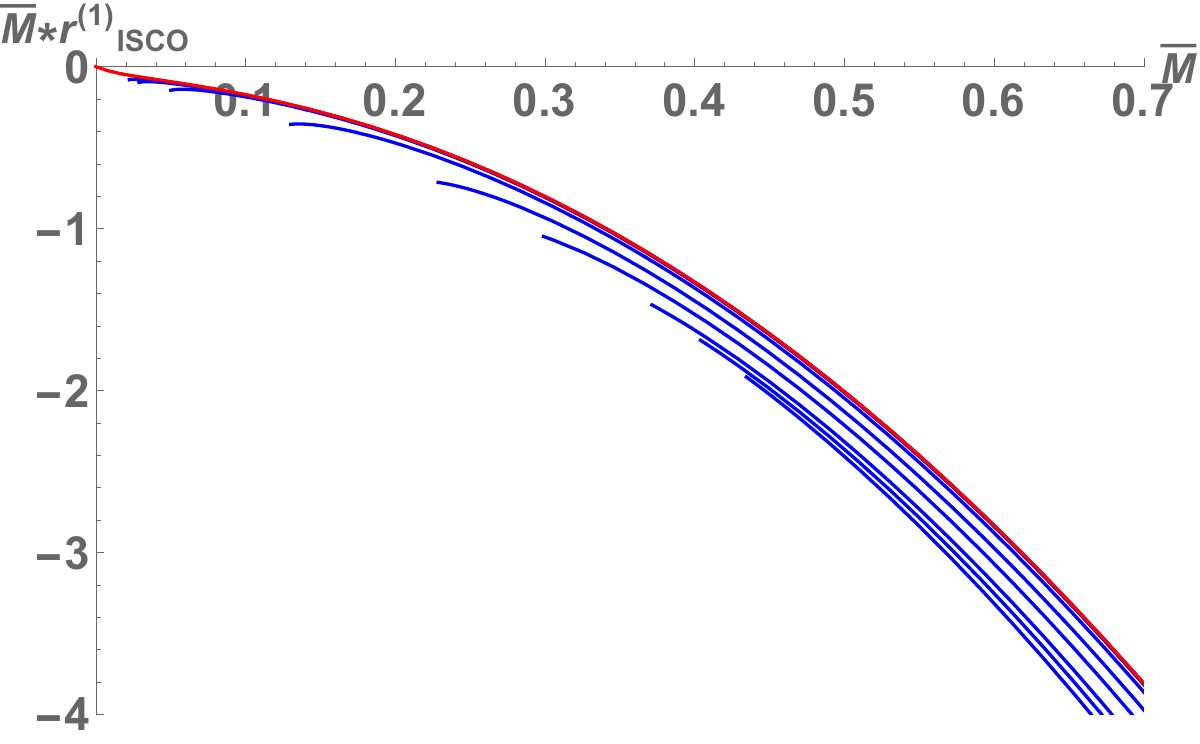}
		
		\label{fig:}
	\end{subfigure}\vspace{10mm}
	
	\begin{subfigure}{.5\textwidth}
		\includegraphics[width=8.6cm]{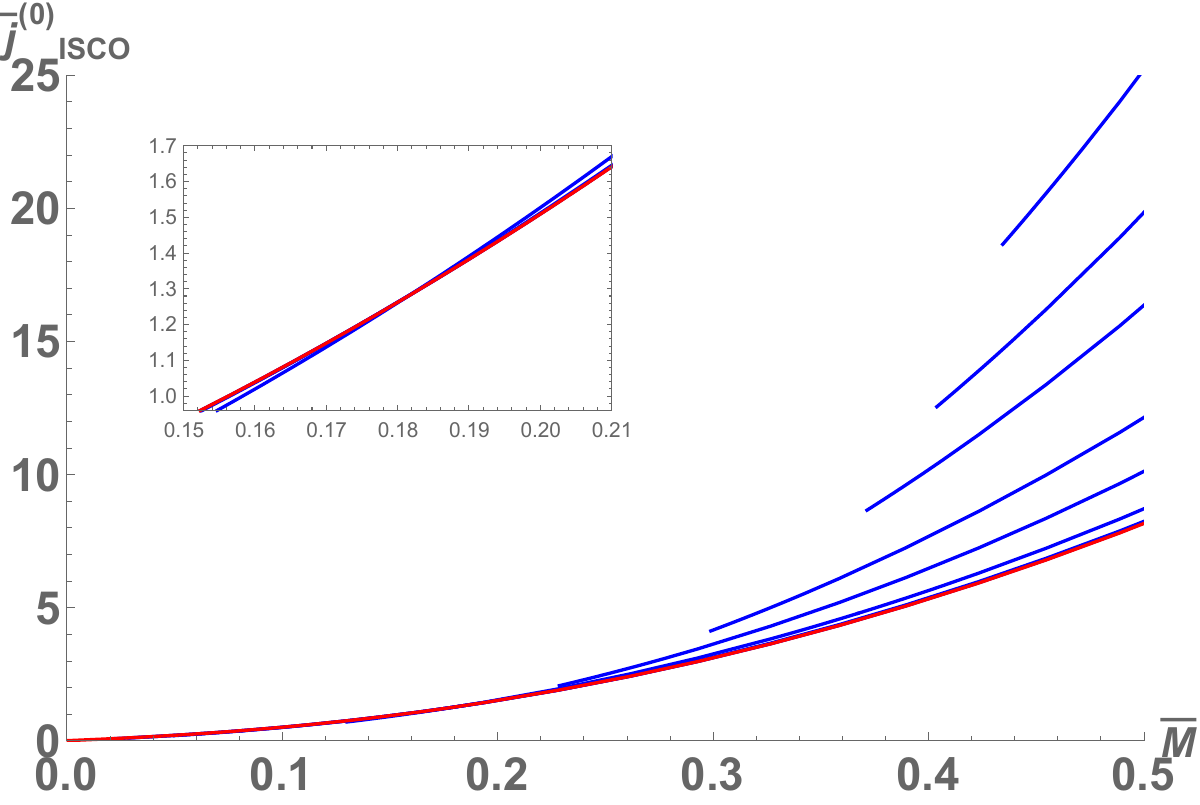}
		
		\label{fig:}
	\end{subfigure}\hfill%
	\begin{subfigure}{.5\textwidth}
		\includegraphics[width=8.6cm]{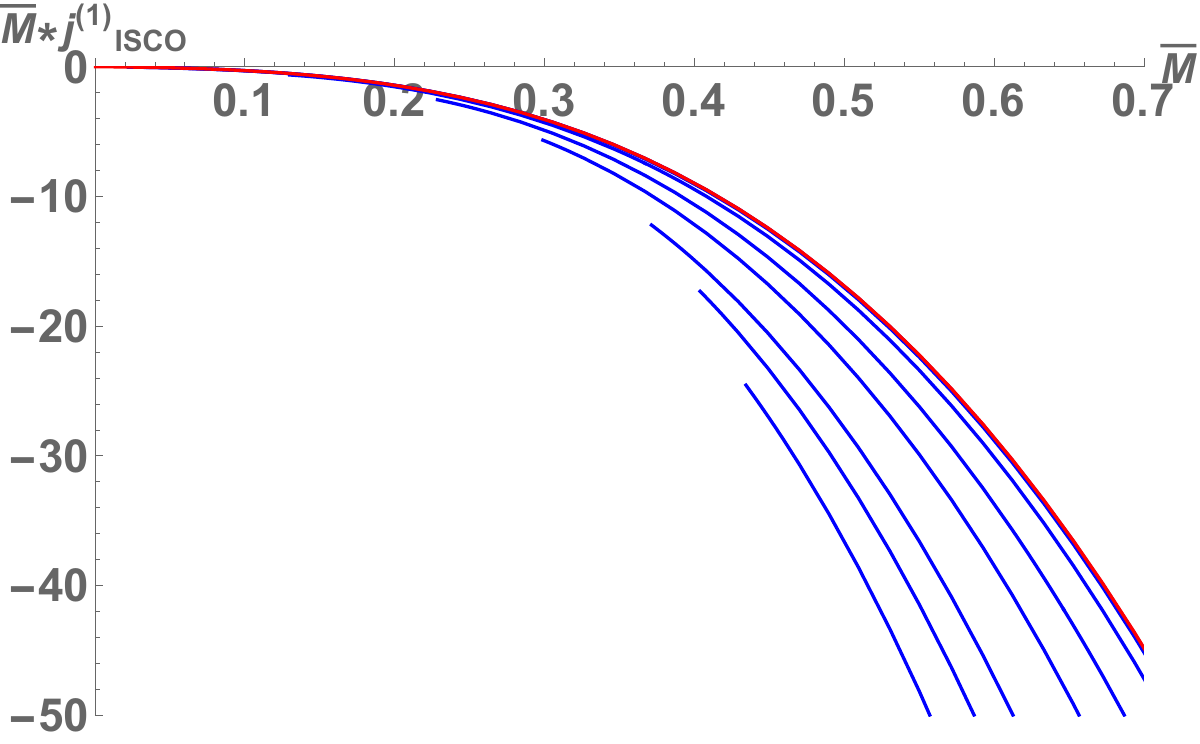}
		
		\label{fig:}
	\end{subfigure}\vspace{10mm}
	
	\begin{subfigure}{.5\textwidth}
		\includegraphics[width=8.6cm]{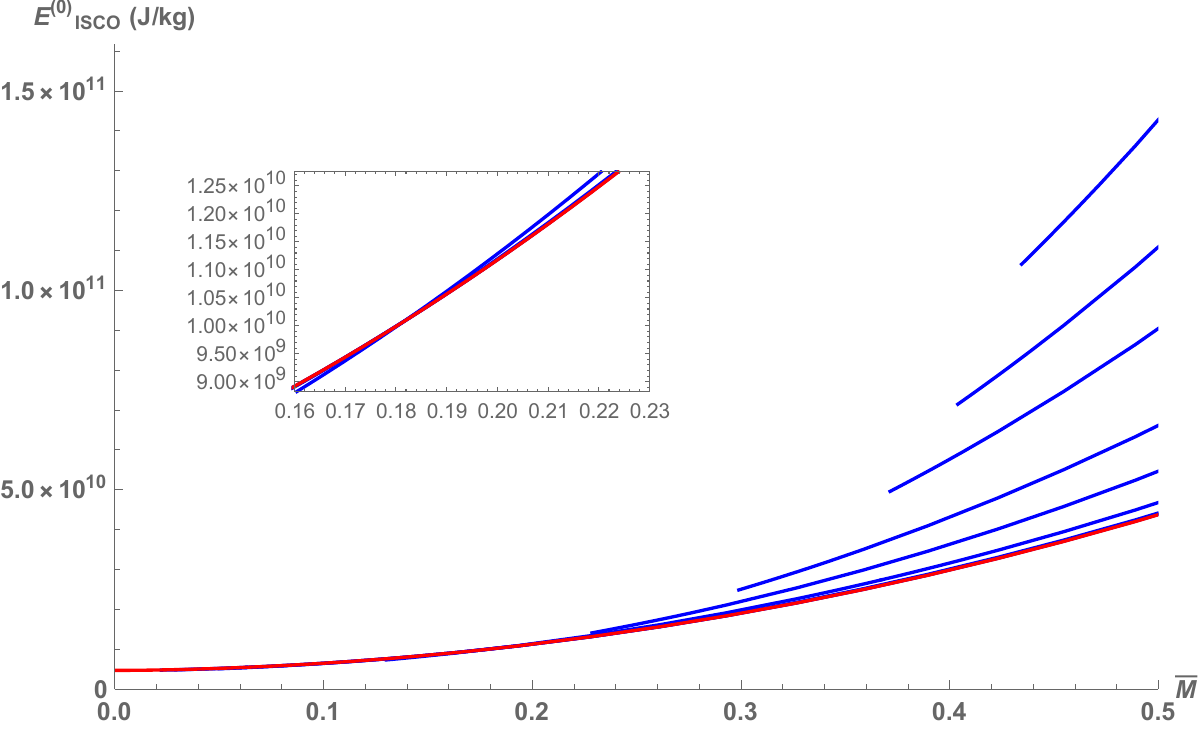}
		
		\label{fig:}
	\end{subfigure}\hfill%
	\begin{subfigure}{.5\textwidth}
		\includegraphics[width=8.6cm]{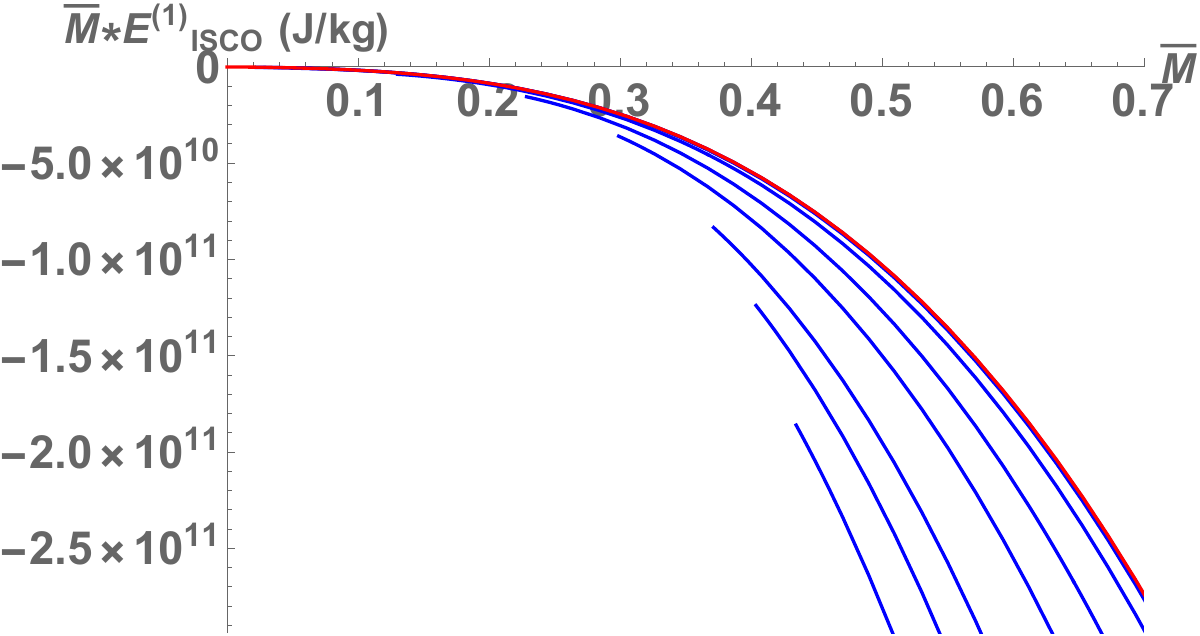}
		
		\label{fig:}
	\end{subfigure}
	
	\caption[ISCO parameters in AdS when $\alpha$ is small.]{In the regime $\alpha \sim \frac{\alpha_C}{100}$ we see the 4DEGB solution cross over the GR solution (in red) as mass increases for the 0th order ISCO parameters, shown in the left column. In the above plots we plot this behaviour in an asymptotically anti-de Sitter spacetime whith $\frac{\alpha}{\alpha_C}$ = 0.002, 0.0033, 0.01, 0.066, 0.2, 0.33, 0.5, 0.5833, 0.66 from left to right (in blue).}
	\label{fig:crossovers}
\end{figure*}

\begin{figure*}
	\begin{subfigure}{0.5\textwidth}
		\includegraphics[width=8.6cm]{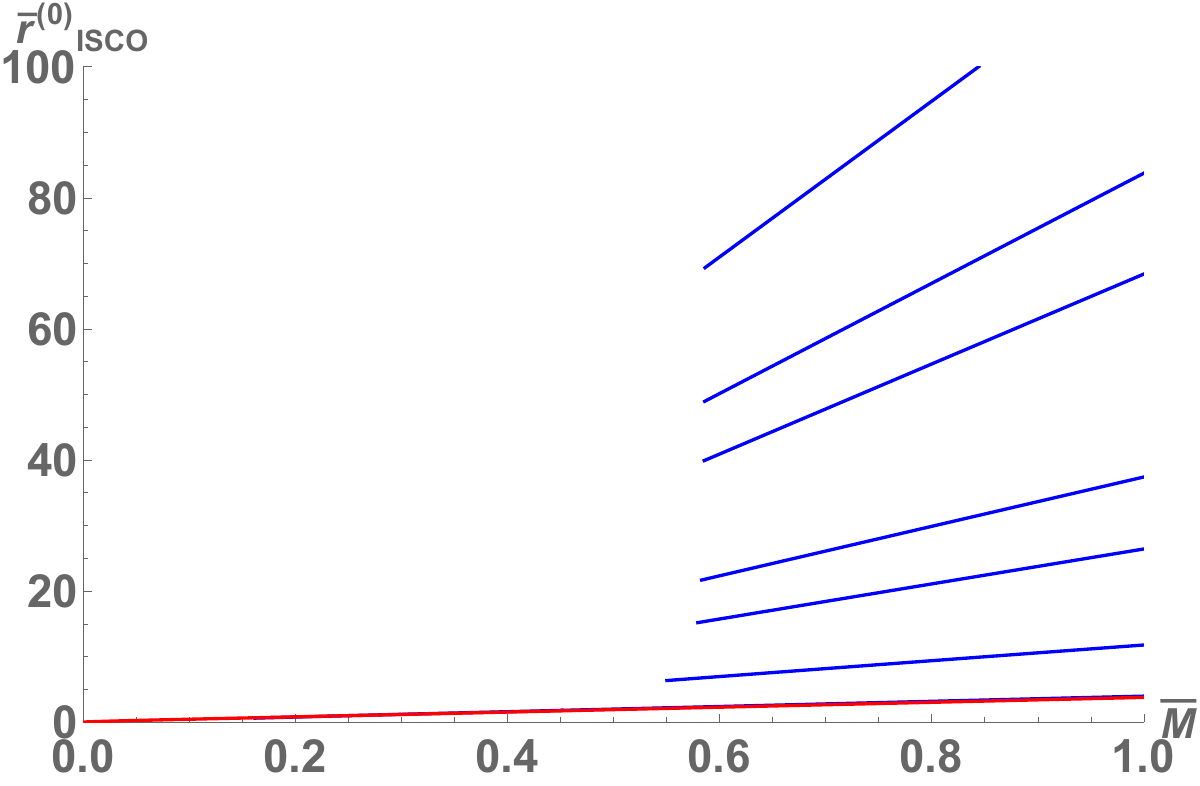}
		
		\label{fig:}
	\end{subfigure}\hfill%
	\begin{subfigure}{.5\textwidth}
		\includegraphics[width=8.6cm]{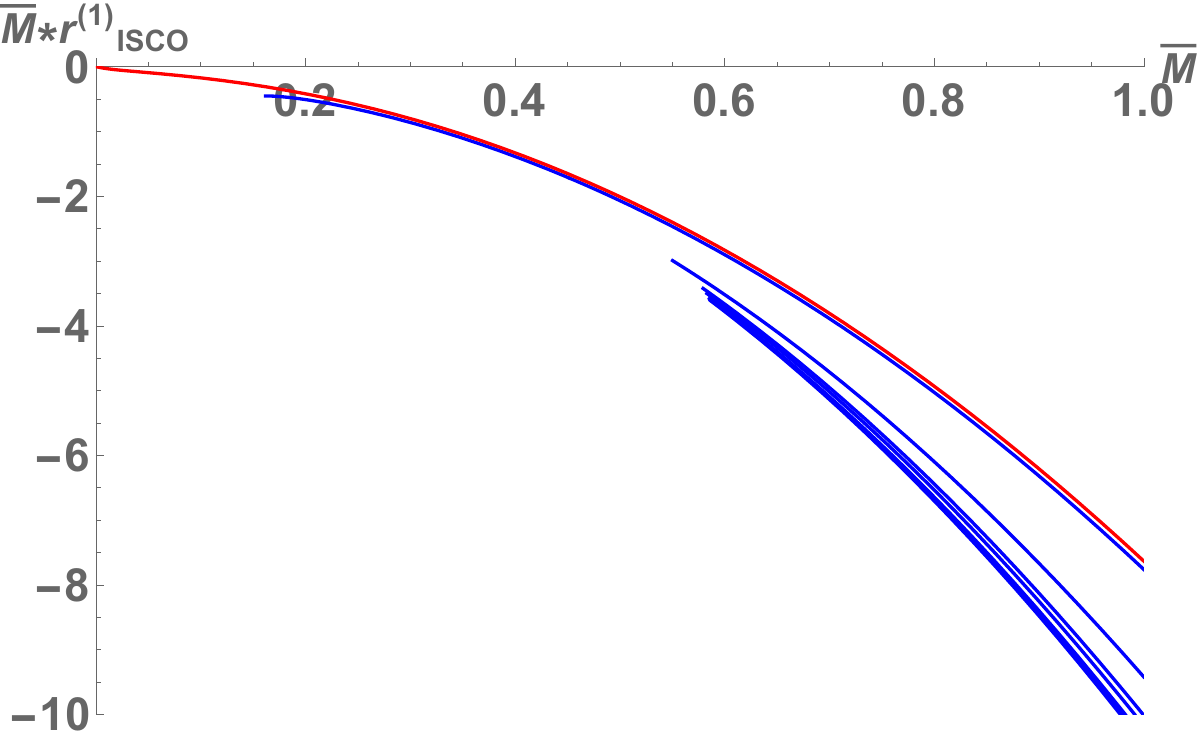}
		
		\label{fig:}
	\end{subfigure}\vspace{10mm}
	
	\begin{subfigure}{.5\textwidth}
		\includegraphics[width=8.6cm]{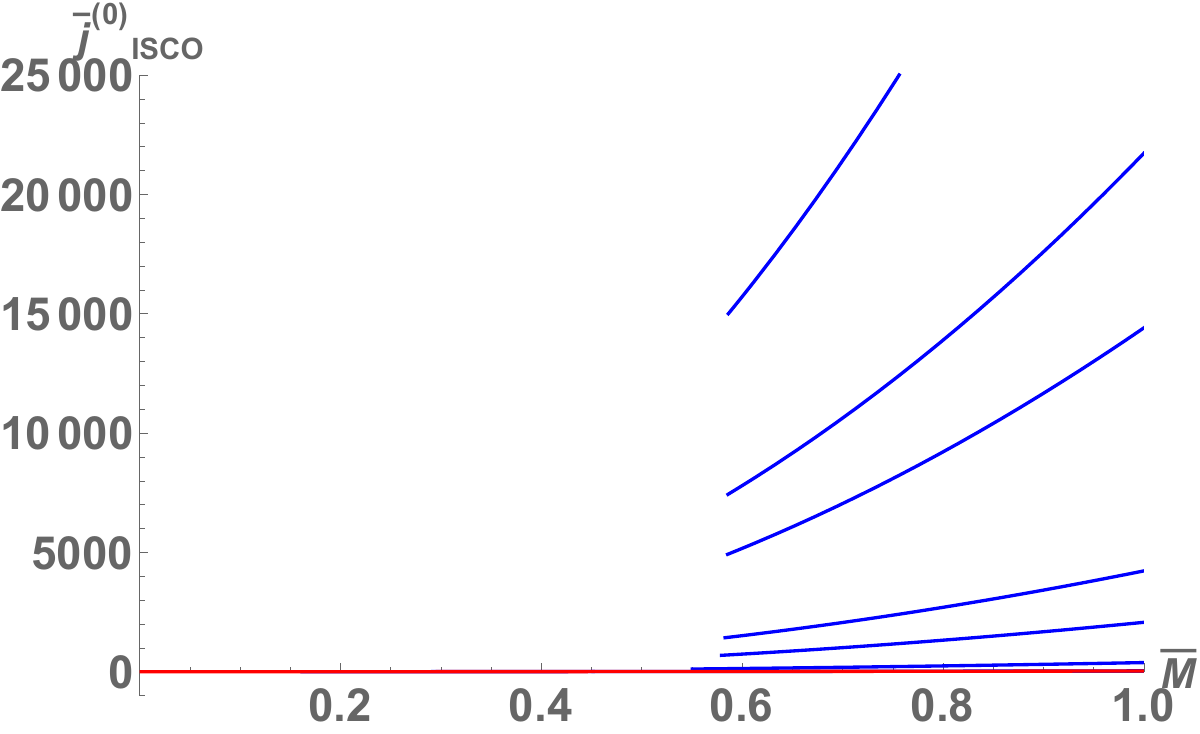}

		\label{fig:}
	\end{subfigure}\hfill%
	\begin{subfigure}{.5\textwidth}
		\includegraphics[width=8.6cm]{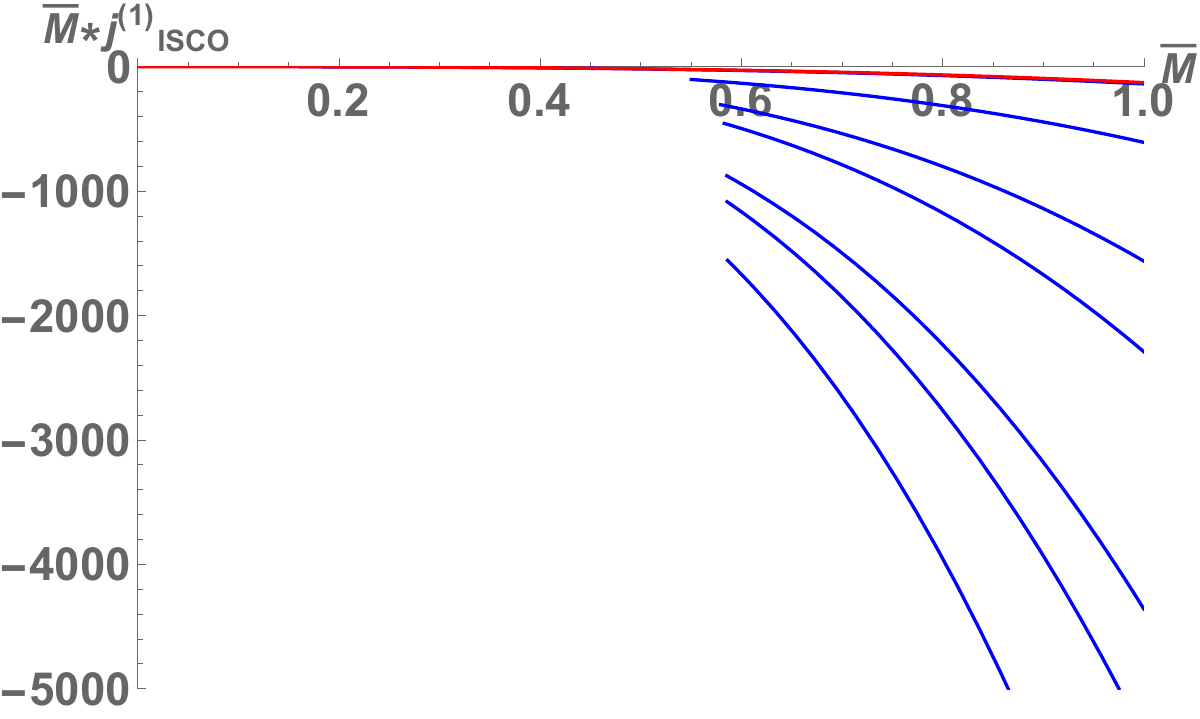}

		\label{fig:}
	\end{subfigure}\vspace{10mm}
	
	\begin{subfigure}{.5\textwidth}
		\includegraphics[width=8.6cm]{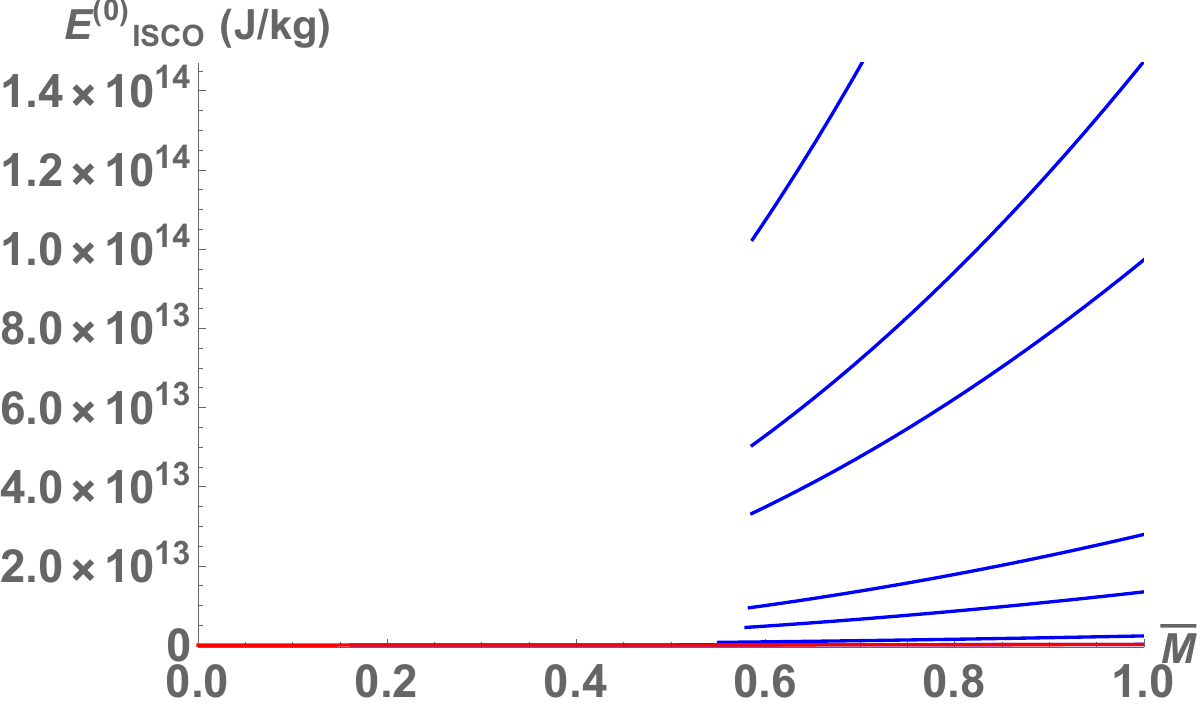}
		
		\label{fig:}
	\end{subfigure}\hfill%
	\begin{subfigure}{.5\textwidth}
		\includegraphics[width=8.6cm]{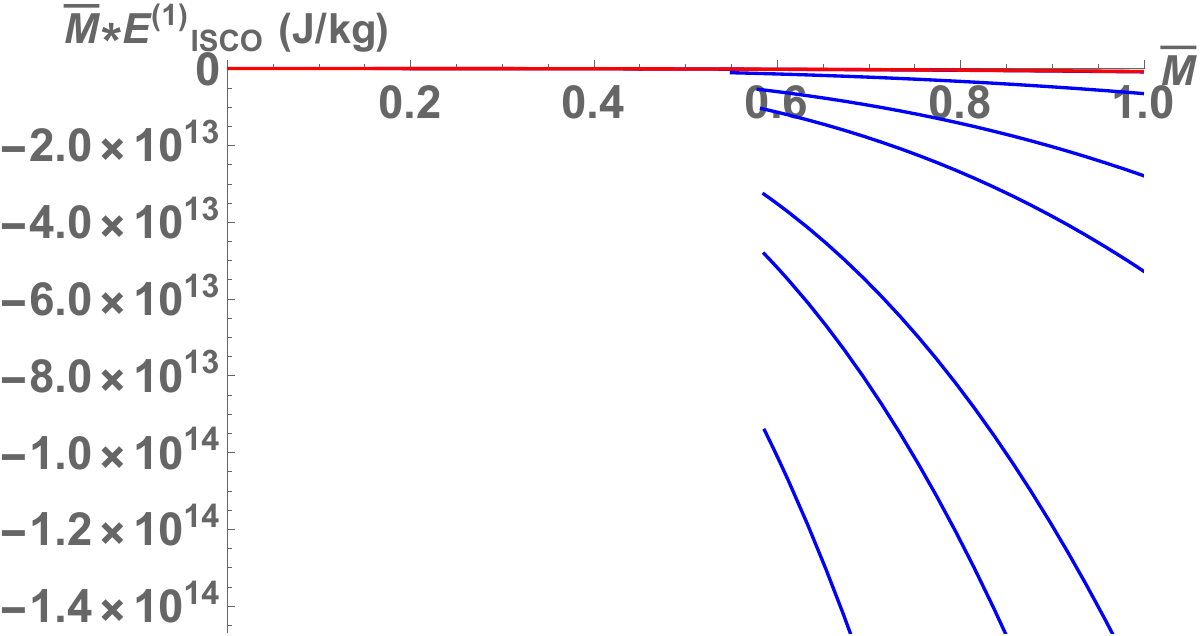}
		
		\label{fig:}
	\end{subfigure}
	\caption[ISCO parameters in AdS.]{Plots of the ISCO parameters in an asymptotically andi-de Sitter spacetime. The leftmost column contains the 0th order (static) terms, whereas the rightmost column contains the leading order corrections due to rotational effects. In all cases the red line represents the solution from GR (ie. when $\alpha=0$), and the blue lines represent the 4DEGB solutions for  $\frac{\alpha}{\alpha_C}$ = 0.1, 0.9,  0.98,  0.99,  0.997, 0.998,  0.999 from left to right.}
	\label{fig:iscoads}
\end{figure*}

In figures \ref{fig:crossovers} and \ref{fig:iscoads} we plot the results for the ISCO parameters when $\Lambda = -\frac{3}{L^2}$. Recalling the AdS criticality condition   \eqref{eq:adsalphacrit}, we 
cover a range from  $\alpha = 0$ (GR) to $\alpha \rightarrow \alpha_C$. When $\alpha$ is outside this allowed range we find solution sets for the ISCO which lie in the region of parameter space with no physical event horizon. 


When $\alpha$ is small, all 0th order ISCO parameters (figure \ref{fig:crossovers}) start below the GR solution and become larger at some constant critical mass. We can find this crossover point for $\bar{r}^{(0)}_{ISCO}$ by expanding the nontrivial contribution to $V_{\rm{eff}}^{\prime \prime (0)}(r)$ as a power series in $\bar{\alpha}$ (where $V''_{\rm{eff}}(r) = V_{\rm{eff}}^{\prime \prime (0)}(r) + a V_{\rm{eff}}^{\prime \prime (1)}(r)$), since $\bar{r}^{(0)}_{ISCO}$ is uniquely decided by this parameter.  In doing so we find
\begin{widetext}
\begin{align}
V_{\rm{eff}}^{\prime \prime (0)}(r) &= \frac{2 \bar{r}^{(0)}_{ISCO}}{3 \bar{M} -\bar{r}^{(0)}_{ISCO} } \Bigg[6 \bar{M}^2  + \bar{M} \bar{r}^{(0)}_{ISCO} (15 \bar{r}^{(0)}_{ISCO}{}^2-1))
- 4 \bar{r}^{(0)}_{ISCO}{}^4\Bigg]
\label{eq:Vppexp0 expansion} \\
&+ \frac{1}{\bar{r}^{(0)}_{ISCO}{}^2(\bar{r}^{(0)}_{ISCO}-3 \bar{M})^2 } \Bigg[ -72 \bar{M}^4 + 8 \bar{r}^{(0)}_{ISCO}{}^8 - 48 \bar{M}^3 (\bar{r}^{(0)}_{ISCO} + 3 \bar{r}^{(0)}_{ISCO}{}^3) + \bar{M}^2(32 \bar{r}^{(0)}_{ISCO}{}^2 + 48 \bar{r}^{(0)}_{ISCO}{}^4 \nonumber\\
&\qquad\qquad\qquad \qquad\qquad\qquad\qquad + 90\bar{r}^{(0)}_{ISCO}{}^6) + \bar{M}(4 \bar{r}^{(0)}_{ISCO}{}^5 - 66\bar{r}^{(0)}_{ISCO}{}^7\Bigg]\alpha + \mathcal{O}(\bar{\alpha}^2) \nonumber
\end{align}
The crossover point should occur when the leading order $\bar{\alpha}$ contribution vanishes and the equation reduces to its GR equivalent. This is done by first fixing the mass in terms of the GR ISCO radius (since, at the crossover point, $\bar{r}^{(0)}_{ISCO} = \bar{r}^{(0)GR}_{ISCO}$). This relation can be found by solving the 0th order coefficient of equation \eqref{eq:Vppexp0 expansion}: 
\begin{equation}\label{eq:m gr ads isco}
\begin{aligned}
&\bar{M} = \frac{1}{12}\Big(\bar{r}^{(0)GR}_{ISCO}-15(\bar{r}^{(0)GR}_{ISCO})^3 +\sqrt{(\bar{r}^{(0)GR}_{ISCO})^2+66(\bar{r}^{(0)GR}_{ISCO})^4 + 225 (\bar{r}^{(0)GR}_{ISCO})^6}\Big).
\end{aligned}
\end{equation}

This fixed mass is then substituted into the leading order term in equation \eqref{eq:Vppexp0 expansion}. Setting this equal to 0 and solving, we find the crossover occurs at $\bar{r}^{(0)GR}_{ISCO} = 0.965679$ which corresponds to a crossover mass of
\begin{equation}
\bar{M}_{\rm{x}}^{\bar{r}^{(0)}_{ISCO}} = 0.247935.
\end{equation}

For $\bar{j}^{(0)}_{ISCO}$ we have an analytic expression that is not too complicated, and can be directly expanded in a small $\alpha$ power series as follows:
\begin{align}
&\bar{j}^{(0)}_{ISCO} = \frac{\sqrt{-(\bar{r}^{(0)GR}_{ISCO})^2 \left(\bar{M}+(\bar{r}^{(0)GR}_{ISCO})^3\right)}}{\sqrt{3 \bar{M}-\bar{r}^{(0)GR}_{ISCO}}} \\
&\quad 
-\frac{\left(2 \bar{M}-(\bar{r}^{(0)GR}_{ISCO})^3\right) \sqrt{-(\bar{r}^{(0)GR}_{ISCO})^2 \left(\bar{M}+(\bar{r}^{(0)GR}_{ISCO})^3\right)}}{2 \left((\bar{r}^{(0)GR}_{ISCO})^3 (3 \bar{M}-\bar{r}^{(0)GR}_{ISCO})^{3/2} \left(\bar{M}+(\bar{r}^{(0)GR}_{ISCO})^3\right)\right)} \left(6 \bar{M}^2-3 \bar{M} (\bar{r}^{(0)GR}_{ISCO})^3-4 \bar{M} \bar{r}^{(0)GR}_{ISCO}-(\bar{r}^{(0)GR}_{ISCO})^4\right) \bar{\alpha } +\mathcal{O}\left(\bar{\alpha} ^{2}\right)
\nonumber
\end{align}
\end{widetext}
suggesting that the results will cross over when $\left(2 \bar{M}-(\bar{r}^{(0)GR}_{ISCO})^3\right) = 0$. We can solve this by replacing $\bar{M}$ in this simple expression with equation \eqref{eq:m gr ads isco}, yielding a crossover point of $\bar{r}^{(0)GR}_{ISCO} = \frac{1}{\sqrt{2}}$, or equivalently
\begin{equation}
\bar{M}_{\rm{x}}^{\bar{j}^{(0)}_{ISCO}} = \frac{1}{4 \sqrt{2}} \approx 0.1768.
\end{equation}

For $\bar{E}^{(0)}_{ISCO}$, the crossover condition is identical to that for $\bar{j}^{(0)}_{ISCO}$: $\left(2 \bar{M}-(\bar{r}^{(0)GR}_{ISCO})^3\right) = 0$. Because of this is is clear that
\begin{equation}
\bar{M}_{\rm{x}}^{\bar{E}^{(0)}_{ISCO}} =\bar{M}_{\rm{x}}^{\bar{j}^{(0)}_{ISCO}} = \frac{1}{4 \sqrt{2}} \approx 0.1768.
\end{equation}

These values for $\bar{M}_{\rm{x}}$ all agree with the expected values from a visual inspection of figure \ref{fig:crossovers}. On the other hand, the leading order ISCO parameters are smaller (more negative) than their GR counterpart. 

As $\alpha$ approaches criticality (figure \ref{fig:iscoads}), we see an extreme departure from the GR results for almost all of the AdS ISCO solutions. No crossover behaviour is observed, and the equations become very sensitive to changes in $\alpha$ in this region of parameter space (however, the value of $\bar{M}_{min}$ is nearly constant). When $\alpha/\alpha_C \sim 0.90$ we start to see this extreme sensitivity, with $\bar{M}_{min} = 0.544$ when $\bar{\alpha} = \bar{\alpha}_c$.

\subsubsection{dS ($\mathbf{\Lambda > 0}$)}

The ISCO solutions in de Sitter space are particularly nuanced. We begin with a thorough investigation of the non-rotating ISCO solutions in GR ($\alpha=0$) before moving forward with the 4DEGB solutions. \newline

\begin{figure*}
\begin{subfigure}{.30\textwidth}
	\includegraphics[width=1\textwidth]{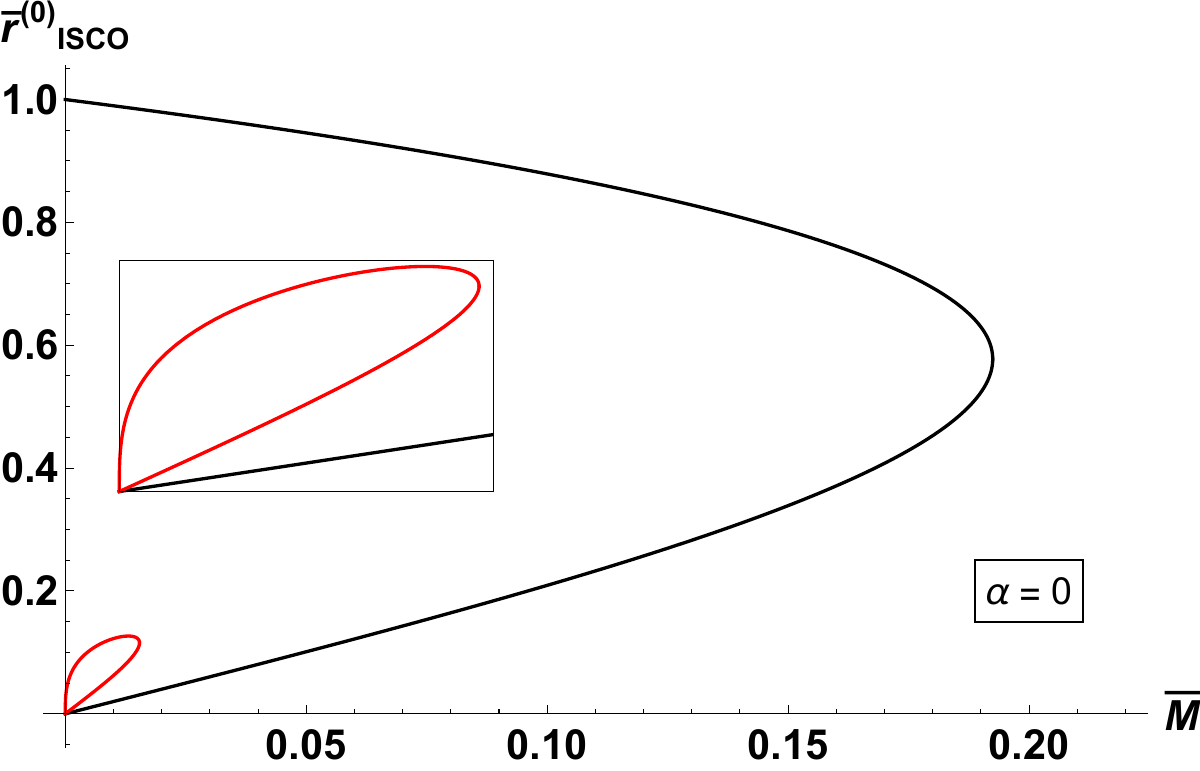}
	
	\caption[The ISCO radius in dS for GR.]{The 0th order (static) ISCO/OSCO radius (red) and the black hole horizons (black) for GR ($\alpha=0$) plotted as a function of mass in an asymptotically de Sitter spacetime. A non-zero, positive $\Lambda$ forces the ISCO solution to curl back into a loop, which has two stable radii (the larger OSCO and smaller ISCO)
 for each fixed mass (and also has a maximum mass at  which the OSCO and ISCO merge; above this there is no ISCO).}
	\label{fig:einposlamisco}
\end{subfigure}\hfill
\begin{subfigure}{.30\textwidth}
	\includegraphics[width=1\textwidth]{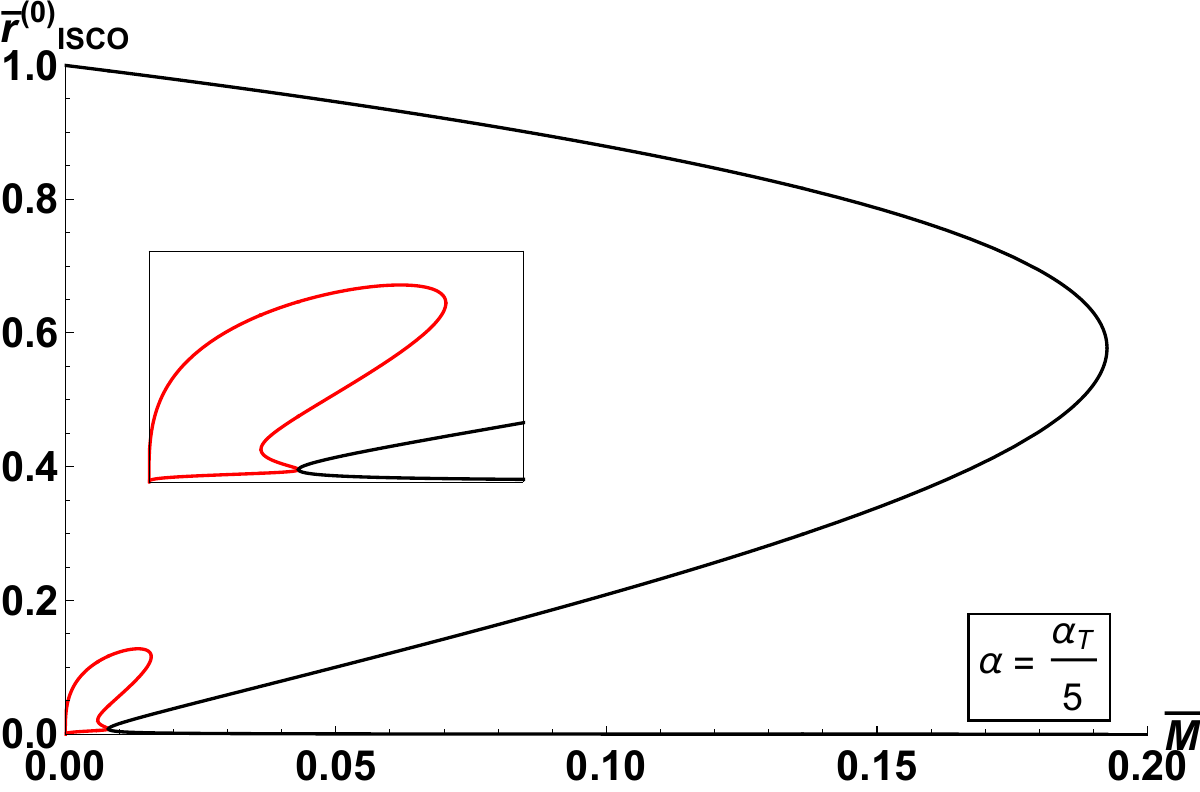}
	
	\caption[The ISCO/OSCO radius in dS for small $\alpha$.]{The 0th order (static) ISCO/OSCO radius (red) and the black hole horizons (black) for $\alpha=\frac{\alpha_T}{5}$) plotted as a function of mass when in an asymptotically de Sitter spacetime. The introduction of a non-zero Gauss-Bonnet coupling constant introduces an inner black hole horizon, enforcing a minimum mass as well as the maximum mass from the de Sitter GR case.}
	\label{fig:einposlamiscodeviation}
\end{subfigure}\hfill
\begin{subfigure}{.30\textwidth}\hfill
	\includegraphics[width=1\textwidth]{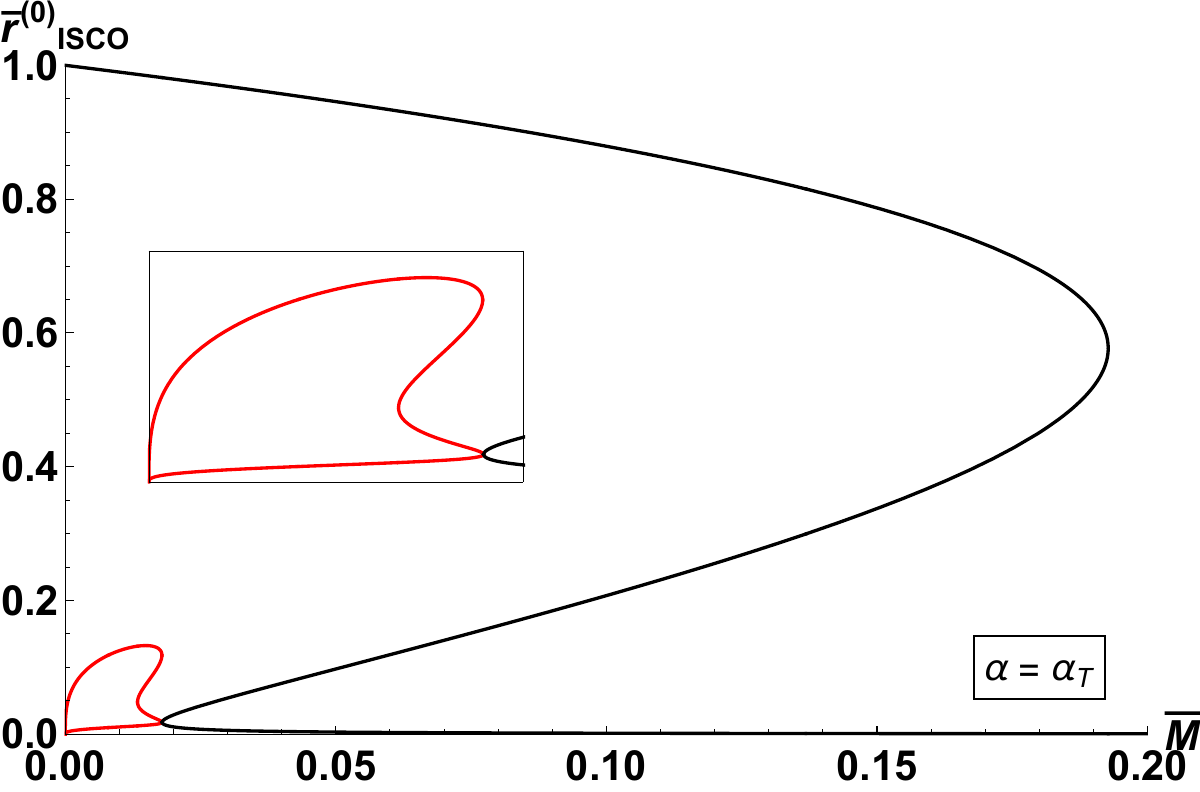}
	
	\caption[The ISCO radius in dS for $\alpha = \alpha_T$.]{The 0th order (static) ISCO/OSCO radius (red) and the black hole horizons (black) for $\alpha=\alpha_T$) plotted as a function of mass in an asymptotically de Sitter spacetime. By this point the entire ISCO/OSCO solution set lies in the region of parameter space that describes a naked singularity.}
	\label{fig:einposlamiscodeviationalphaT}
\end{subfigure}
\caption[Progression of the dS ISCO radius as $\alpha$ increases.]{Depiction of the 0th order  ISCO/OSCO radii for $\alpha=0$ (left), $\alpha=\frac{\alpha_T}{5}$ (middle), and $\alpha= \alpha_T$ 
(right). The lower turning point of the horizon curve corresponds to the minimal mass; the upper turning point corresponds to the maximal mass.}
\label{dSISCO}
\end{figure*}

\begin{figure}
    \centering
    \includegraphics[width=8cm]{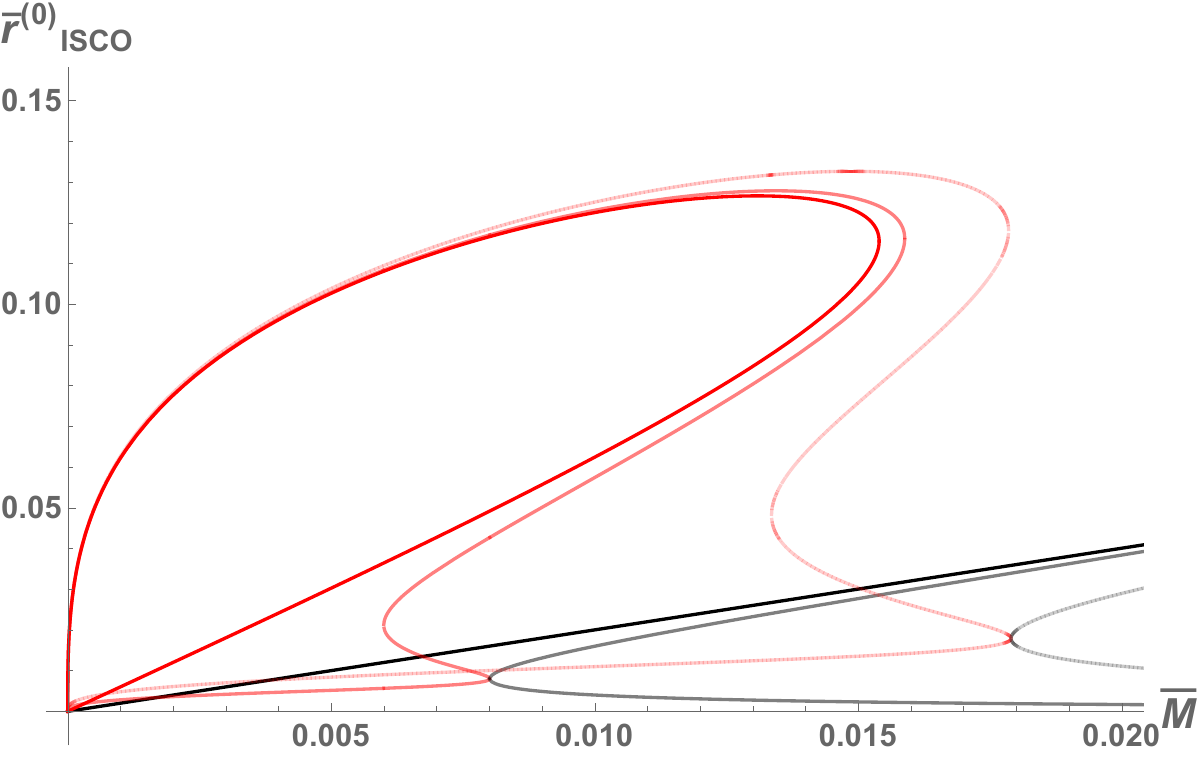}
    \caption{ Depiction of the 0th order  ISCO radius for $\alpha=0$, $\alpha=\frac{\alpha_T}{5}$, and $\alpha= \alpha_T$ (from most to least opaque, respectively). These plots are the same as those shown in figure \ref{dSISCO}, superimposed on one another. Increasing the 4DEGB coupling causes the 0th order ISCO radius to decrease, and the 0th order OSCO to increase.}\label{fig:dsiscoprogression}
\end{figure}

The Einstein de Sitter ISCO solutions are plotted in figures \ref{fig:einposlamisco} and \ref{fig:dsiscoprogression} alongside the corresponding black hole horizons for a positive cosmological constant. We immediately observe that for non-zero positive $\Lambda$, a turning point appears for $M = \frac{2}{75 \sqrt{\Lambda}}$. 
There are  no ISCOs for $\frac{2}{75 \sqrt{\Lambda}} < M < \frac{1}{3\sqrt{\Lambda }}$, the latter value being the Nariai upper mass limit  $\frac{1}{3 \sqrt{\Lambda }}$, obtained by  setting $\alpha=0$ in \eqref{eq:mmaxminpos}.

For   $M < \frac{2}{75 \sqrt{\Lambda}}$, there is at most one stable orbit and two unstable orbits at any given mass less than this value, depending on the value of $l_z$. Requiring \eqref{ISCOcond} and
setting $V''(r)=0$ we find two distinct solutions
for the ISCO depending of the magnitude of $|l_z|$.
The larger of these is the  outer innermost stable orbit (or OSCO) \cite{boonserm_2020}. Since this turning point always occurs at $M < M_{max}$, the ISCO and OSCO solutions within the allowed mass range are physical. For spacetimes with  masses 
that do not permit  stable orbits, the effective potential always has at least one local maxima, corresponding to an \textit{unstable} orbit - this happens via the merging of the outer and inner stable orbits and occurs due to the balance of gravity with a positive cosmological constant. These unstable orbits beyond the ISCO turnaround point are only noted for pedagogical reasons and are not shown in any plots.

When a non-zero Gauss-Bonnet coupling constant is introduced, many of our ISCO solutions are no longer physical due to the presence of a minimum mass. This is clear from figure \ref{fig:dsiscoprogression} (and
the middle diagram in figure \ref{dSISCO}), which plots $\bar{r}_{\text{ISCO}}^{(0)}$ (red curve) as a function of $M$ for different values of $\alpha$.  We see for $\alpha=0$ that all points on the double-valued ISCO curve are physically allowed, whereas for $\alpha > 0$, all points to the left of $M=M_{min}$ (indicated by the lower turning point of the black horizon curve) are unphysical. This is because the $M < M_{min}$ region of parameter space represents a naked singularity due to a change in horizon structure (see figures \ref{fig:horizonsds} and \ref{fig:einposlamiscodeviation}). 
Increasing $\alpha$ causes the 0th order ISCO radius to decrease, and the 0th order OSCO to increase.

Finally, at some transitory value ($\alpha = \alpha_T$), the turning point corresponding to the merger of the ISCO and OSCO 
 (which is also its maximum mass point) equals  the minimum mass value, shown in the rightmost diagram \ref{fig:einposlamiscodeviationalphaT} in figure \ref{dSISCO}. We find that $\alpha_T = \frac{\alpha_C}{260} =  \frac{1}{1040 \Lambda}$. For $\alpha > \alpha_T$ no physical ISCOs or OSCOs exist;  all putative solutions correspond to spacetimes with naked singularities surrounded by a cosmological horizon. When $\alpha = \alpha_T$, there is a single ISCO, which occurs at the critical mass point 
\begin{equation}
	M_{crit} |_{\alpha=\alpha_T}=\frac{ \sqrt{263-259 \sqrt{\frac{259}{260}}}}{3 \sqrt{520\Lambda }} \approx \frac{1}{\sqrt{1000 \Lambda}}
\end{equation}
from  \eqref{eq:mmaxminpos}. \newline

\begin{figure*}
	\begin{subfigure}{.5\textwidth}
		\includegraphics[width=8.6cm]{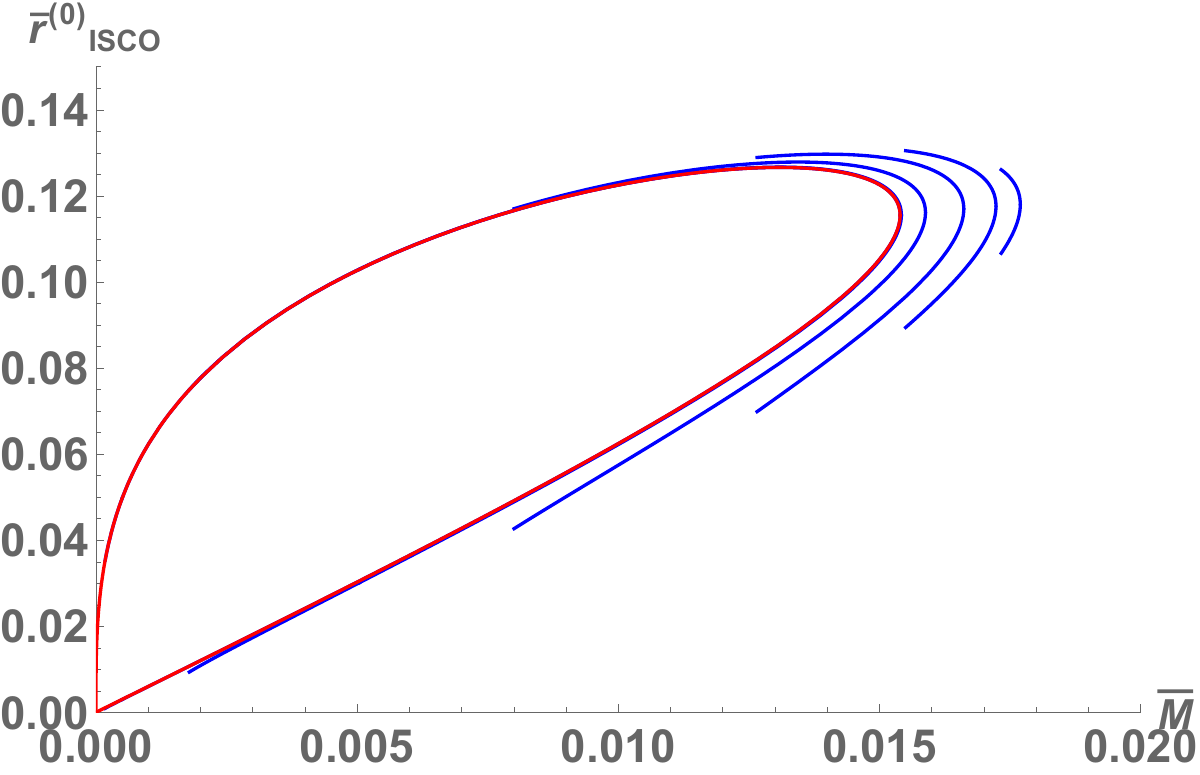}
		
		\label{fig:}
	\end{subfigure}\hfill%
	\begin{subfigure}{.5\textwidth}
		\includegraphics[width=8.6cm]{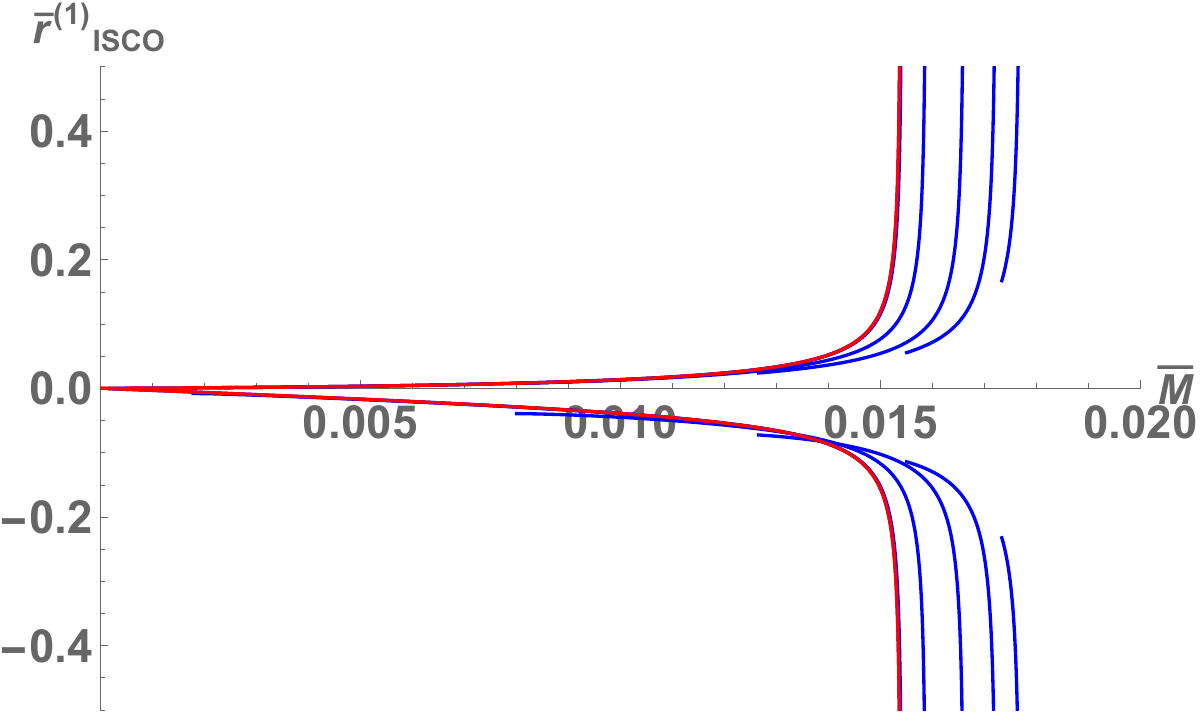}
		
		\label{fig:}
	\end{subfigure}\vspace{10mm}
	
	\begin{subfigure}{.5\textwidth}
		\includegraphics[width=8.6cm]{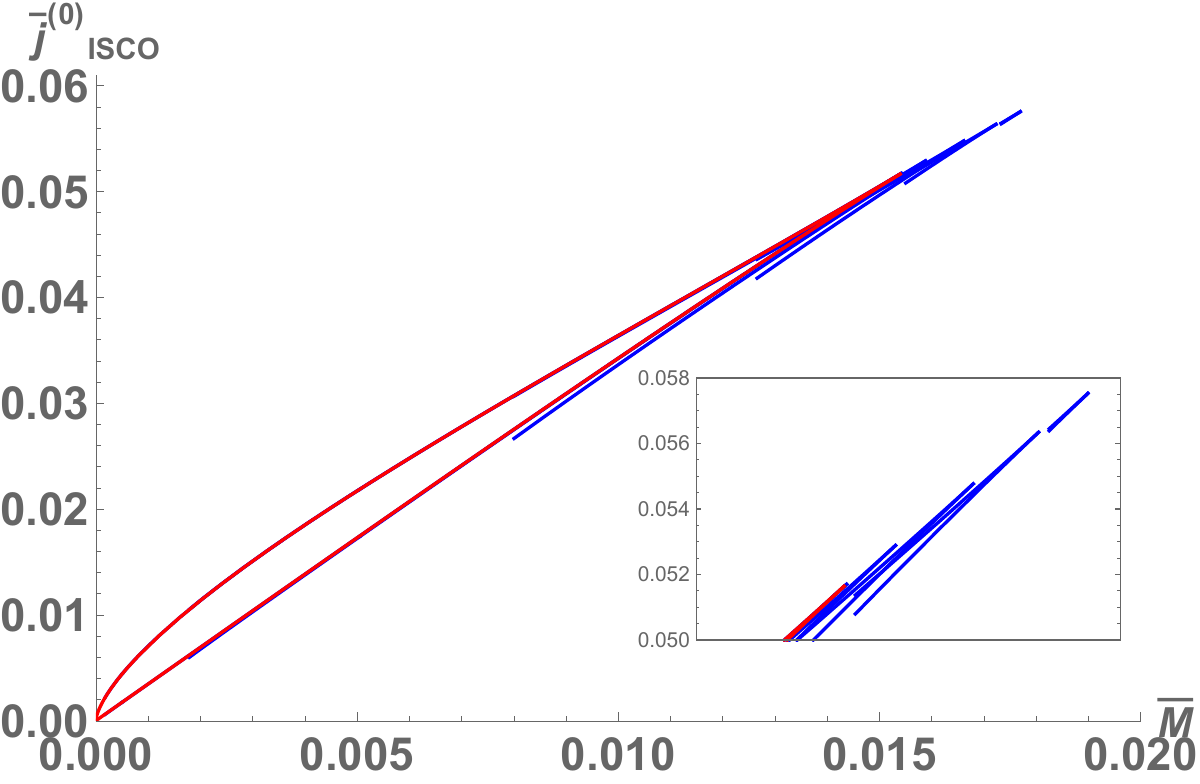}
		
		\label{fig:}
	\end{subfigure}\hfill%
	\begin{subfigure}{.5\textwidth}
		\includegraphics[width=8.6cm]{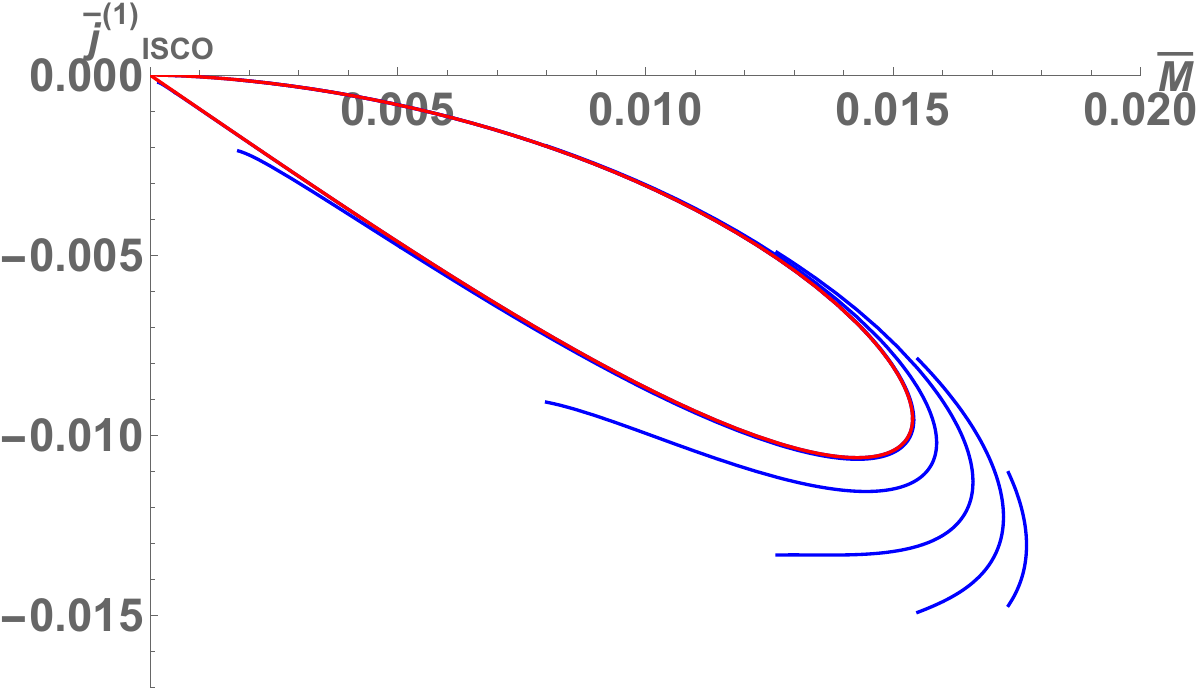}
		
		\label{fig:}
	\end{subfigure}\vspace{10mm}
	
	\begin{subfigure}{.5\textwidth}
		\includegraphics[width=8.6cm]{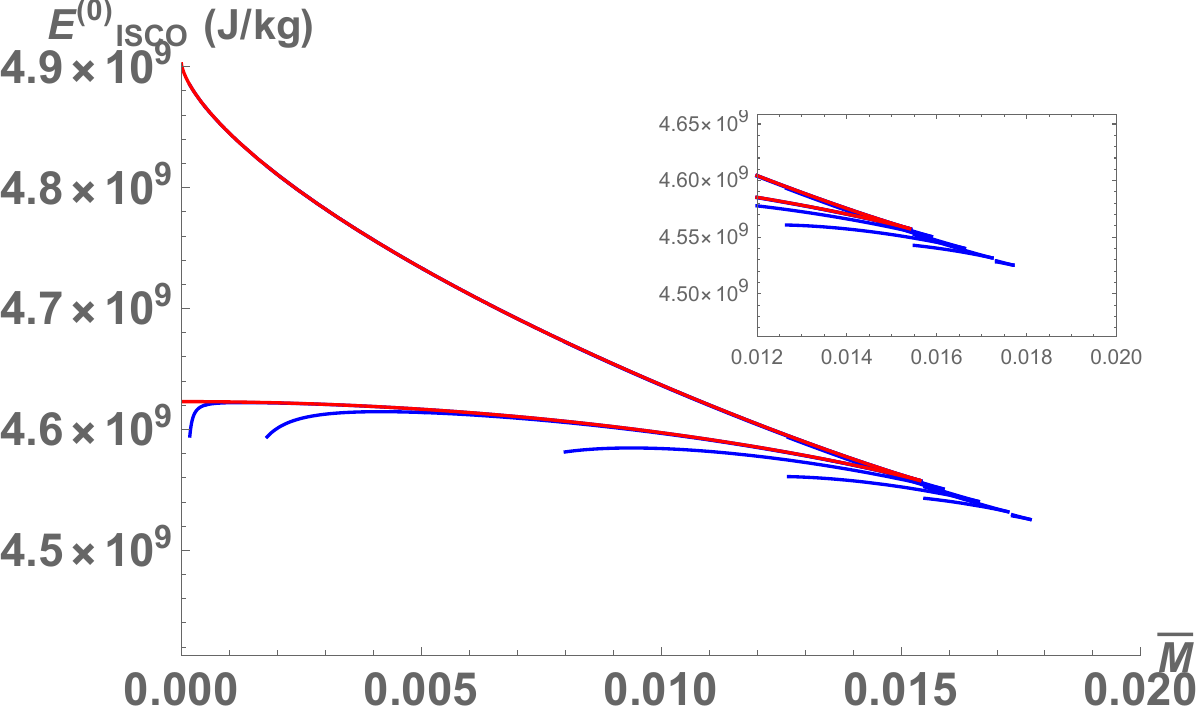}
		
		\label{fig:}
	\end{subfigure}\hfill%
	\begin{subfigure}{.5\textwidth}
		\includegraphics[width=8.6cm]{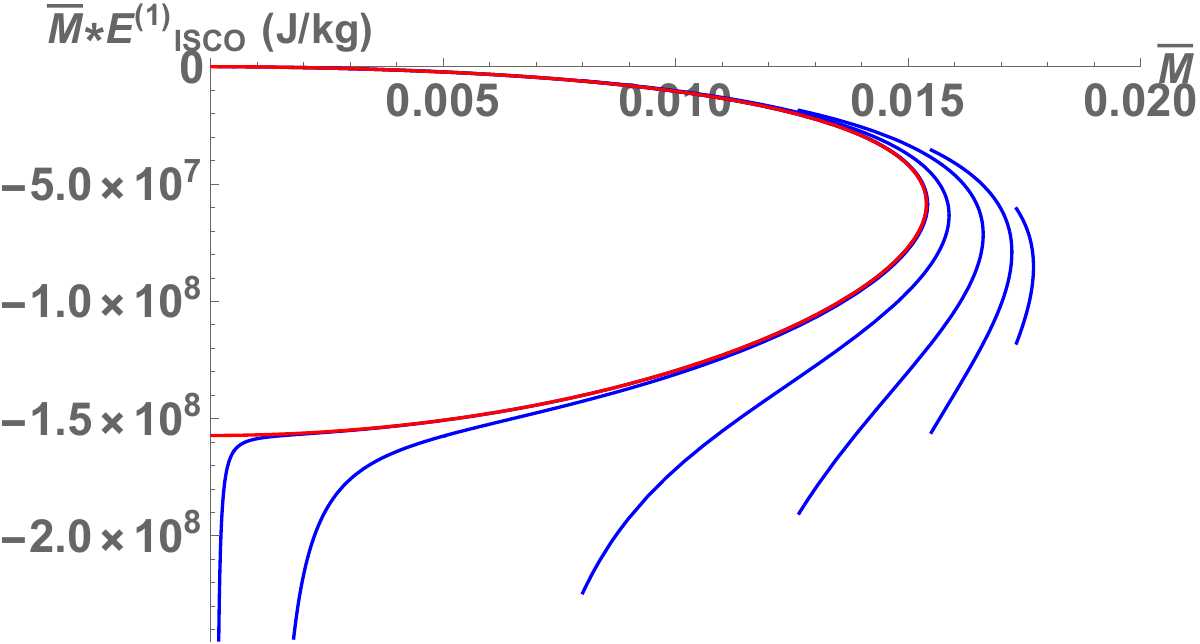}
		
		\label{fig:}
	\end{subfigure}
	\caption[ISCO parameters in dS.]{Plots of the ISCO parameters in an asymptotically de Sitter spacetime. The leftmost column contains the 0th order (static) terms, whereas the rightmost column contains the leading order corrections due to rotational effects. In all cases the red line represents the solution from GR (ie. when $\alpha=0$), and the blue lines represent the 4DEGB solutions for $\frac{\alpha}{\alpha_T}$ = 0.0001, 0.01, 0.2, 0.5, 0.75, 0.9375 from left to right.}
	\label{fig:dsisco}
\end{figure*}

Armed with this knowledge, we solve numerically the ISCO equations once again for a positive cosmological constant with $\alpha$ varying from $0$ to $\alpha_T$. These results are shown in figure \ref{fig:dsisco}. Again, the main feature of interest in asymptotically de Sitter space is the existence of two discrete stable orbits for a fixed black hole mass. All 0th order ISCO parameters are double-valued
(hence yielding an ISCO and an OSCO), with the 4DEGB case having a cutoff at $M_{min}$ as noted above, and turning around at $M_{crit}$.  The 1st order ISCO parameters are also double-valued and confined between $M_{min}$ and $M_{crit}$, with the actual values depending on $\alpha$. The smallest and largest allowed 0th order ISCO radii at any given mass in the 4DEGB theory are respectively smaller and larger than their GR counterparts. As $\alpha \rightarrow \alpha_T$, the 4DEGB ISCO parameters occur at values of the mass larger than the GR upper limit 
$M =\frac{2}{75 \sqrt{\Lambda}}$, hence having no overlap in allowed mass with GR.

\subsection{Null Geodesics: Photon Rings}\label{subsec:photonrings}

Now we turn our attention to how slow rotation in the 4DEGB theory deforms the photon rings of black holes. The photon ring is defined by null geoedsic orbits ($\xi = 0$) of constant $r$, which we take to  lie in the equatorial plane (ie. $x = 0$) without loss of generality. Rather than work with conserved quantities $E$ and $j$ as in the previous section, we instead follow \cite{adair2020} and consider the problem in terms of angular velocity ($\omega = d\phi/dt$), which is conserved along the null trajectory. From equation \ref{eq:effpot} we know that a photon outside of our black hole will be subject to an effective potential
\begin{equation}
	V_{\text{ph}} = \frac{f(r) \ell_z^2 - r^2 E^2}{2 r^2} - a \frac{P(r) \ell_z E}{r^2}.  
\end{equation}
 
For circular orbits (ie. $\dot{r} = 0$), the location of the photon ring is determined by
\begin{eqnarray}
	V_{\text{ph}}(r) = 0 \qquad	V'_{\text{ph}}(r) = 0
\end{eqnarray}
where $\omega = \frac{d \phi}{d t} = \frac{\dot{\phi}}{\dot{t}}$ is conserved along the photon trajectory, and we have let $E = 1$ without loss of generality by rescaling the affine parameter. These end up being equivalent to the following two conditions:
\begin{eqnarray}
	f(r_{pr}) - r_{pr}^2 \omega^2  -  4 a P(r_{pr}) \omega = 0 \\
	f'(r_{pr}) - 2r_{pr} \omega^2 -  4 a P'(r_{pr}) \omega = 0
\end{eqnarray}
which can be solved analytically to leading order in $a$ by writing
\begin{equation}\label{eq:ps perturbative}
	r_{pr} = r_{pr}^{(0)} + a r_{pr}^{(1)} \qquad
	\omega_{pr} = \omega_{pr}^{(0)} + a \omega_{pr}^{(1)}.
\end{equation}

With this, we find that
\begin{equation}
	f'(r_{pr}^{(0)}) = 2 \frac{ f( r_{pr}^{(0)})}{r_{pr}^{(0)}}, \;\;\; \omega_{pr}^{(0)} = \pm \frac{\sqrt{f(r_{pr}^{(0)})}}{r_{pr}^{(0)}}
\end{equation}
and
\begin{equation}
	r_{pr}^{(1)} = \pm \frac{2 r_{pr}^{(0)3}\sqrt{f(r_{pr}^{(0)})} p'(r_{pr}^{(0)})}{r_{pr}^{(0)2} f''(r_{pr}^{(0)}) - 2 f(r_{pr}^{(0)})}, \;\;\; \omega_{pr}^{(1)} = - p(r_{pr}^{(0)}).
\end{equation}
where the plus/minus signs correspond to prograde/retrograde motion. The above expressions can easily be solved numerically
for $r_{pr}^{(0)}$, $ r_{pr}^{(1)}$, $\omega_{pr}^{(0)}$, and  $\omega_{pr}^{(1)}$, which is done in the following subsections for $\Lambda = 0$ as well as dS/AdS space.

One interesting combination of quantities to consider is $\frac{\omega_+}{|\omega_-|}$. In GR it is known that this ratio is controlled by the black hole spin parameter alone \cite{adair2020,cano2019} in asymptotically flat space (or the spin parameter and black hole mass in asymptotically dS/AdS space), whereas in the 4DEGB theory we also have dependence on the higher order coupling. In principle this feature could be useful in constraining the 4DEGB coupling constant via an independent black hole spin/mass measurement. \newline



We conclude this section by investigating the stability of the photon ring orbits. Of course they are unstable, although we can better understand this instability by calculating the associated Lyapunov exponent \cite{adair2020,mashhoon1985}, which provides a measure of the growth of the photon orbit instability as a function of time. More precisely, under the Eikonal approximation the Lyapunov exponent comes from the imaginary part of the quasi-normal mode frequencies of the photon orbit \cite{mashhoon1985} (while the real part is related to the angular velocity of the unstable null orbits). It can be proven \cite{cardoso2009} that for any static, spherically symmetric spacetime
\begin{equation}
	\omega_\mathrm{QNM} = \omega_{pr}^{(0)} \ell - i(n + \frac{1}{2})|\lambda |
\end{equation}
where $\omega_\mathrm{QNM}$ is a quasi-normal mode frequency, and $\lambda$ is the Lyapunov exponent associated with the orbit. We can extract its value by again following the work of \cite{adair2020,mashhoon1985}. We begin with differentiating the geodesic equation \eqref{photogeo} with respect to the affine parameter
\begin{eqnarray}
	\frac{d}{ds} \left( \frac{1}{2} \dot{r}^2 + V_{\text{ph}}(r) = 0 \right) \Rightarrow 
	\frac{d^2r}{ds^2} + 
	\frac{d V_{\text{ph}}(r)}{dr} = 0 \label{eq:ly_difeq}
\end{eqnarray}
and then perturb  \eqref{eq:ly_difeq} according to
\begin{equation}
\begin{aligned}
	r(t) &= r_{pr}^{(0)} \left( 1 + \epsilon F(t) \right), \; s(t) = \frac{t}{\beta} + \epsilon G(t), \; \\
 \phi (t) &= |\omega_\pm|(1 + \epsilon H(t))
 \end{aligned}
\end{equation}
where $\epsilon$ is a small parameter controlling the perturbation strength
and $\beta = \frac{dt}{ds}\vert_{r\to r_{pr}}$ is a parameter relating coordinate time to $s$ in the absence of a perturbation. We require the system to have vanishing perturbation at $t = 0$. Expanding \eqref{eq:ly_difeq} to first order in $\epsilon$ yields
\begin{equation}
	\beta^2 F''(t) +  F(t) V_{\text{ph}}''(r_{pr}^{(0)})  = 0
\end{equation}
which has the solution
\begin{equation}
	F(t) \propto \sinh \left( t \lambda\right)
\end{equation}
where 
$$\lambda^2 =  -\frac{V''(r_{ps}^{(0)})}{\beta^2} \coloneqq U \; .
$$
As with everything else up to this point, we write the solution as a zeroth-order term plus a linear rotational correction:
\begin{equation}
	\lambda = \lambda^{(0)} + a \lambda^{(1)}.
\end{equation}

Since the perturbative and non-perturbative parts should hold true independently (and since $V''(r_{ps}^{(0)}) = V_{0}'' + a V_{1}''$), we can break the relation between $\lambda$ and $V''(r_{ps}^{(0)})$ into two equations, namely: 
\begin{eqnarray}
    \lambda_0 = \sqrt{U_0} \\
    \lambda_1 = \frac{U_1}{2 \lambda_0}
\end{eqnarray}
where $U_0$ and $U_1$ are the 0th and 1st order terms of a power series representation of $U$ in the rotation parameter $a$, respectively.

\subsubsection{$\mathbf{\Lambda = 0}$}

\begin{figure*}
	\begin{subfigure}{.5\textwidth}
		\includegraphics[width=8.6cm]{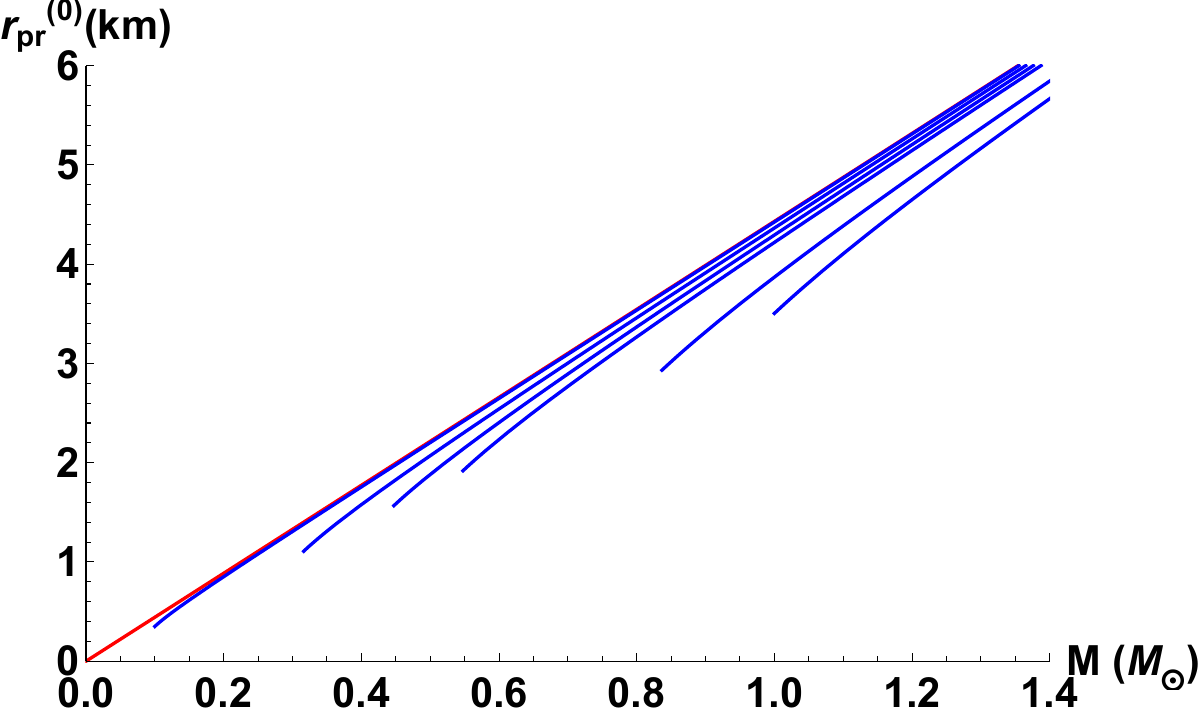}
		
		\label{fig:}
	\end{subfigure}\hfill%
	\begin{subfigure}{.5\textwidth}
		\includegraphics[width=8.6cm]{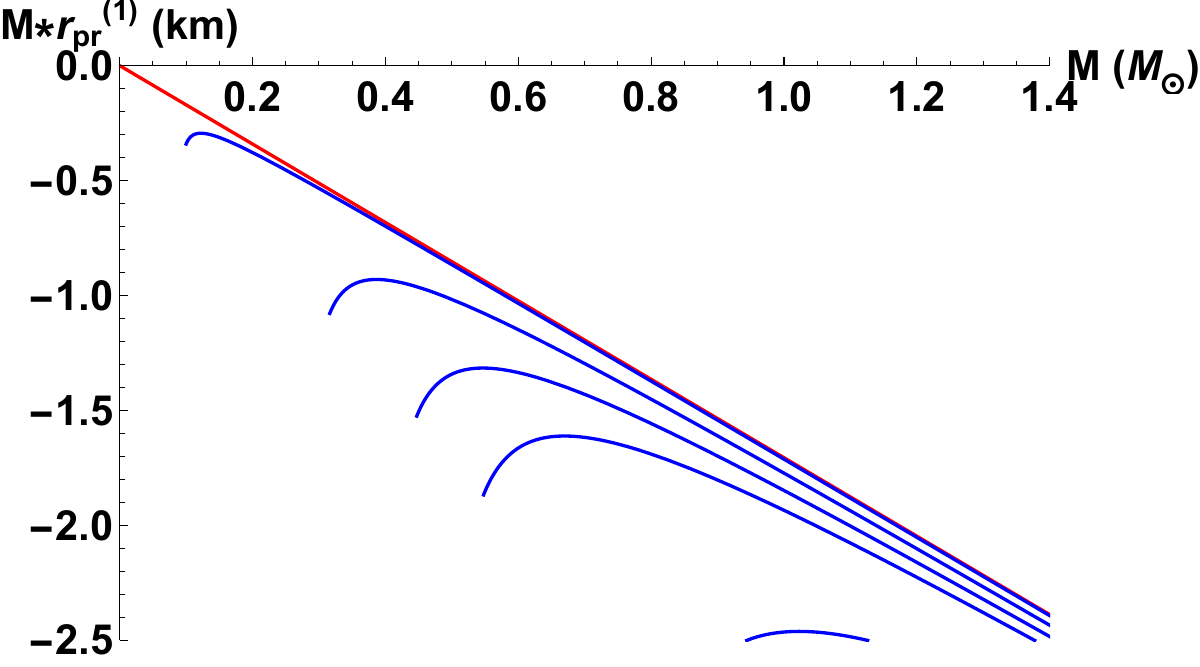}
		
		\label{fig:}
	\end{subfigure}\vspace{10mm}
	
	\begin{subfigure}{.5\textwidth}
		\includegraphics[width=8.6cm]{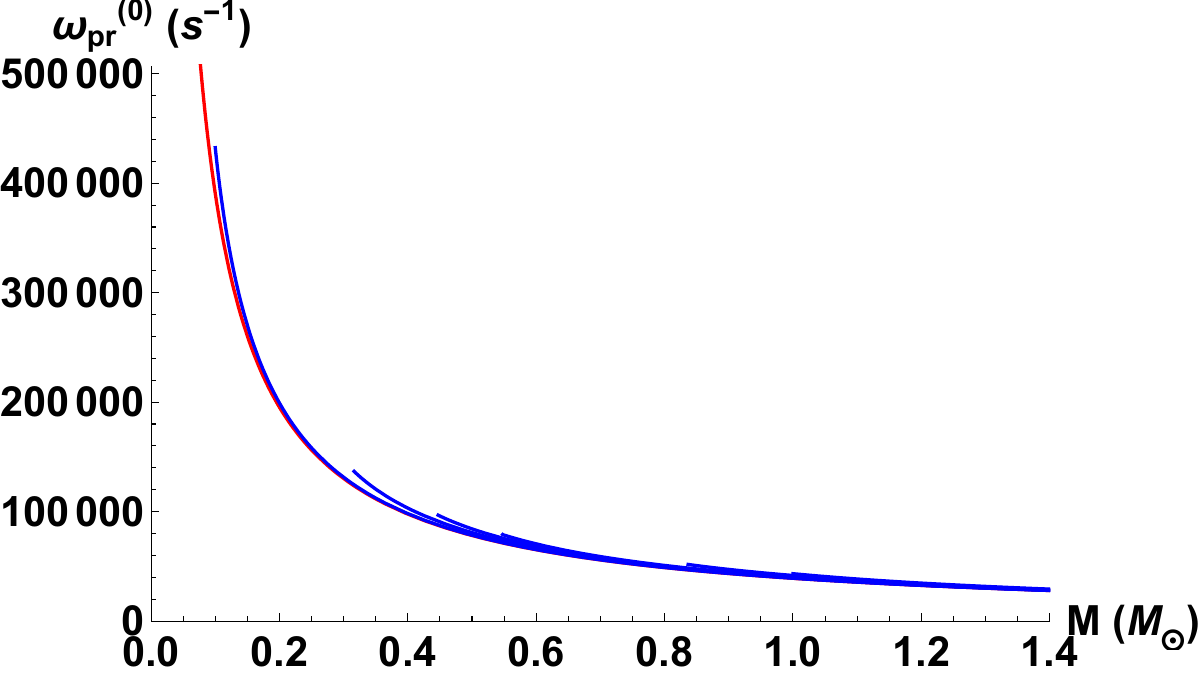}
		
		\label{fig:}
	\end{subfigure}\hfill%
	\begin{subfigure}{.5\textwidth}
		\includegraphics[width=8.6cm]{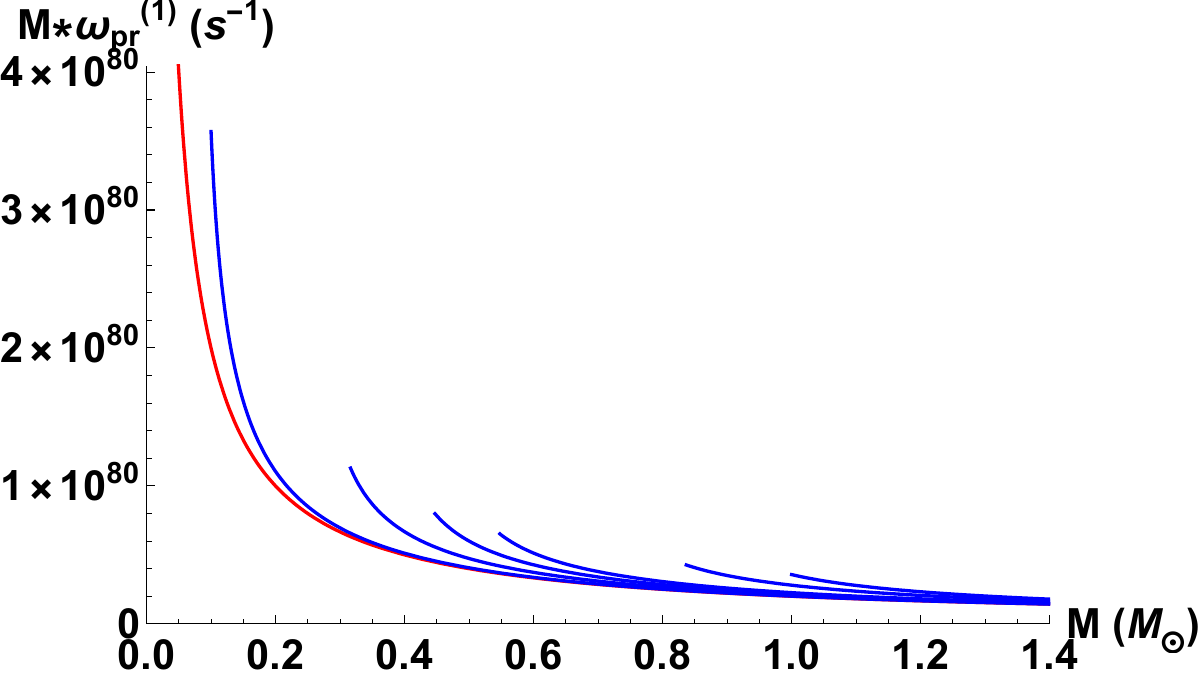}
		
		\label{fig:}
	\end{subfigure}\vspace{10mm}
	\caption[Photon ring solutions when $\Lambda = 0$.]{Photon ring solutions when $\Lambda = 0$. In all cases the red line represents the GR solution ($\alpha=0$), and the blue lines represent the 4DEGB solutions for $\frac{\alpha}{M_{\odot}^2}$ = 0.01, 0.1, 0.2, 0.3, 0.7, 1 (from left to right).}
	\label{fig:ps l0}
\end{figure*}

\begin{figure*}
\subcaptionbox[0th order Lyapunov exponent for the photon ring when $\Lambda = 0$.]{0th order Lyapunov exponent  correction for the photon ring when $\Lambda = 0$. The red curve shows the GR ($\alpha = 0$) result, whereas the blue curves represent the 4DEGB results for $\frac{\alpha}{M_{\odot}^2}$ = 0.01, 0.05, 0.1, 0.2, 0.3, 0.7, 1. \label{fig:lam0 l0}}[0.43\textwidth]
{\includegraphics[valign=t,width=0.43\textwidth]{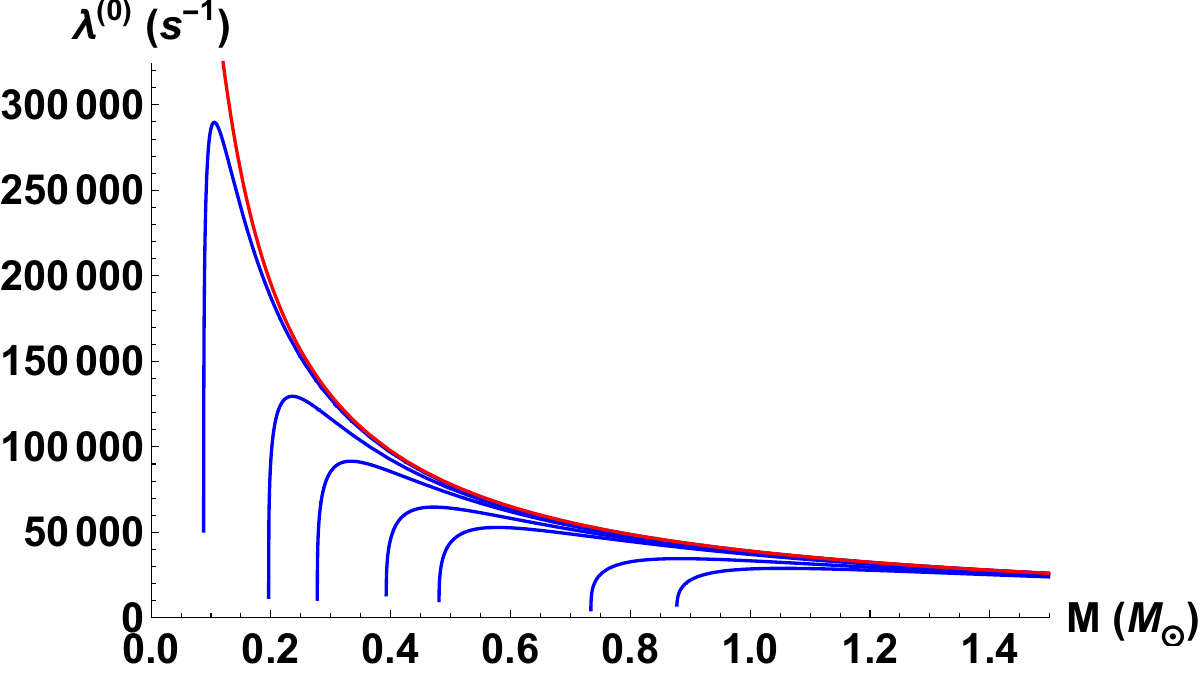}}
\hspace{0.05\textwidth} 
\subcaptionbox[Leading order Lyapunov exponent correction for the photon ring when $\Lambda = 0$.]{Leading order Lyapunov exponent correction for the photon ring when $\Lambda = 0$. The red curve shows the GR ($\alpha = 0$) result, whereas the blue curves represent the 4DEGB results for $\frac{\alpha}{M_{\odot}^2}$ = 0.01, 0.05, 0.1, 0.2, 0.3, 0.7, 1. \label{fig:lam1 l0}}[0.43\textwidth]
{\includegraphics[valign=t,width=0.43\textwidth]
{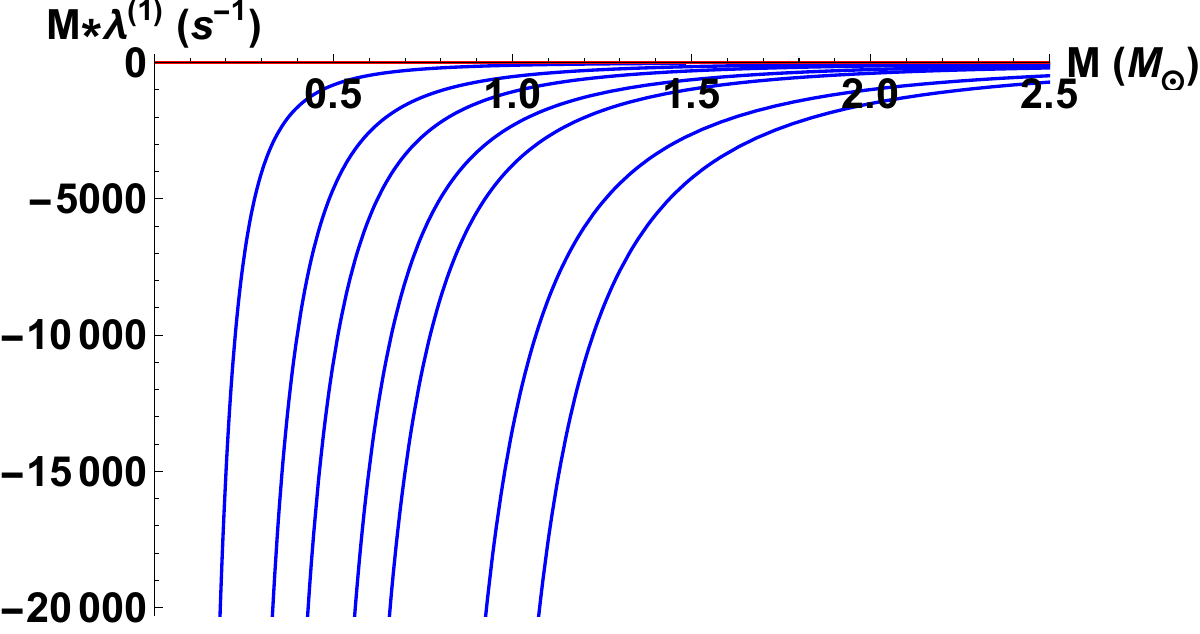}
}
\caption{Lyapunov exponents when $\Lambda = 0$.}
\end{figure*}

In this subsection we solve for the photon ring in asymptotically flat space. The solutions to equation \eqref{eq:ps perturbative} are plotted in figure \ref{fig:ps l0} alongside the results for GR. As the mass for a given 4DEGB solution approaches its minimum allowed mass, the results increasingly depart from those in GR (whereas they become virtually indistinguishable for larger masses). All curves are similar in form to the GR results, apart from this low mass behaviour. It is worth noting that the reason for the large values on the $y$-axes of figures \ref{fig:lam0 l0} and \ref{fig:lam1 l0} are due to the unit conversion from inverse solar masses to inverse seconds. If instead the quantity $\lambda/\Omega$ is plotted (see \cite{cardoso2009,giri2022}), results on the order of unity are observed.

\begin{figure}
	\includegraphics[width=8cm]{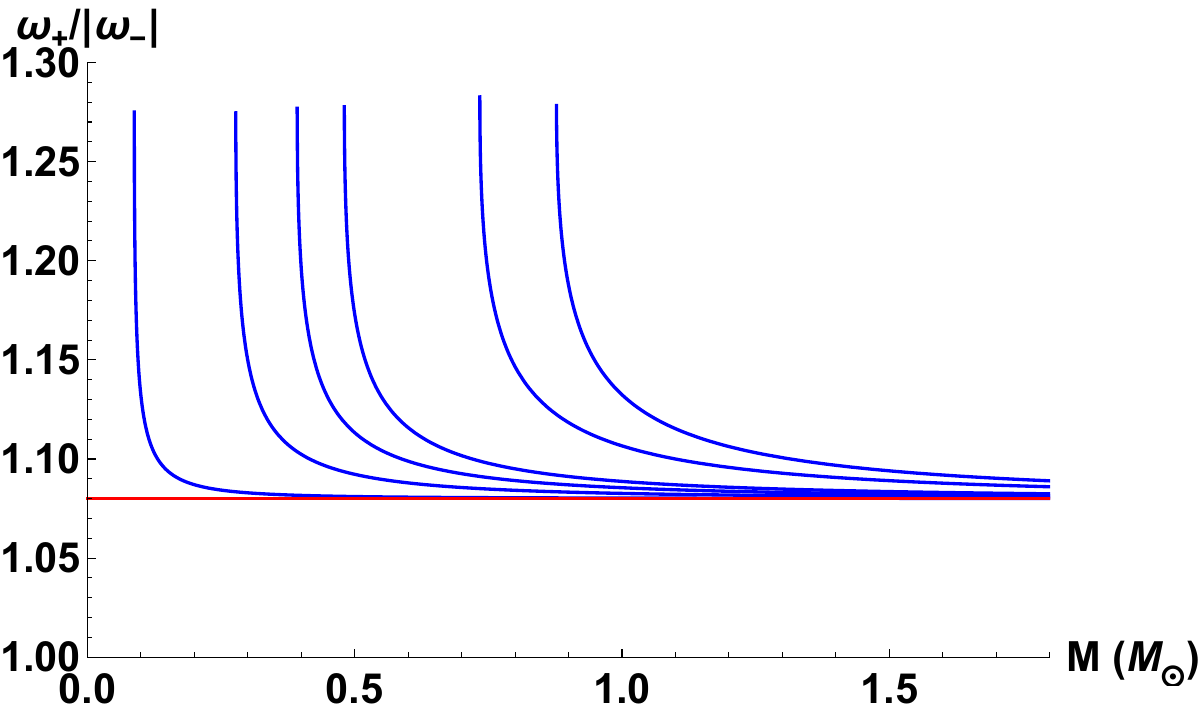}
	\centering
	\caption[Ratio of $\frac{\omega_+}{|\omega_-|}$ when $\Lambda = 0$.]{Ratio of $\frac{\omega_+}{|\omega_-|}$ in GR (red) plotted against the same ratio in the 4DEGB theory (blue) when $\Lambda = 0$ for $\frac{\alpha}{M_{\odot}^2}$ = 0.01, 0.1, 0.2, 0.3, 0.7, 1 (from left to right). In this figure we have fixed $\chi = 0.1$.}
	\label{fig:ratio l0}
\end{figure}

In asymptotically flat spacetimes, the ratio $\Omega \coloneqq \frac{\omega_+}{|\omega_-|}$ ends up being a function of $M$, $\alpha$, and $\chi$. The numerical solutions of this ratio for $\Lambda = 0$ are plotted in figure \ref{fig:ratio l0} alongside the corresponding quantity from GR. We see that the GR results are constant with respect to mass, whereas the 4DEGB results change rapidly as mass gets small. Note that the greatest departure from the GR results occurs again in the low mass region near the minimum, and convergence with the GR solution is observed as mass gets large.  

Similarly, in figures \ref{fig:lam0 l0} and \ref{fig:lam1 l0}, we find that the stability of the photon ring in the 4DEGB theory is very similar to GR unless the mass is near its 
minimum value, in which case the value of $\lambda$ drops considerably. While the leading order corrections to the Lyapunov exponent vanish in Einstein's theory, we find a nonzero negative value that converges to GR in the large mass limit. Since both the 0th and leading order contributions to $\lambda$ in the 4DEGB theory are large and negative near $M_{\rm{min}}$, near-minimal mass objects should have significantly less unstable photon rings than the values predicted by GR.


\subsubsection{AdS ($\mathbf{\Lambda < 0}$)}\label{subsec:adsphotonring}

\begin{figure*}
	\begin{subfigure}{.5\textwidth}
		\includegraphics[width=8.6cm]{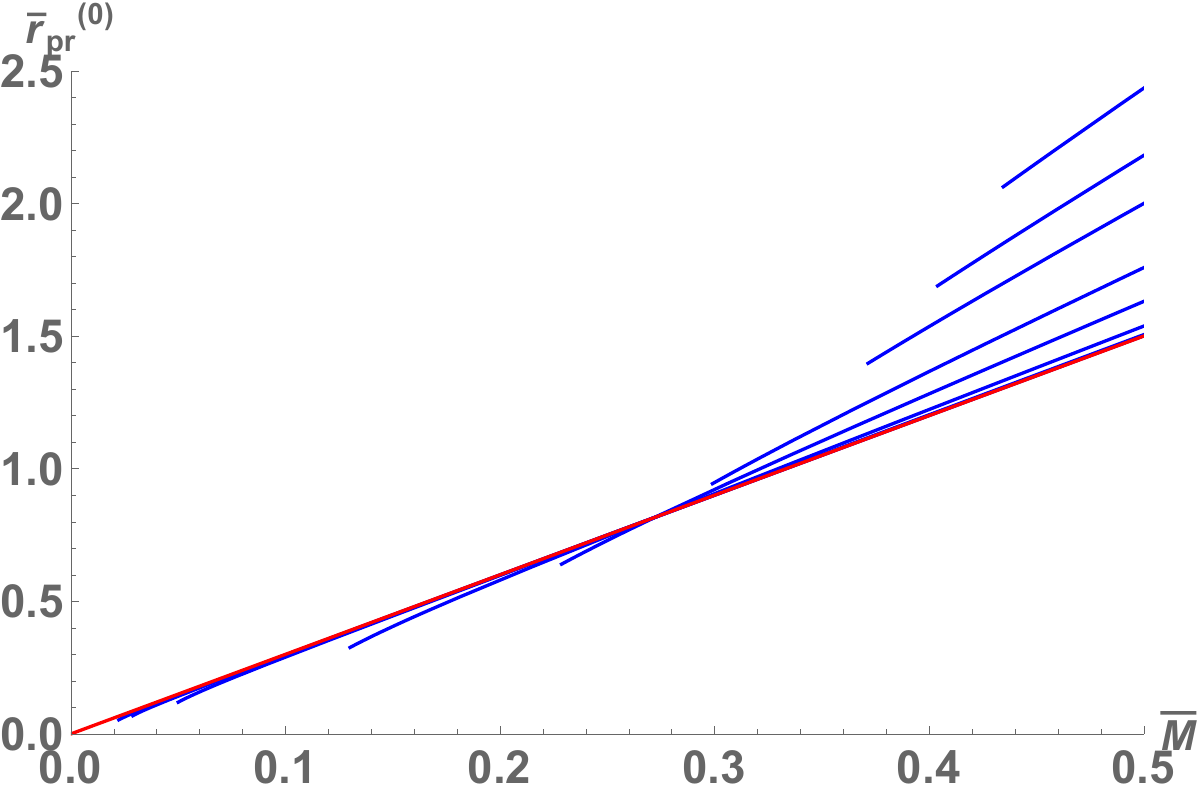}
		
		\label{fig:}
	\end{subfigure}\hfill%
	\begin{subfigure}{.5\textwidth}
		\includegraphics[width=8.6cm]{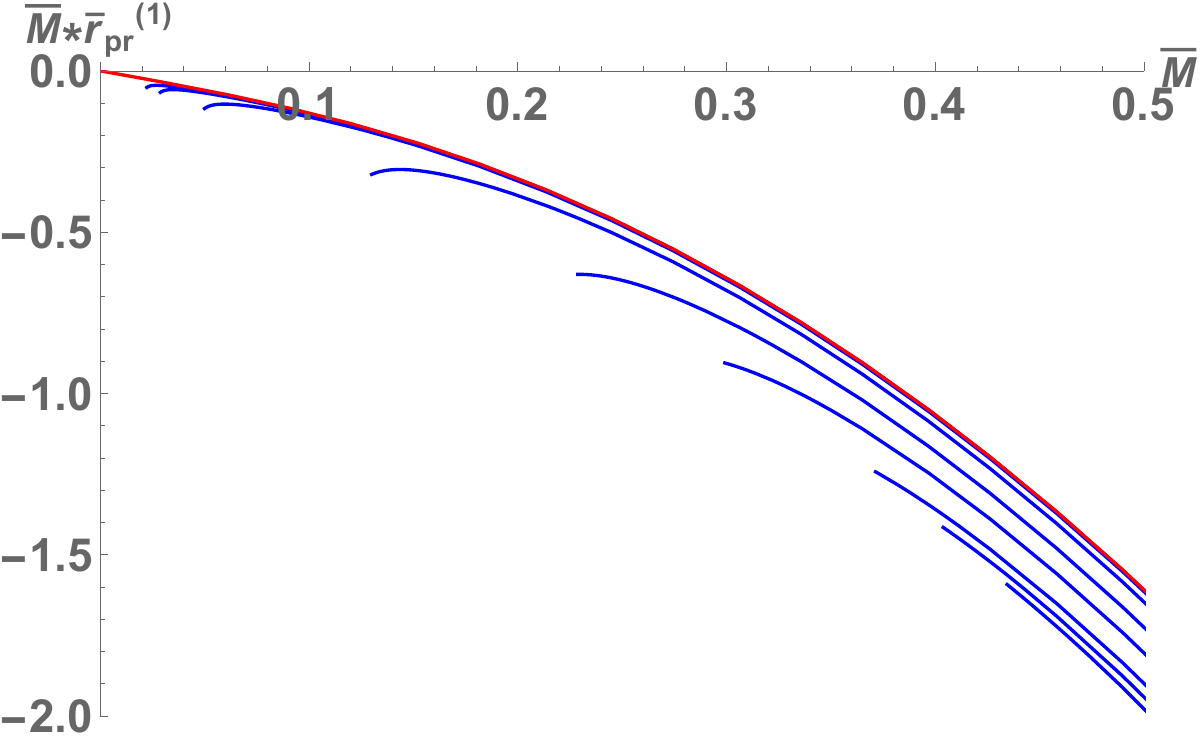}
		
		\label{fig:}
	\end{subfigure}\vspace{10mm}
	
	\begin{subfigure}{.5\textwidth}
		\includegraphics[width=8.6cm]{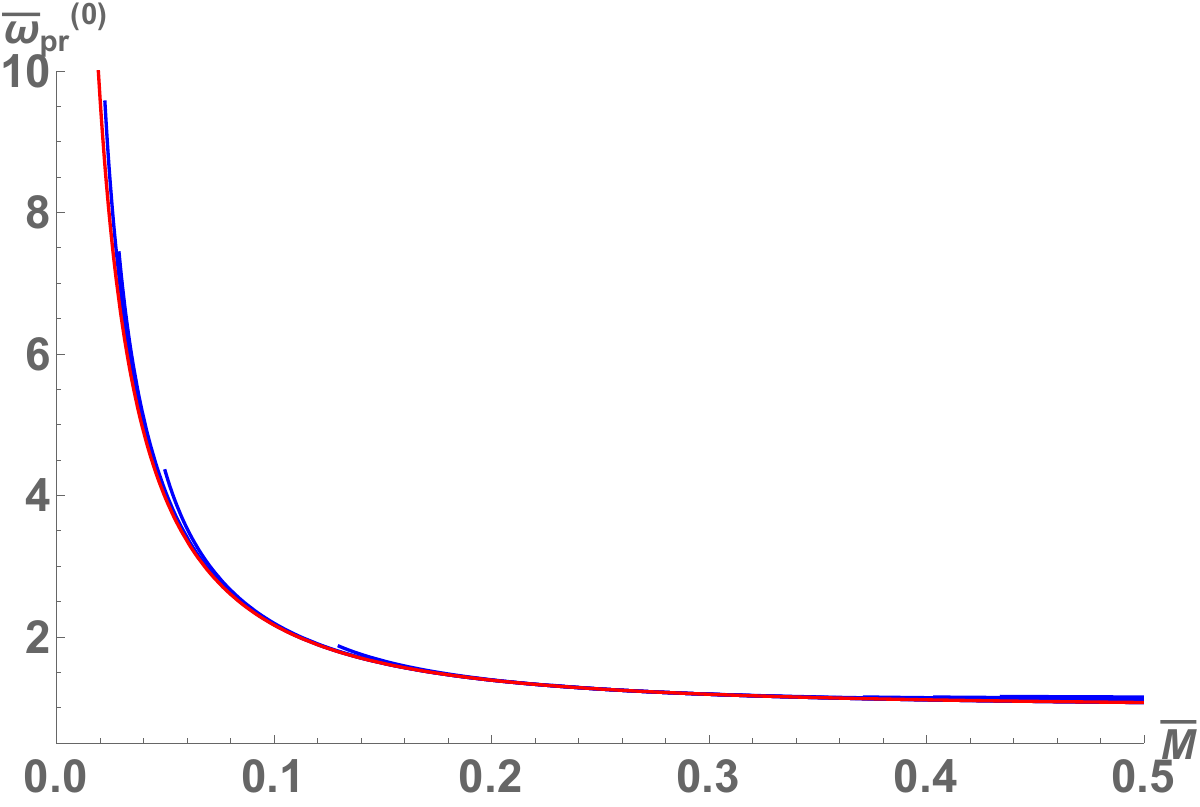}
		
		\label{fig:}
	\end{subfigure}\hfill%
	\begin{subfigure}{.5\textwidth}
		\includegraphics[width=8.6cm]{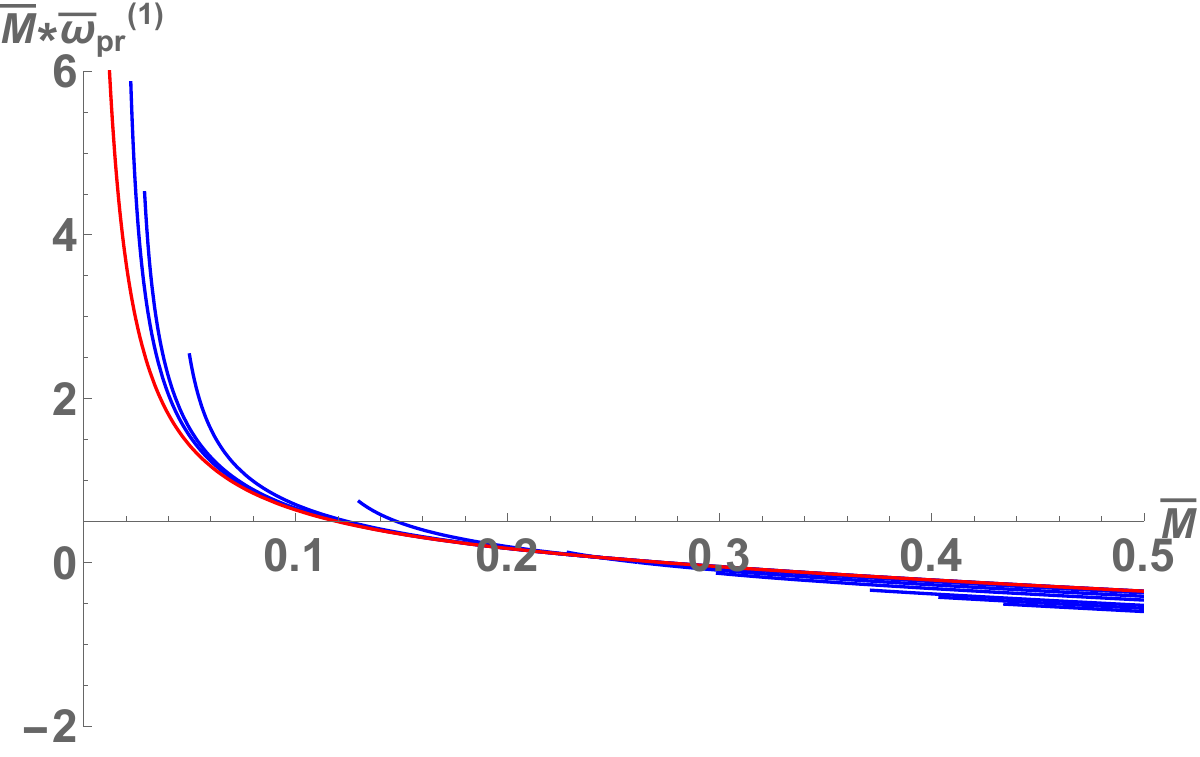}
		
		\label{fig:}
	\end{subfigure}\vspace{10mm}
	\caption[Photon ring solutions in AdS when $\alpha$ is small.]{Photon ring solutions in an asymptotically anti-de Sitter spacetime for $\frac{\alpha}{\alpha_C}$ = 0.002, 0.0033, 0.01, 0.066, 0.2, 0.33, 0.5, 0.5833, 0.66 (from left to right in blue) against the GR ($\alpha=0$) solution in red.}
	\label{fig:ps ads smallalpha}
\end{figure*}

\begin{figure*}
	\begin{subfigure}{.5\textwidth}
		\includegraphics[width=8.6cm]{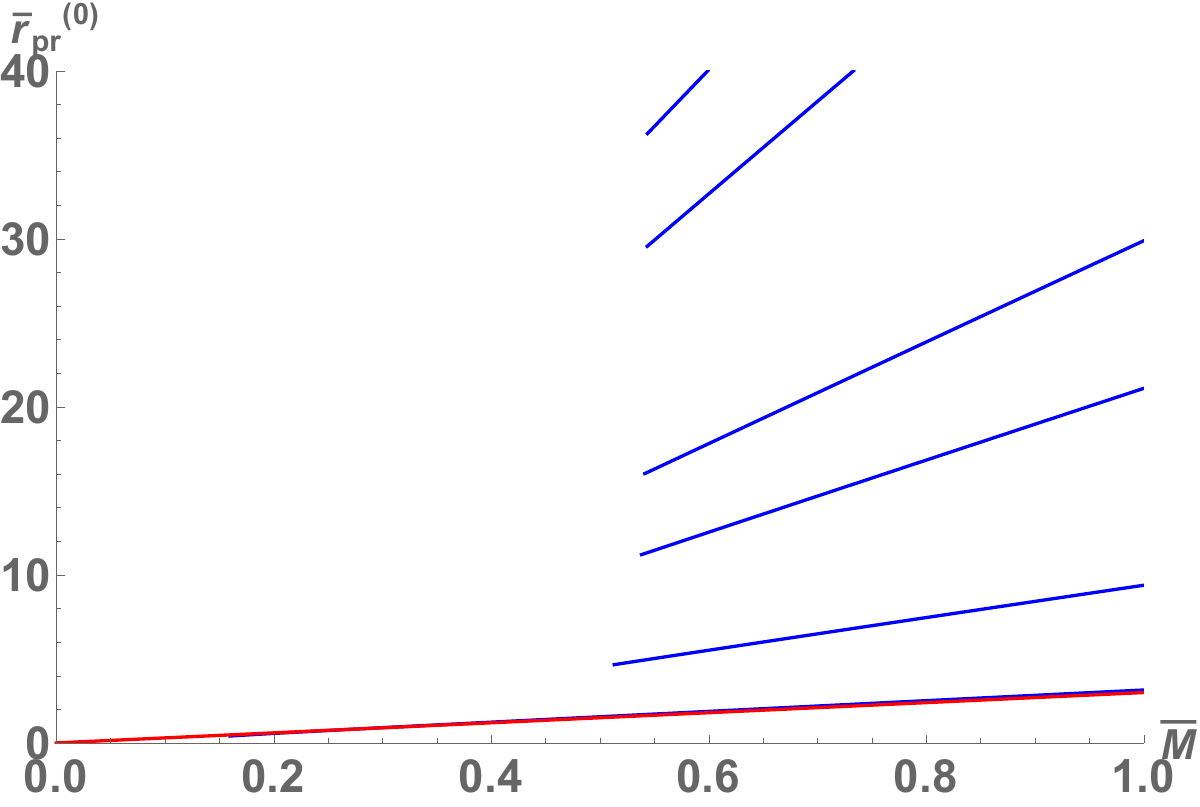}
		
		\label{fig:}
	\end{subfigure}\hfill%
	\begin{subfigure}{.5\textwidth}
		\includegraphics[width=8.6cm]{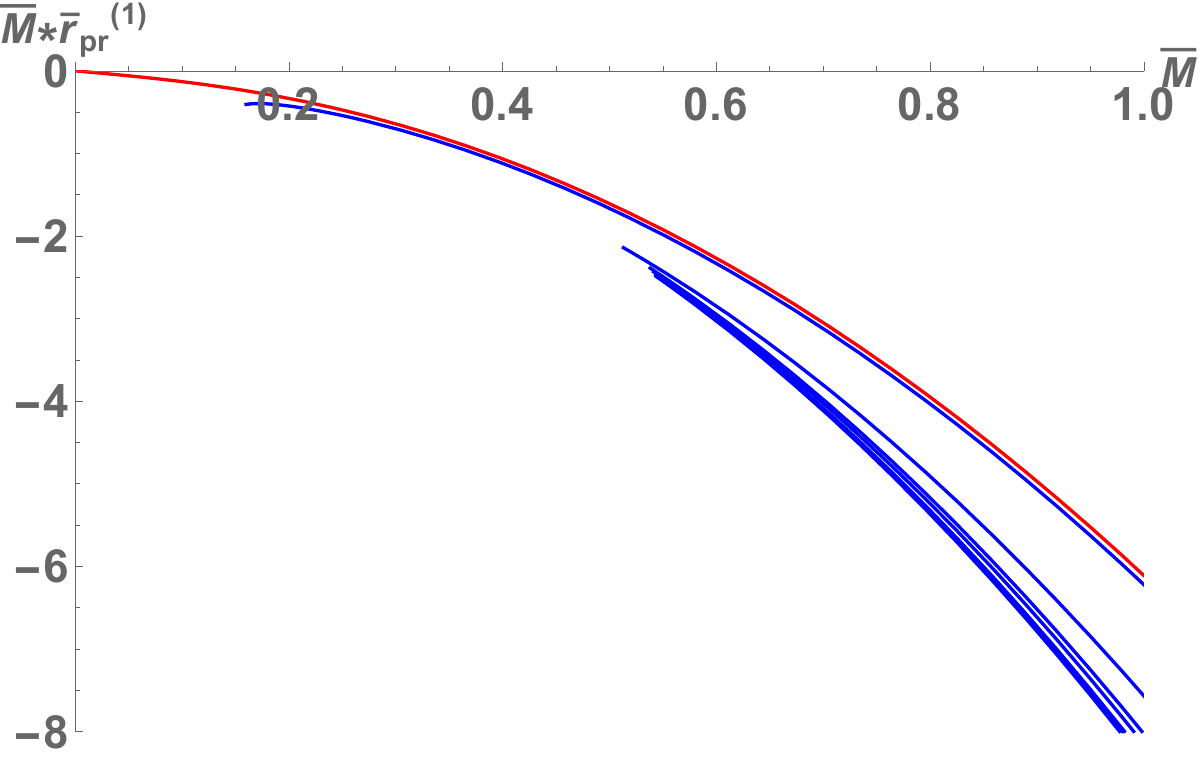}
		
		\label{fig:}
	\end{subfigure}\vspace{10mm}
	\begin{subfigure}{.5\textwidth}
		\includegraphics[width=8.6cm]{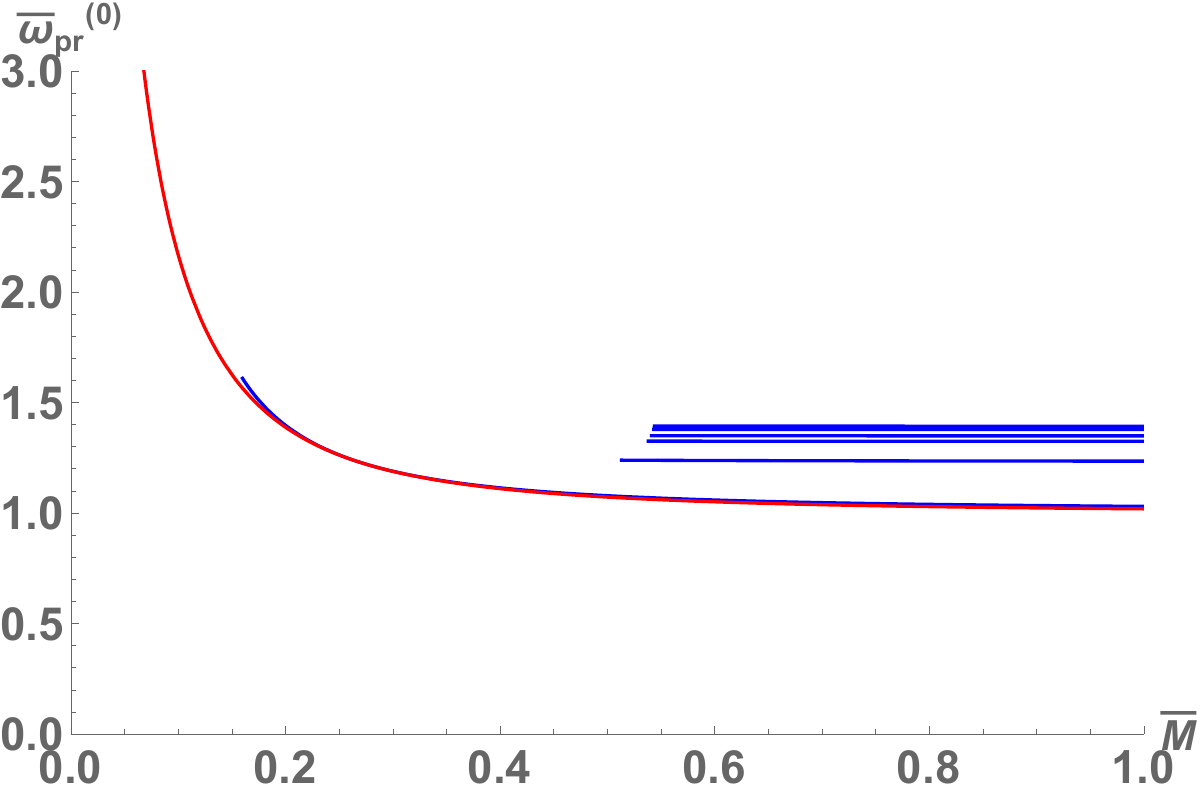}
		
		\label{fig:}
	\end{subfigure}\hfill%
	\begin{subfigure}{.5\textwidth}
		\includegraphics[width=8.6cm]{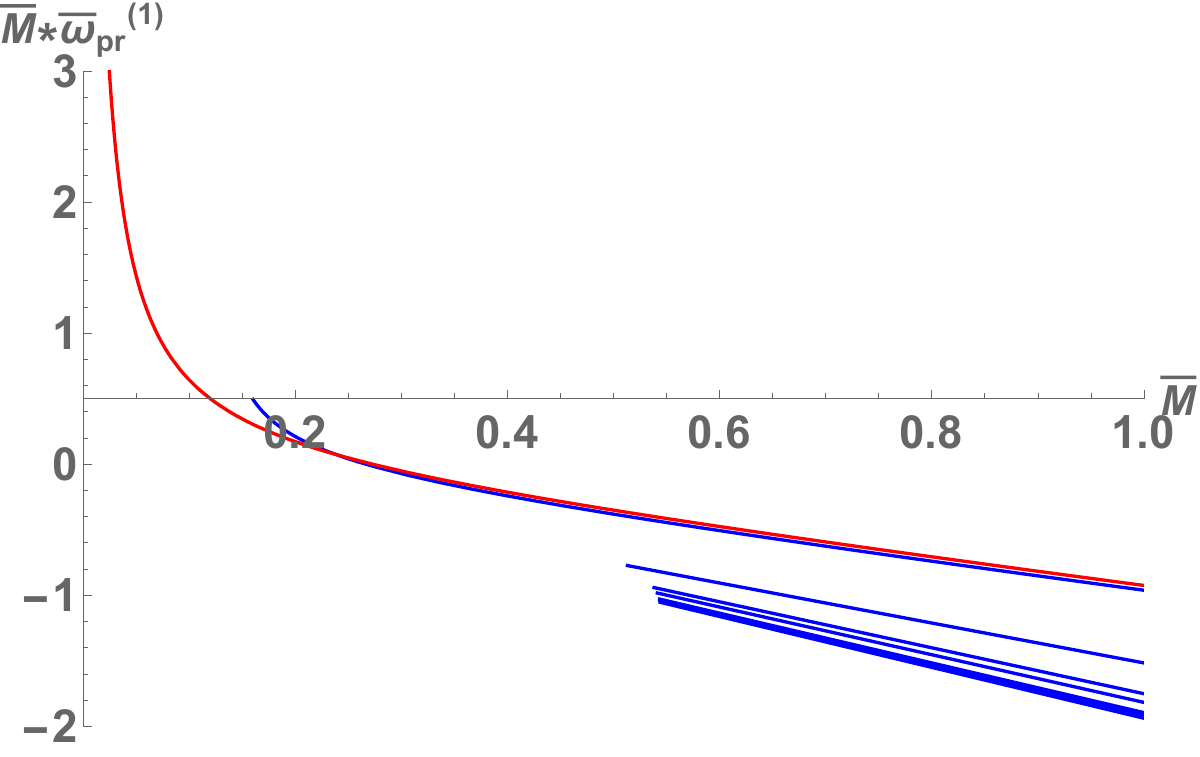}
		
		\label{fig:}
	\end{subfigure}\vspace{10mm}
	\caption[Photon ring solutions in AdS when $\alpha$ is near criticality.]{Photon ring solutions in the asymptotically anti-de Sitter case when $\alpha$ is near its critical value (ie. $\frac{\alpha}{\alpha_C}$ = 0.1, 0.9, 0.98, 0.99, 0.997, 0.998, 0.999) from left to right in blue against the GR ($\bar{\alpha}=0$) solution in red.}
	\label{fig:ps ads}
\end{figure*}

Here we solve for the photon ring parameters again, this time in asymptotically AdS space. The relevant solutions to \eqref{eq:ps perturbative} are plotted in figures \ref{fig:ps ads smallalpha} and \ref{fig:ps ads} alongside the analogous results from GR. In the small alpha regime we again observe a ``crossover" behaviour at some critical mass for two of the four photon ring parameters (with an additional parameter just touching the GR solution rather than crossing over). The radial location of the photon sphere is uniquely determined by the equation
\begin{equation}
f'(\bar{r}^{(0)}_{ISCO}) - 2\frac{f(\bar{r}^{(0)GR}_{ISCO})}{\bar{r}^{(0)GR}_{ISCO}} = 0.
\end{equation}

As before we expand this equation to leading order in $\bar{\alpha}$:
\begin{equation}
\left(\frac{3 \bar{M}}{\bar{r}^{(0)GR}_{ISCO}}-1\right)-\frac{6 \bar{M} \left(2 \bar{M}-(\bar{r}^{(0)GR}_{ISCO})^3\right)}{(\bar{r}^{(0)GR}_{ISCO})^4} \bar{\alpha} +\mathcal{O}\left(\bar{\alpha} ^2\right) = 0
\end{equation}
and fix the black hole mass in terms of its GR definition (by solving the 0th order coefficient):
\begin{equation}
\bar{M} = \frac{1}{3} \bar{r}^{(0)GR}_{pr}.
\end{equation}

We can then substitute this fixed mass into the leading order coefficient to find the point of interest, yielding a crossover at $\bar{r}^{(0)}_{pr} = \sqrt{\frac{2}{3}}$, or equivalently
\begin{equation}
\bar{M}_{\rm{x}}^{\bar{r}^{(0)}_{pr}} = \sqrt{\frac{2}{27}}
\end{equation}
which matches what is expected from figure \ref{fig:ps ads smallalpha}. For $\bar{\omega}^{(0)}_{pr}$, rather than a crossover we have a point at which the solution touches the GR curve before increasing again. For this point the analysis is identical, and we find
\begin{equation}
\bar{M}_{\rm{x}}^{\bar{\omega}^{(0)}_{pr}} = \bar{M}_{\rm{x}}^{\bar{r}^{(0)}_{pr}} = \sqrt{\frac{2}{27}}.
\end{equation}

In the regime of large $\alpha$ (ie. $\frac{\alpha}{\alpha_C} \sim \frac{9}{10}$, see figure \ref{fig:ps ads}) the solutions become very sensitive to small changes in the 4DEGB coupling, and again we see non-rotating radius results that depart  dramatically from GR at all mass scales in this regime (similar to what was discussed for the AdS ISCO). The other three parameters are bounded in this large $\alpha$ limit, and change very little with respect to varying $\alpha$. 


\begin{figure}
	\includegraphics[width=8cm]{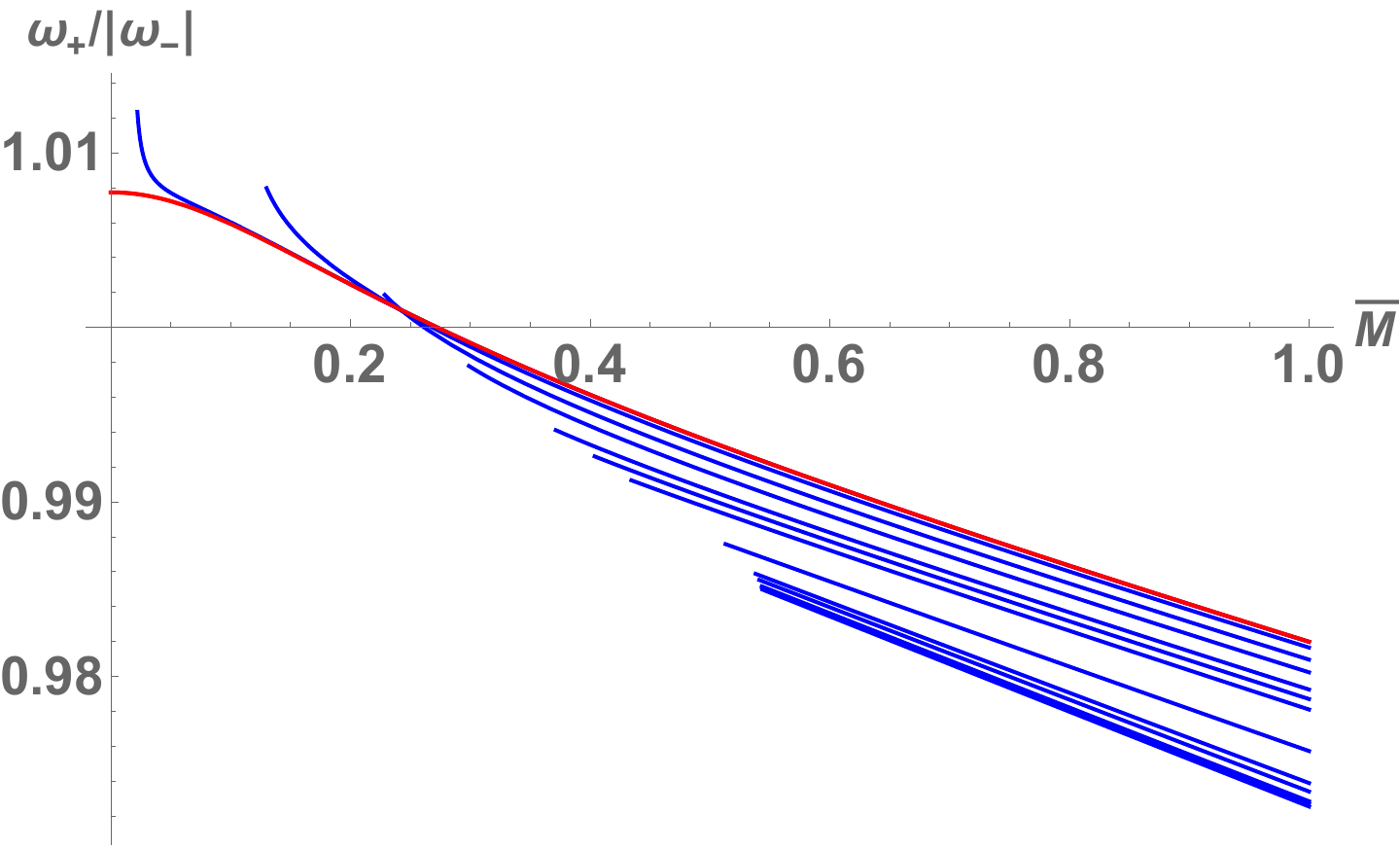}
	\centering
	
	\caption[Ratio of $\frac{\omega_+}{|\omega_-|}$ in AdS.]{Ratio of $\frac{\omega_+}{|\omega_-|}$ in an asymptotically anti-de Sitter spacetime. The GR solution (red) is plotted against the same ratio in the 4DEGB theory (blue) for $\frac{\alpha}{\alpha_C}$ = 0.002, 0.066, 0.2, 0.33, 0.5, 0.5833, 0.66, 0.9, 0.98, 0.99, 0.997, 0.998, 0.999. In this figure we have fixed $\chi = 0.01$.}
	\label{fig:ratio ads}
\end{figure}

In spacetimes with a nonzero cosmological constant, the ratio $\Omega \coloneqq \frac{\omega_+}{|\omega_-|}$ again ends up being a function of $\bar{M}$, $\bar{\alpha}$, and $\chi$. The numerical solutions of this ratio for AdS space are plotted in figure \ref{fig:ratio ads} where we hold $\chi = 0.01$ constant. When $\bar{M}$ is small (and thus $\alpha$ is small), this ratio is larger for the 4DEGB theory than it is for GR. As the fixed mass increases, the curves cross and $\Omega_{GR} > \Omega_{4DEGB}$. Since the 4DEGB theory has a minimum mass that  depends on $\alpha$, only the black holes corresponding to $\alpha/\alpha_C \lessapprox \frac{1}{5}$ will start above the GR curve, whereas solutions with a larger coupling constant are not defined in the small mass regime. In order to find which combinations of $\bar{M}$ and $\chi$ in the 4DEGB theory will give a ratio that agrees with GR (ie. the crossing point), we expand $\Omega_{4DEGB}$ to 1st order in $\alpha$: 
\begin{equation}
\begin{aligned}\label{eq:ratioadscross}
 &\Omega_{4DEGB} = \Omega_{GR}  \\
 &+\frac{  \left(27 \bar{M}^2 \left(729 \bar{M}^4+864 \bar{M}^2-28\right)-32\right) \chi}{\bar{M}^2 \sqrt{81 \bar{M}^2+3} \left(\sqrt{3} \left(27 \bar{M}^2-2\right) \chi+9 \sqrt{27 \bar{M}^2+1}\right)^2} \bar{\alpha}\\
 &+ \mathcal{O}(\bar{\alpha}^2).
 \end{aligned}
\end{equation}
Requiring that the contribution linear in $\bar{\alpha}$ go to 0 implies a crossover mass of $\bar{M} = 0.234239$ which can also be seen via inspection of figure \ref{fig:ratio ads}.


\begin{figure*}
\subcaptionbox[0th order Lyapunov exponent for the photon ring in an AdS when $\alpha$ is small.]{0th order Lyapunov exponent  correction for the photon ring in an asymptotically anti-de Sitter spacetime when $\alpha$ is small. The red curve shows the GR ($\alpha = 0$) result, whereas the blue curves represent the 4DEGB results for $\frac{\alpha}{\alpha_C}$ = 0.002, 0.0033, 0.01, 0.066, 0.2, 0.33, 0.5, 0.5833, 0.66. \label{fig:lam0 ads smallalpha}}[0.43\textwidth]
{\includegraphics[valign=t,width=0.43\textwidth]{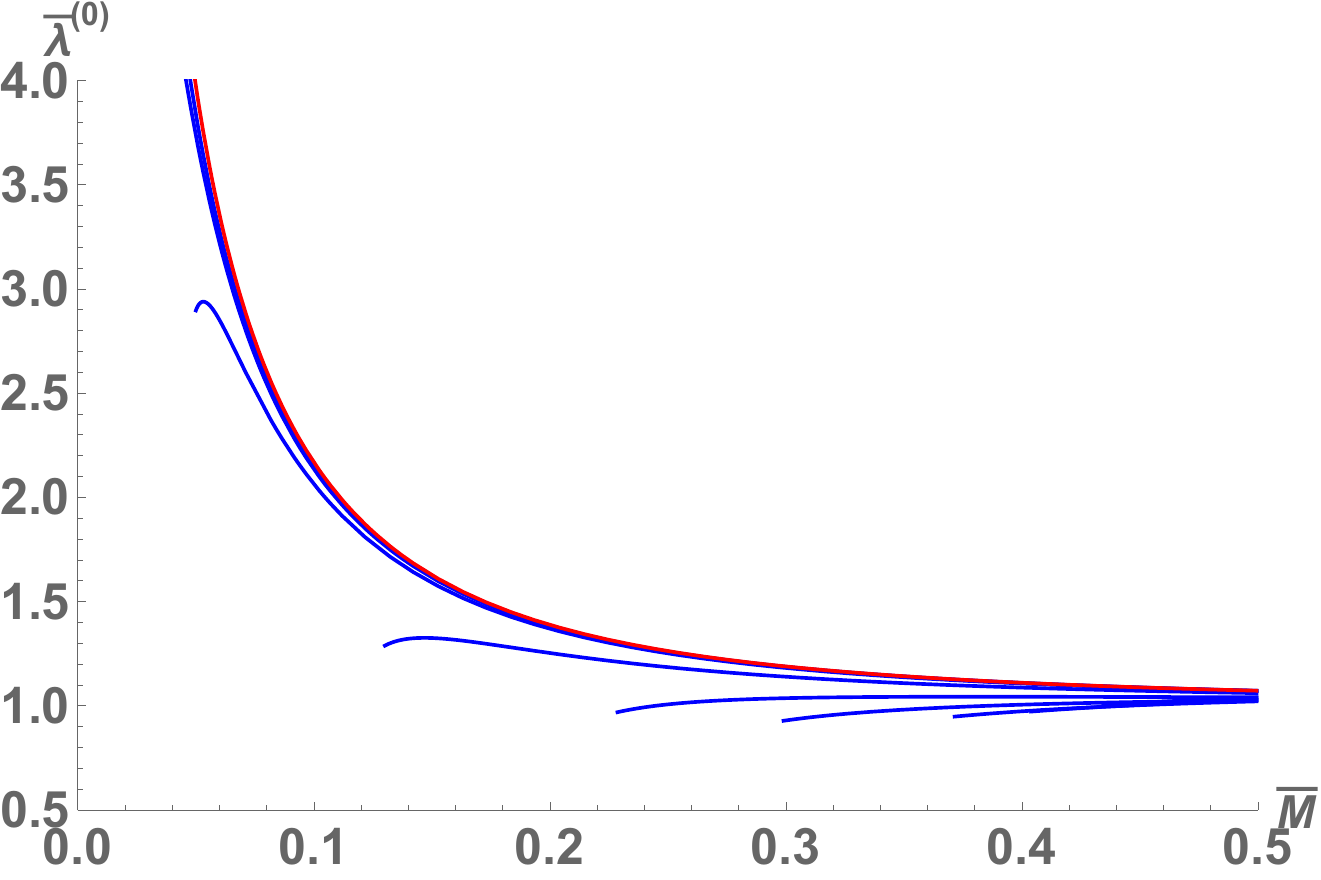}}
\hspace{0.05\textwidth} 
\subcaptionbox[Leading order Lyapunov exponent correction for the photon ring when $\alpha$ is small and $\Lambda < 0$.]{Leading order Lyapunov exponent correction for the photon ring when $\alpha$ is small and $\Lambda < 0$. The red curve shows the GR ($\alpha = 0$) result, whereas the blue curves represent the 4DEGB results for $\frac{\alpha}{\alpha_C}$ = 0.002, 0.0033, 0.01, 0.066, 0.2, 0.33, 0.5, 0.5833, 0.66. \label{fig:lam1 ads smallalpha}}[0.43\textwidth]
{\includegraphics[valign=t,width=0.43\textwidth]
{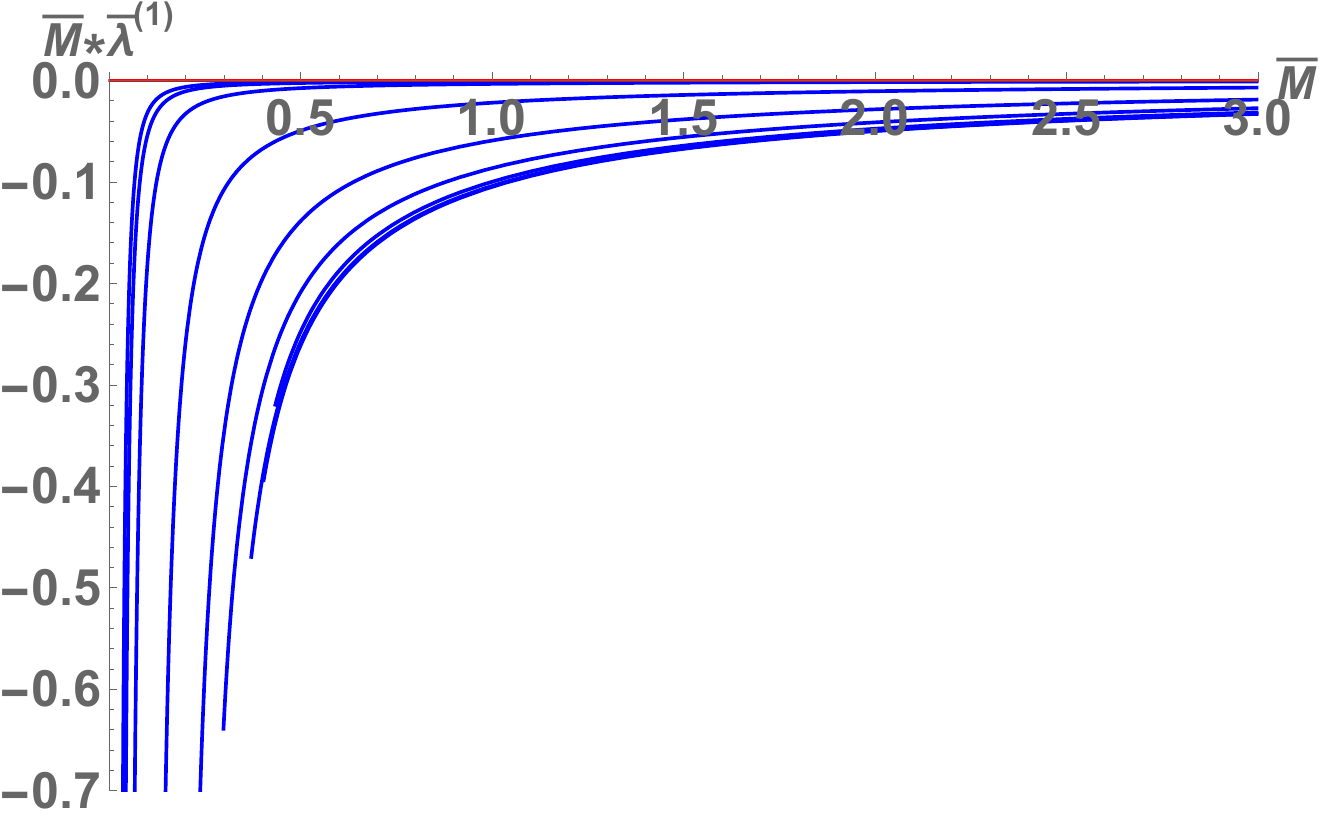}
}
\caption[Lyapunov exponents in AdS when $\alpha$ is small.]{Lyapunov exponents when $\Lambda < 0$ and $\alpha$ is small.}
\end{figure*}

\begin{figure*}
\subcaptionbox[0th order Lyapunov exponent for the photon ring as $\alpha$ approaches criticality in AdS.]{0th order Lyapunov exponent for the photon ring in an asymptotically anti-de Sitter spacetime as $\alpha$ approaches criticality. The red curve shows the GR ($\alpha = 0$) result, whereas the blue curves represent the 4DEGB results for $\frac{\alpha}{\alpha_C}$ = 0.1, 0.3, 0.5, 0.9, 0.98, 0.99, 0.999. \label{fig:lam0 ads corr}}[0.43\textwidth]
{\includegraphics[valign=t,width=0.43\textwidth]{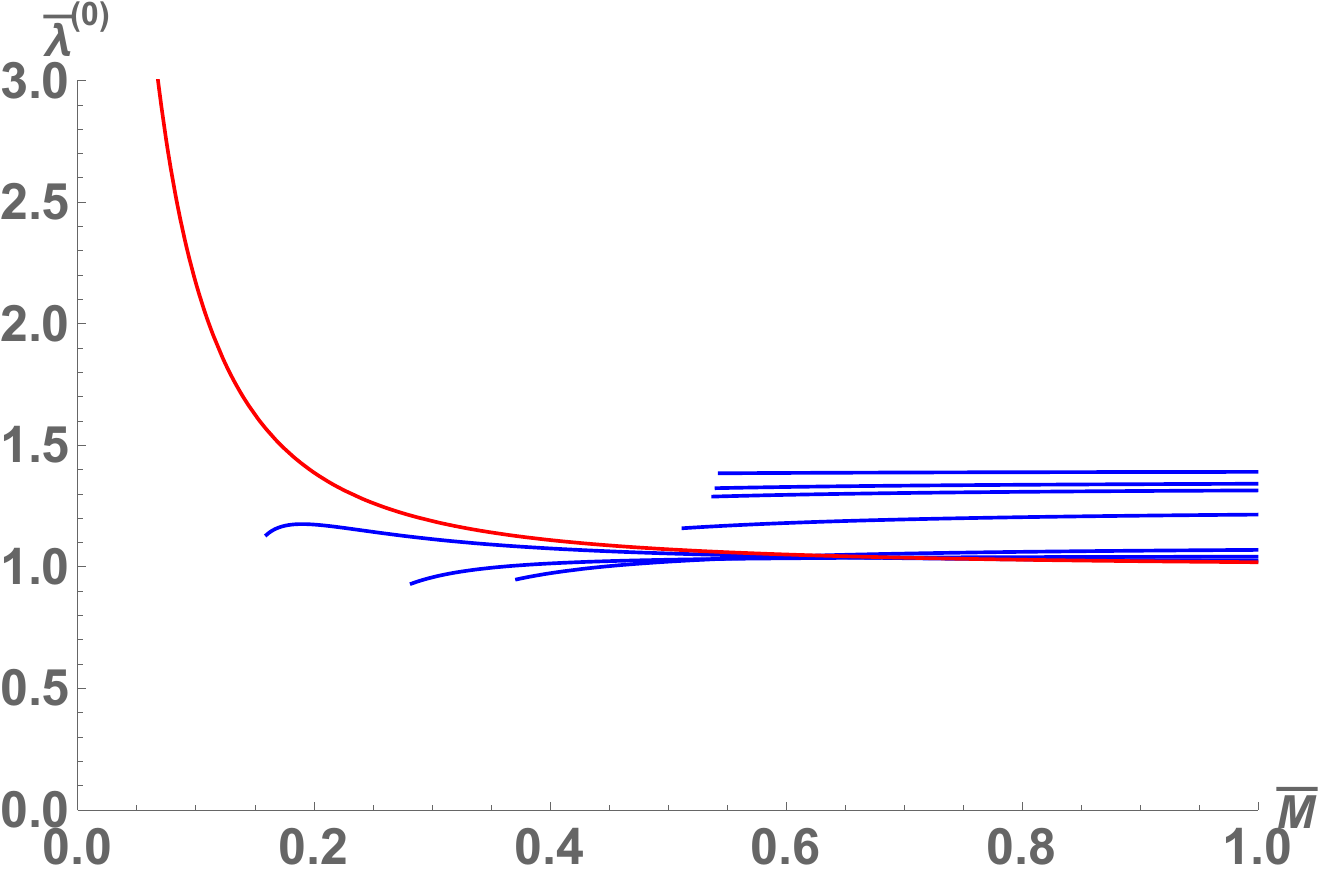}}
\hspace{0.05\textwidth} 
\subcaptionbox[Leading order Lyapunov exponent correction for the photon ring in AdS as $\alpha$ approaches criticality.]{Leading order Lyapunov exponent correction for the photon ring in an asymptotically anti-de Sitter spacetime as $\alpha$ approaches criticality. The red curve shows the GR ($\alpha = 0$) result, whereas the blue curves represent the 4DEGB results for $\frac{\alpha}{\alpha_C}$ = 0.1, 0.3, 0.5, 0.9, 0.98, 0.99, 0.999. \label{fig:lam1 ads corr}}[0.43\textwidth]
{\includegraphics[valign=t,width=0.43\textwidth]
{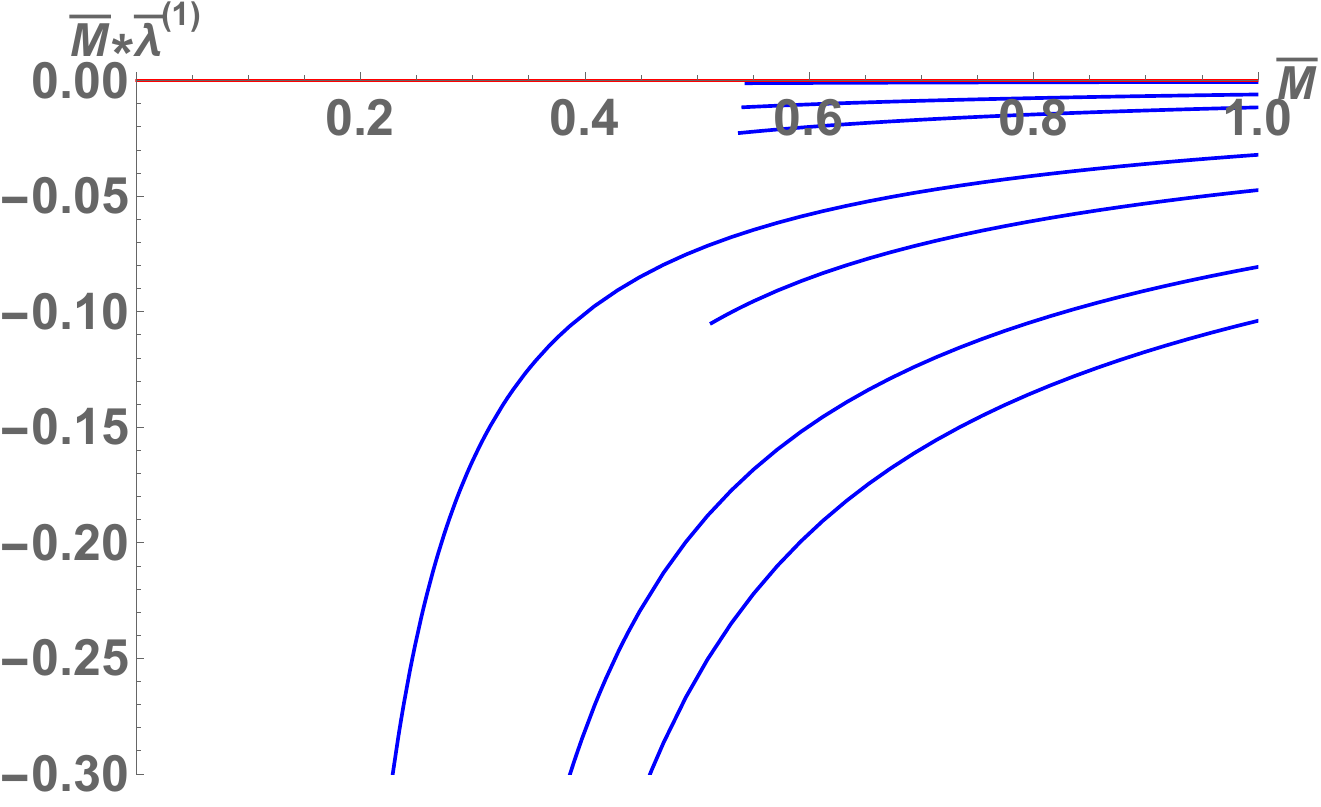}
}
\caption[Lyapunov exponents in AdS when $\alpha$ is near criticality.]{Lyapunov exponents when $\Lambda < 0$ and $\alpha$ is near criticality.}
\end{figure*}

When $\alpha$ is small we again see values for the Lyapunov exponent that are similar in form to GR until the minimum mass is approached (see figures \ref{fig:lam0 ads smallalpha} and \ref{fig:lam1 ads smallalpha}). We see  that the departure in this regime is less dramatic than what was seen in the asymptotically flat case. However, as $\alpha$ approaches criticality (figures \ref{fig:lam0 ads corr} and \ref{fig:lam1 ads corr}), we notice a change in direction in these results with respect to increasing $\alpha$. The 0th order contribution for large enough  $\alpha$ will either cross over or start above the GR results (ie. more unstable), and the leading order corrections end up converging back to GR near $\alpha_C$, with even the point at $M_{\rm{min}}$ not diverging too far from the red curve.


\subsubsection{dS ($\mathbf{\Lambda > 0}$)}\label{subsec:dsphotonring}

\begin{figure*}
	\begin{subfigure}{.46\textwidth}
		\includegraphics[width=1\textwidth]{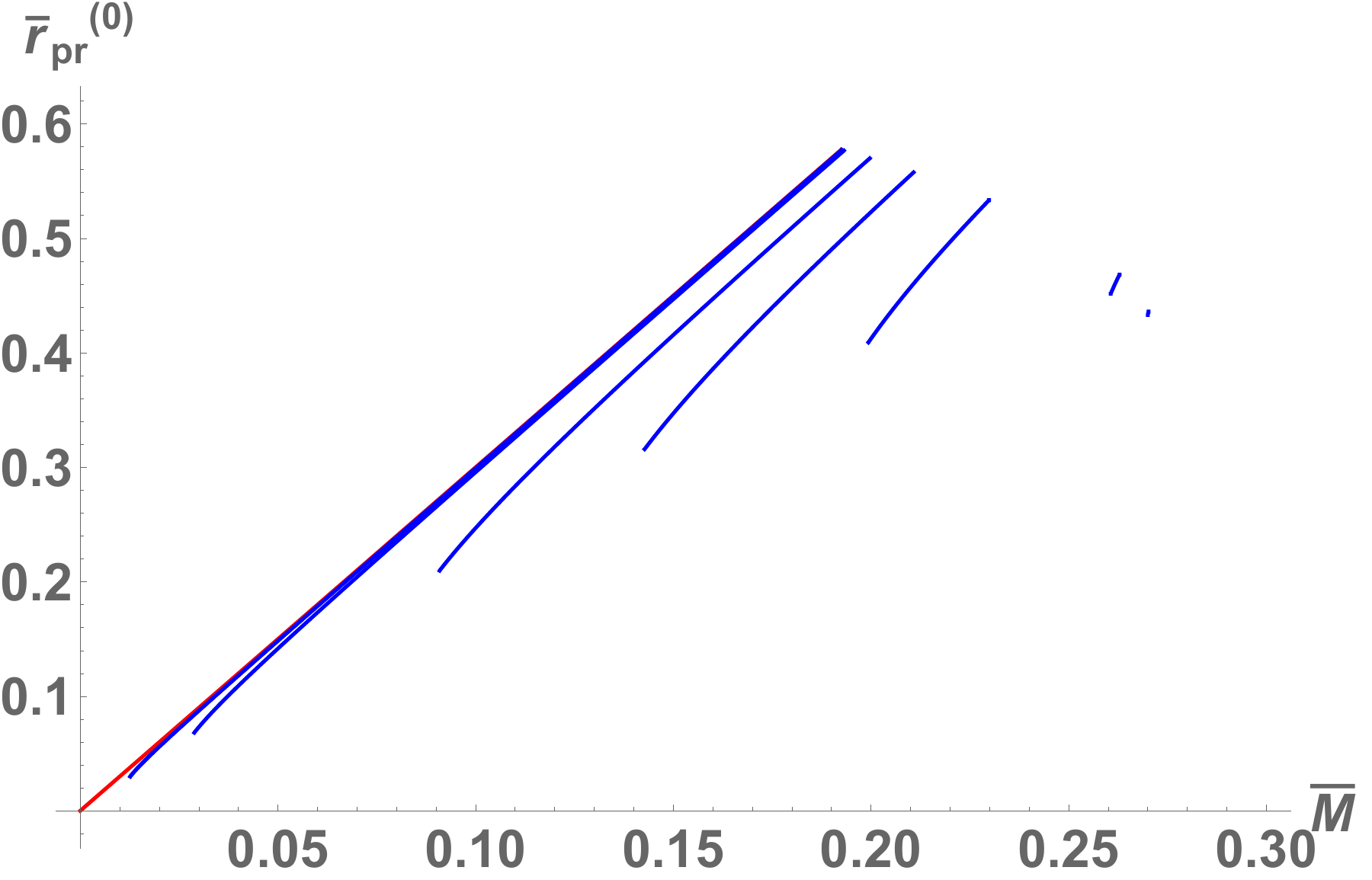}
		
		\label{fig:}
	\end{subfigure}\hfill%
	\begin{subfigure}{.46\textwidth}
		\includegraphics[width=1\textwidth]{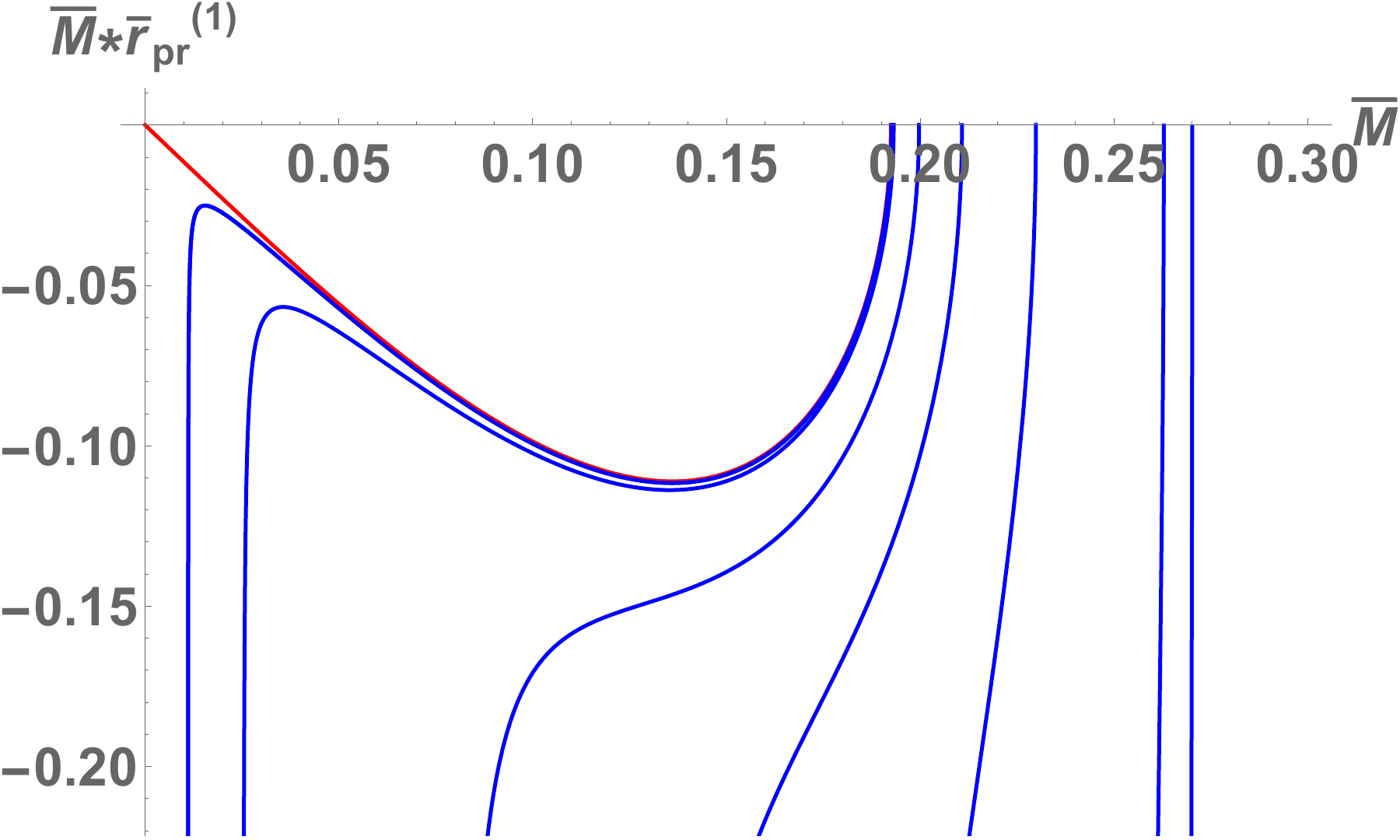}
		
		\label{fig:}
	\end{subfigure}\vspace{10mm}
	
	\begin{subfigure}{.46\textwidth}
		\includegraphics[width=1\textwidth]{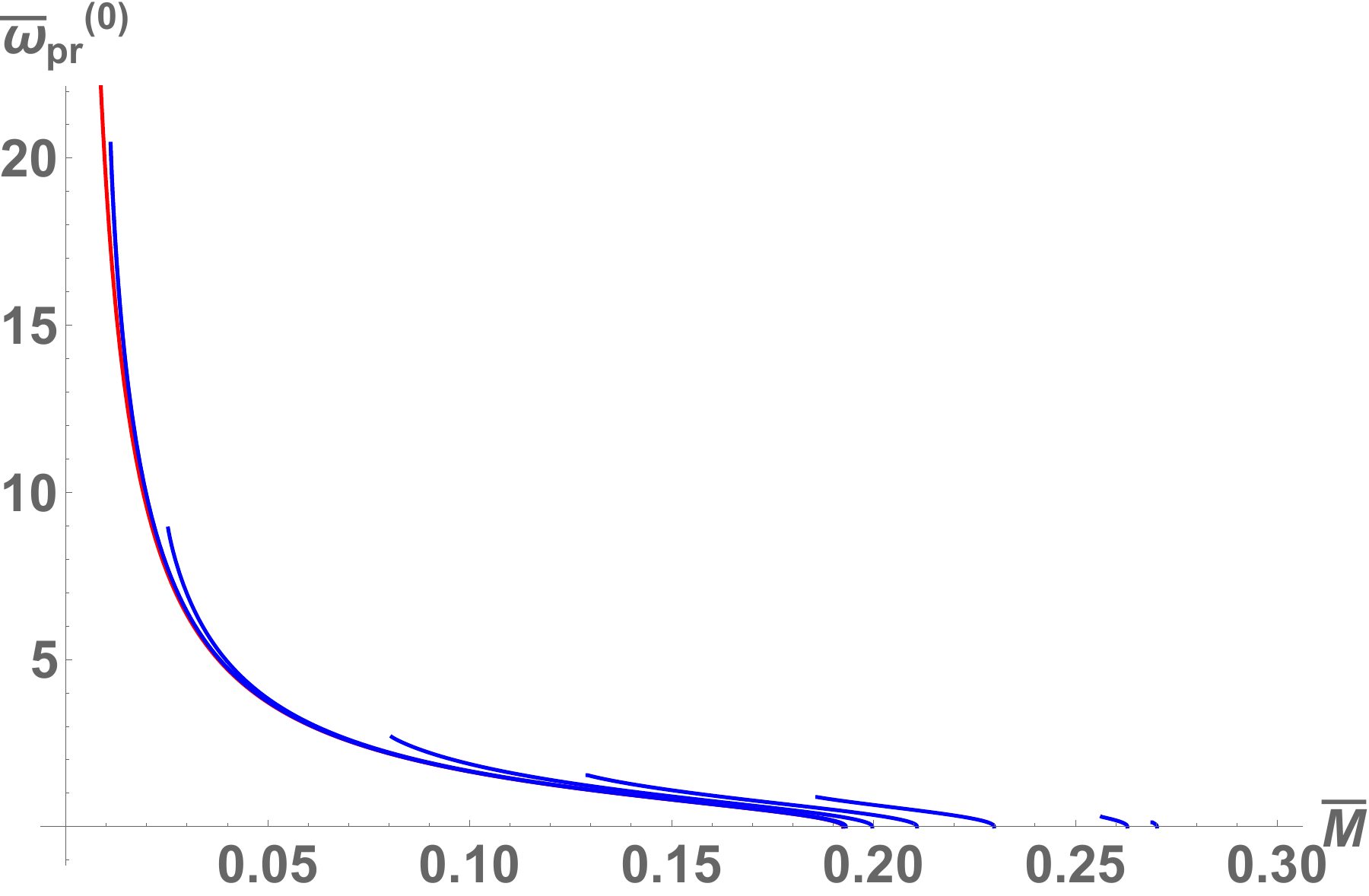}
		
		\label{fig:}
	\end{subfigure}\hfill%
	\begin{subfigure}{.46\textwidth}
		\includegraphics[width=1\textwidth]{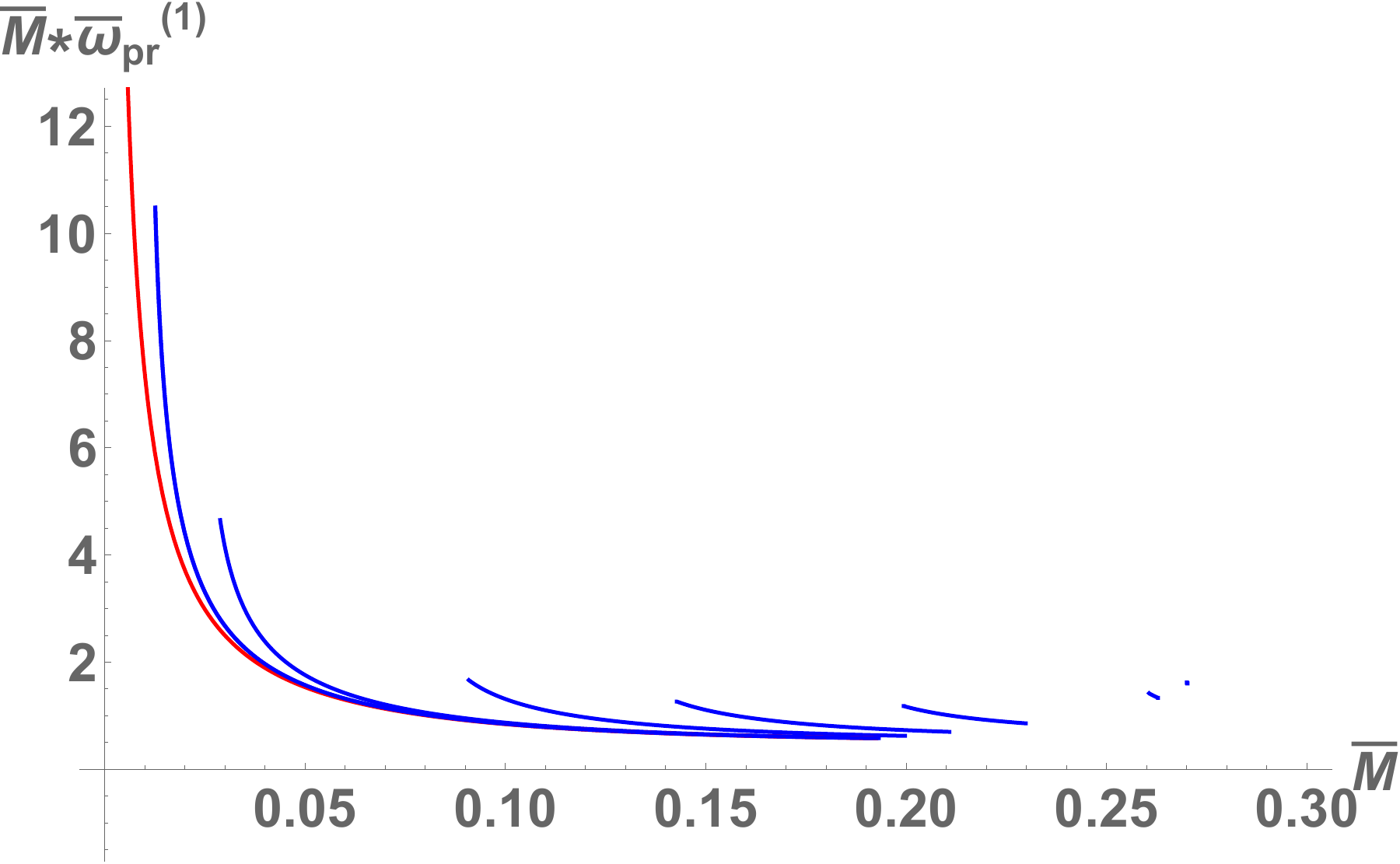}
		
		\label{fig:}
	\end{subfigure}\vspace{10mm}
	\caption[Photon ring solutions in dS.]{Photon ring solutions in an asymptotically de Sitter spacetime for $\alpha$ = $0.5\alpha_T$, $0.01\alpha_C$, $0.1\alpha_C$, $0.25\alpha_C$, $0.5\alpha_C$, $0.9\alpha_C$, $0.98\alpha_C$ (from left to right in blue) against the GR ($\alpha=0$) solution in red.}
	\label{fig:ps ds}
\end{figure*}

Finally, we solve for the photon ring parameters one last time in asymptotically de Sitter space. The relevant solutions to  \eqref{eq:ps perturbative} are plotted in figure \ref{fig:ps ds}. When $\alpha$ is small, again we see the largest departure from the GR case for small masses, and convergence as mass becomes large. Larger $\alpha$ again corresponds to the pushing of the allowed region of parameter space to the right, until eventually there is no longer any overlap in allowed mass with the GR results. The general forms of these curves are all reminiscent of the Einstein result when sufficiently far from the minimum mass and critical $\alpha$.

\begin{figure}
	\includegraphics[width=8cm]{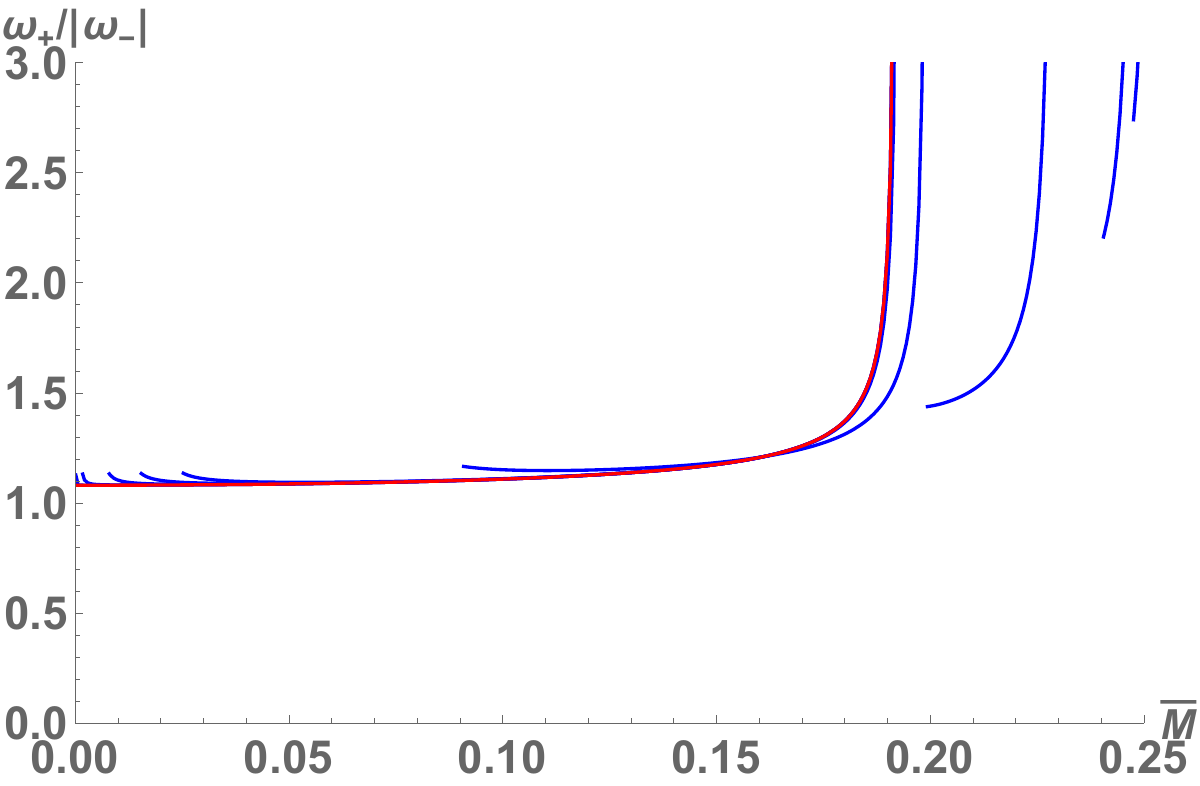}
	\centering
	
	\caption[Ratio of $\frac{\omega_+}{|\omega_-|}$ in dS.]{Ratio of $\frac{\omega_+}{|\omega_-|}$ in GR (red) plotted against the same ratio in the 4DEGB theory (blue) for
(from left to right) $\alpha$ = 0.0001$\alpha_T$, 0.01$\alpha_T$, 0.2$\alpha_T$, 0.75$\alpha_T$, 2$\alpha_T$, 0.1$\alpha_C$, 0.5$\alpha_C$, 0.75$\alpha_C$, 0.8$\alpha_C$ in an asymptotically de Sitter spacetime. In this figure we have fixed $\chi = 0.1$.}
	
	\label{fig:ratio ds}
\end{figure}

The ratio $\Omega \coloneqq \frac{\omega_+}{|\omega_-|}$ is a function of $\bar{M}$, $\bar{\alpha}$, and $\chi$ in the 4DEGB theory whereas the GR solution of course has no $\alpha$ dependence. The numerical solution for this ratio in asymptotically dS space is plotted in figure \ref{fig:ratio ds}. When $\bar{\alpha}$ is small we find that the 4DEGB solution diverges most from GR in the small mass regime, whereas in the large mass regime the result converges exactly with the Einsteinian result. For larger $\bar{\alpha}$ the allowed mass region shrinks and shifts to the right until eventually there is no overlap between the allowed masses of the 4DEGB theory and GR. \newline

In order to better understand the small $\bar{\alpha}$ behaviour, we expand $\Omega_{4DEGB}$ to leading order in $\bar{\alpha}$:
\begin{equation}
\begin{aligned}
&\Omega_{4DEGB} = \Omega_{GR}^{num}\\
 &+\frac{  \left(27 \bar{M}^2 \left(-729 \bar{M}^4+864 \bar{M}^2+28\right)-32\right) \chi}{3 \bar{M}^2 \sqrt{3-81 \bar{M}^2} \left(\left(27 \bar{M}^2+2\right) \chi-3 \sqrt{3-81 \bar{M}^2}\right)^2}\bar{\alpha}\\
 &+ \mathcal{O}(\bar{\alpha}^2).
 \end{aligned}
\end{equation}
With this we see that the small $\alpha$ results should asymptote to those of GR when $\left(27 \bar{M}^2 \left(-729 \bar{M}^4+864 \bar{M}^2+28\right)-32\right)=0$, ie. when $\bar{M} = 0.156121$. This behaviour is manifest in figure \ref{fig:ratio ds}.



\begin{figure*}
\subcaptionbox[0th order Lyapunov exponent for the photon ring in dS.]{0th order Lyapunov exponent  correction for the photon ring in an asymptotically de Sitter spacetime. The red curve shows the GR ($\alpha = 0$) result, whereas the blue curves represent the 4DEGB results for $\alpha$ = $0.5\alpha_T$, $0.01\alpha_C$, $0.1\alpha_C$, $0.25\alpha_C$, $0.5\alpha_C$, $0.9\alpha_C$, $0.98\alpha_C$. \label{fig:lam0 ds}}[0.43\textwidth]
{\includegraphics[valign=t,width=0.43\textwidth]{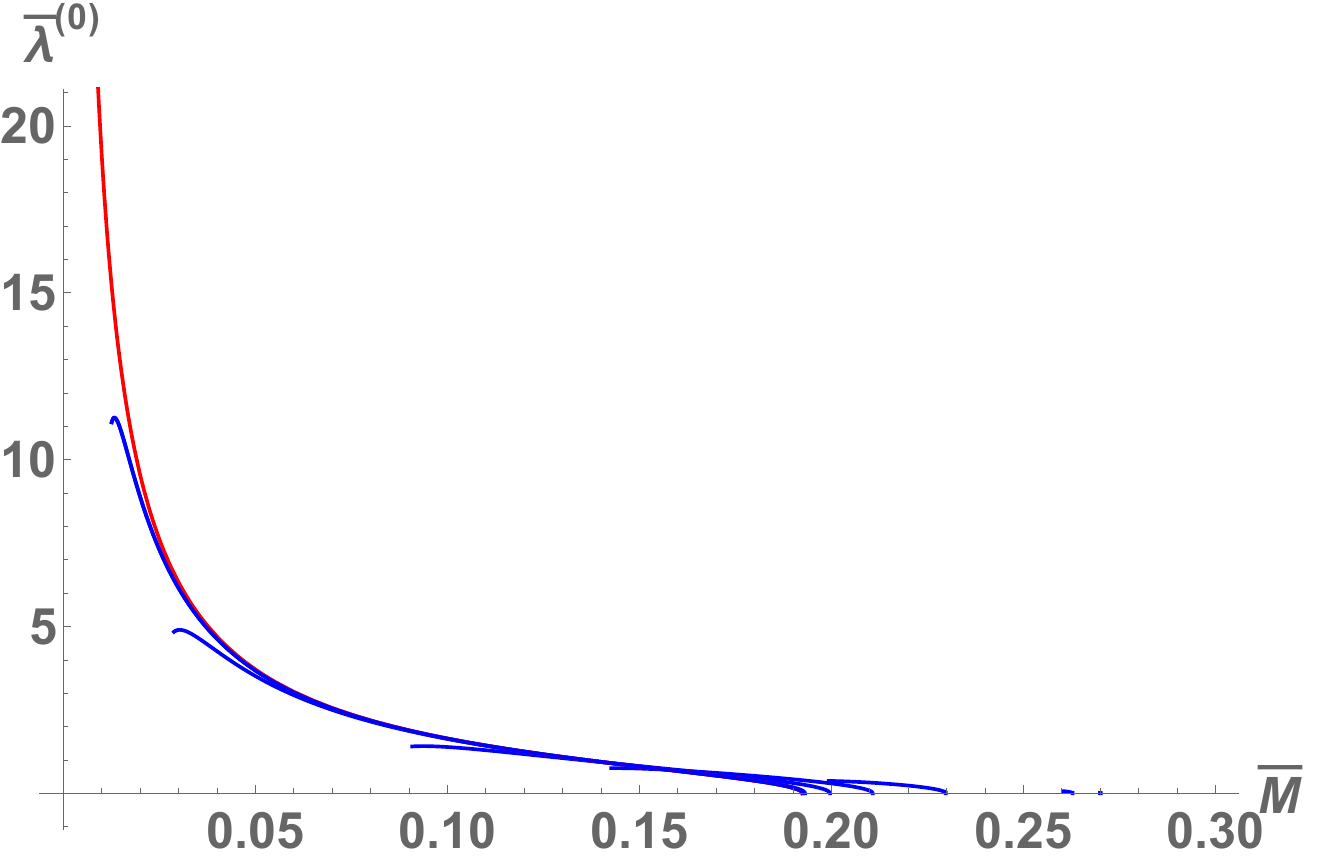}}
\hspace{0.05\textwidth} 
\subcaptionbox[Leading order Lyapunov exponent correction for the photon ring in dS.]{Leading order Lyapunov exponent correction for the photon ring when in an asymptotically de Sitter spacetime. The red curve shows the GR ($\alpha = 0$) result, whereas the blue curves represent the 4DEGB results for $\alpha$ = $0.5\alpha_T$, $0.01\alpha_C$, $0.1\alpha_C$, $0.25\alpha_C$, $0.5\alpha_C$, $0.9\alpha_C$, $0.98\alpha_C$. \label{fig:lam1 ds}}[0.43\textwidth]
{\includegraphics[valign=t,width=0.43\textwidth]
{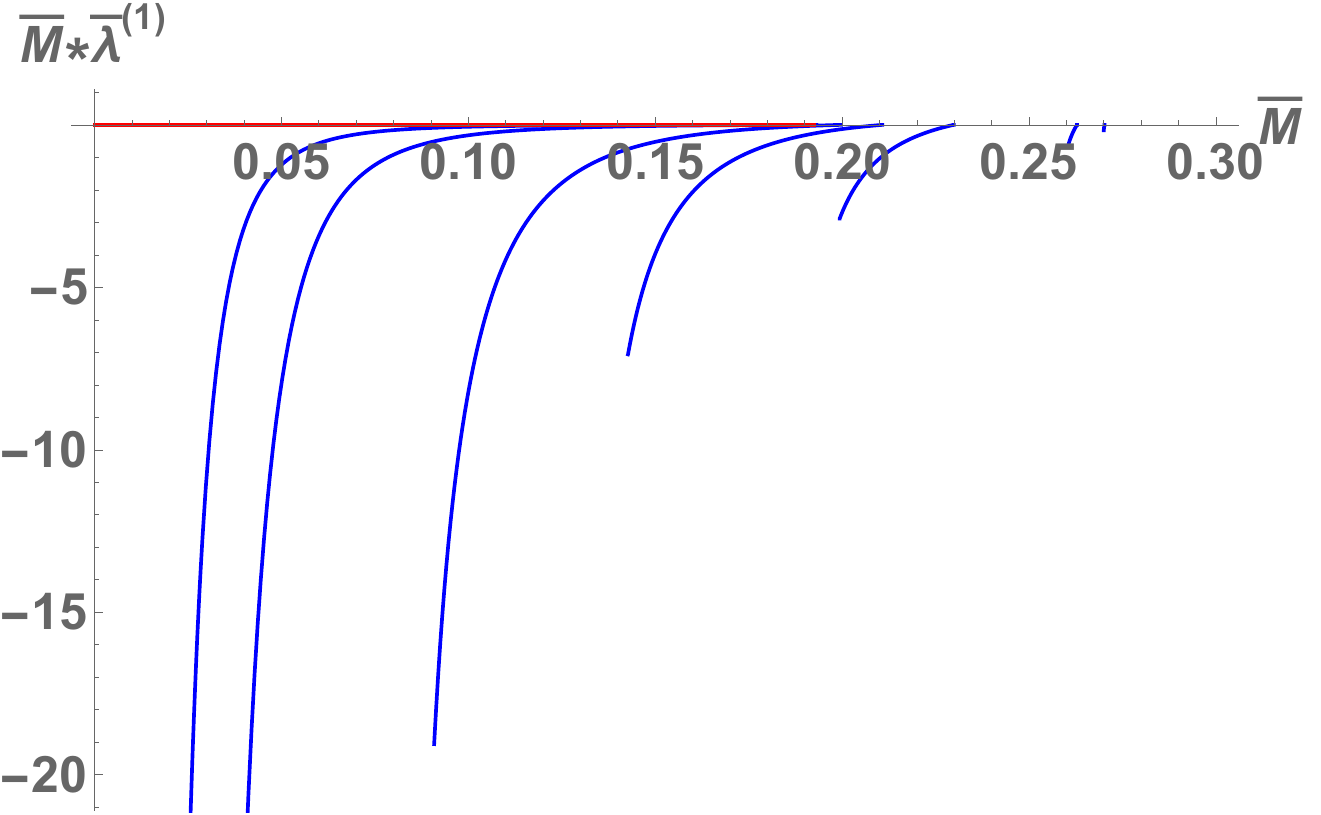}
}
\caption[Lyapunov exponents in dS.]{Lyapunov exponents when $\Lambda > 0$.}
\end{figure*}

The Lyapunov exponents for asymptotically de Sitter space are plotted in figures \ref{fig:lam0 ds} and \ref{fig:lam1 ds} alongside the GR results. Again for small $\alpha$ we find curves for $\lambda^{(0)}$ that are very similar to GR in form, diverging most near the minimum mass limit, with the larger $\alpha$ curves being pushed to a different region of parameter space like usual (and similarly for $\lambda^{(1)}$). The low-mass reduction of instability in the 0th order exponent is comparable the the asymptotically flat case, though less dramatic. The leading order corrections seem to care less about $\alpha$ near $M_{\rm{min}}$, as the stability increases significantly for all curves near this point.



\subsection{Black Hole Shadow}

A detailed, general study of non-equatorial null geodesics for a slowly rotating metric ansatz was carried out in \cite{adair2020}, from which we obtain the following relations:
\begin{equation}\label{eq:bh shadow geodesics}
r^2 \dot{\phi} =\frac{\ell_z}{\sin ^2 \theta}-\frac{a P(r) E}{f(r)}, \; \; r^2 \dot{\theta} =\pm \sqrt{j^2-\frac{\ell_z^2}{\sin ^2 \theta}}
\end{equation}
\begin{equation}
r_{\mathrm{ps}}=r_{\mathrm{ps}}^{(0)}+a r_{\mathrm{ps}}^{(1)}, \quad j_{\mathrm{ps}}^2=\left(j_{\mathrm{ps}}^{(0)}\right)^2+a\left(j_{\mathrm{ps}}^{(1)}\right)^2
\end{equation}
\begin{equation}
r_{\mathrm{ps}}=r_{\mathrm{ps}}^{(0)}+\left.\frac{2 a \ell_z f\left(r P^{\prime}-2 P\right)}{r\left(r^2 f^{\prime \prime}-2 f\right)}\right|_{r=r_{\mathrm{ps}}^{(0)}},
\end{equation}
\begin{equation}\label{eq:j lz}
j_{\mathrm{ps}}^2=\frac{\left(r_{\mathrm{ps}}^{(0)}\right)^2}{f\left(r_{\mathrm{ps}}^{(0)}\right)}+\frac{2 a \ell_z P\left(r_{\mathrm{ps}}^{(0)}\right)}{f\left(r_{\mathrm{ps}}^{(0)}\right)},
\end{equation}
where
\begin{equation}
r_{\mathrm{ps}}^{(0)} f^{\prime}\left(r_{\mathrm{ps}}^{(0)}\right)-2 f\left(r_{\mathrm{ps}}^{(0)}\right)=0.
\end{equation}

With this, similar to the procedure in \cite{adair2020}, we consider an observer placed far away from a black hole, with the spherical coordinates $(r = r_0, \theta = \theta_0, \phi = 0)$ (without loss of generality). One can imagine this observer receiving a photon moving in the direction $dr/dt>0$, defined by the angular momentum parameters $j^2$ and $\ell_z$. Unlike our analysis in section \ref{subsec:photonrings}, $j$ is no longer necessarily aligned with the z-direction, and instead the two are related via \eqref{eq:j lz}. From here, we would like to find the angle between the photon's velocity vector and the plane perpendicular to the $r$-direction at the observer's location. At this point we write the velocity tangent vector as
\begin{equation}
u=-\dot{r} e_r+r_0 \dot{\theta} e_\theta+r_0 \sin \theta_0 \dot{\phi} e_\phi
\end{equation}
where we use the following orthonormal coordinate system for the observer facing the black hole:
\begin{equation}
e_r=-\partial_r, \quad e_\theta=\frac{\partial_\theta}{r_0}, \quad e_\phi=\frac{\partial_\phi}{r_0 \sin \theta_0}.
\end{equation}

\begin{figure}[H]
	\includegraphics[width=8cm]{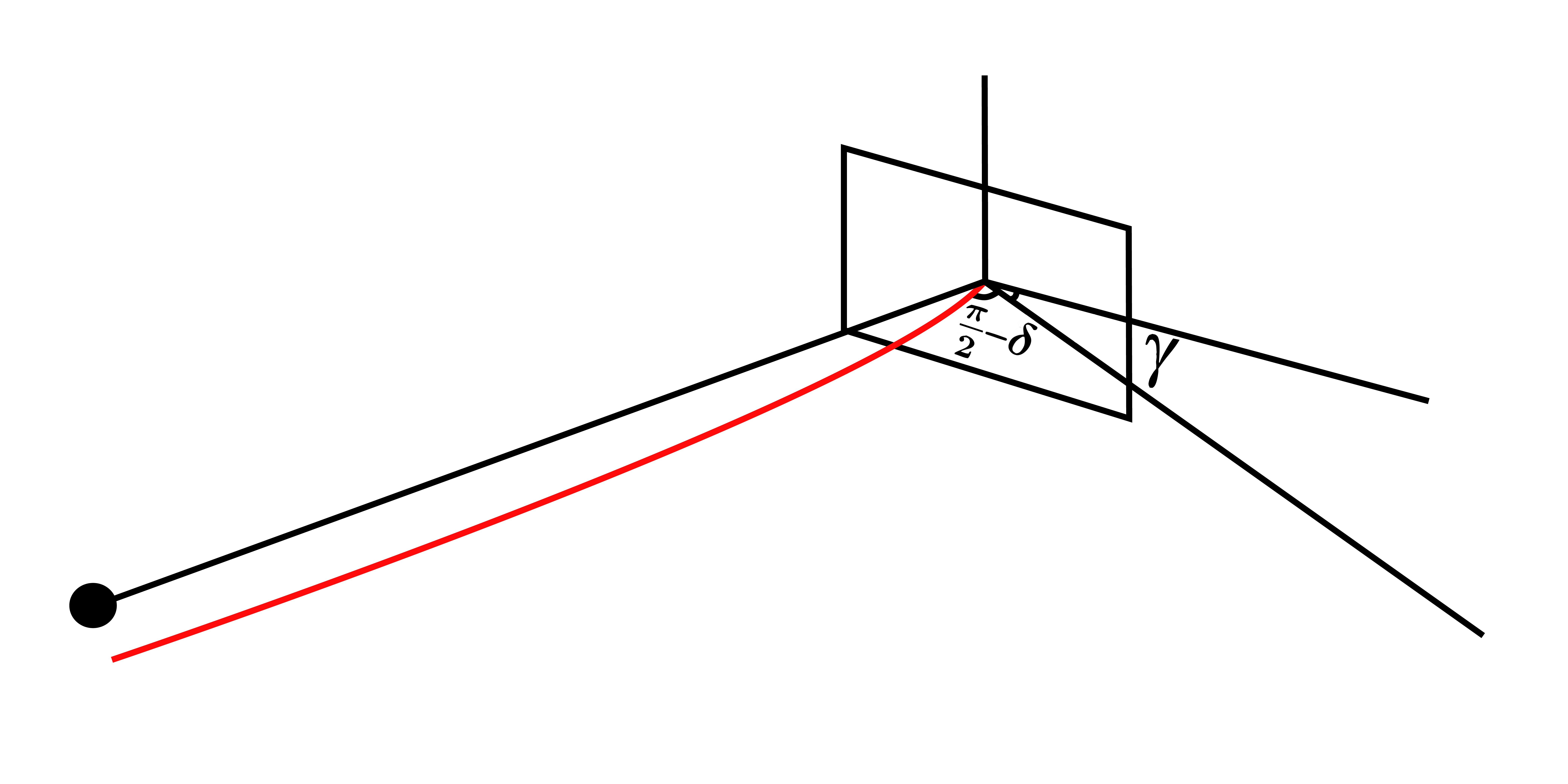}
	\centering
	\caption[Black hole shadow: incoming photon geometry.]{Geometry used in the black hole shadow derivation, with the red line representing an incoming photon.}
	\label{fig:bh shadow geometry}
\end{figure}

We define $\frac{\pi}{2} - \delta$ to be the angle between the tangent vector and the observer's plane (see figure \ref{fig:bh shadow geometry}), which in general has a component below the $x$-$y$ plane.  $\gamma$ is then defined to be the angle the projected vector forms with the direction $e_\phi$. In doing so it is straightforward to show that
\begin{equation}
\sin \delta=r_0 \sqrt{\dot{\theta}^2+\sin ^2 \theta_0 \dot{\phi}^2}, \quad \cos \gamma=\frac{\sin \theta_0 \dot{\phi}}{\sqrt{\dot{\theta}^2+\sin ^2 \theta_0 \dot{\phi}^2}}
\end{equation}
where we have made the assumption that $\dot{r} << r_0 \sqrt{\dot{\theta}^2+\sin ^2 \theta_0 \dot{\phi}^2}$ since the observer is placed at the limit $r_0 \rightarrow \infty$. With this, the tangent vector is effectively re-parameterized:
\begin{equation}
u=-\dot{r} e_r+\sin \delta\left(e_\theta \sin \gamma+e_\phi \cos \gamma\right).
\end{equation}

Invoking equations \eqref{eq:bh shadow geodesics} furthermore allows us to express our angular momentum parameters in terms of these angles and the constants describing our observer's location:
\begin{equation}\label{eq:angmom bhs}
j=r_0 \sin \delta, \quad \ell_z=r_0 \sin \theta_0 \cos \gamma \sin \delta.
\end{equation}
We know that the black hole shadow is determined by those photons passing arbitrarily close to the photon sphere. We take this into account by using relation \eqref{eq:j lz} with \eqref{eq:angmom bhs}, yielding
\begin{equation}
r_0^2 \sin ^2 \delta=\frac{\left(r_{\mathrm{ps}}^{(0)}\right)^2}{f\left(r_{\mathrm{ps}}^{(0)}\right)}+\frac{2 a \sin \theta_0 P \left(r_{\mathrm{ps}}^{(0)}\right)}{f\left(r_{\mathrm{ps}}^{(0)}\right)} \cos \gamma \; r_0 \sin \delta
\end{equation}
which determines the contour $\delta(\gamma)$ of the black hole shadow. Expanding linearly in $a$, the solution reads
\begin{equation}
r_0 \sin \delta=\frac{r_{\mathrm{ps}}^{(0)}}{\sqrt{f\left(r_{\mathrm{ps}}^{(0)}\right)}}+\frac{a \sin \theta_0 P\left(r_{\mathrm{ps}}^{(0)}\right)}{f\left(r_{\mathrm{ps}}^{(0)}\right)} \cos \gamma.
\end{equation}

Finally, since $\delta << 1$ in the large $r_0$ limit, we can set $\sin(\delta) \approx \delta$. Since in the large $r_0$ limit $R \approx r_0 \delta$, we can say
\begin{equation}
R_{sh} \approx \frac{r_{\mathrm{ps}}^{(0)}}{\sqrt{f\left(r_{\mathrm{ps}}^{(0)}\right)}}+\frac{a \sin \theta_0 P\left(r_{\mathrm{ps}}^{(0)}\right)}{f\left(r_{\mathrm{ps}}^{(0)}\right)} \cos \gamma.
\end{equation}
Here the re-scaled equation is obtained simply by replacing unbarred symbols with their barred counterparts. When rotational effects vanish the above expression simply describes a circle of radius
\begin{equation}
R_{sh}^{(0)} = \frac{r_{\mathrm{ps}}^{(0)}}{\sqrt{f\left(r_{\mathrm{ps}}^{(0)}\right)}}.
\end{equation}

The effect of rotation is to offset this image in the $x$-direction of the observer's plane. The center of the circumference is offset by the amount corresponding to $\gamma = 0$, ie:
\begin{equation}
    D_{sh} = a \frac{\sin(\theta_0) P(r_{ps}^{(0)})}{f(r_{ps}^{(0)})}.
\end{equation}
With this it is trivial to plot the contour of the black hole shadow (parameterized by $\gamma:0 \to 2 \pi$), keeping in mind that $x = R_{sh} \rm{cos}\gamma$, $y = R_{sh} \rm{sin}\gamma$. This is done in the following sections for asymptotically flat, anti-de Sitter, and de Sitter spacetimes.



\subsubsection{Asymptotically Flat ($\mathbf{\Lambda = 0}$)}

\begin{figure*}
\subcaptionbox[Black hole shadow radius when $\Lambda = 0$.]{Black hole shadow radius plotted as a function of mass for $\alpha/M_{\odot}^2$ = 0.01, 0.05, 0.1, 0.2, 0.3, 0.7, 1 (in blue from left to right) against the GR results (red) when $\Lambda = 0$. Here we have fixed $\chi = -0.1$ for comparison with \cite{adair2020}, and have also fixed $\theta_0 = \pi/2$. \label{fig:rshadow l0}}
{\includegraphics[width=0.43\textwidth]{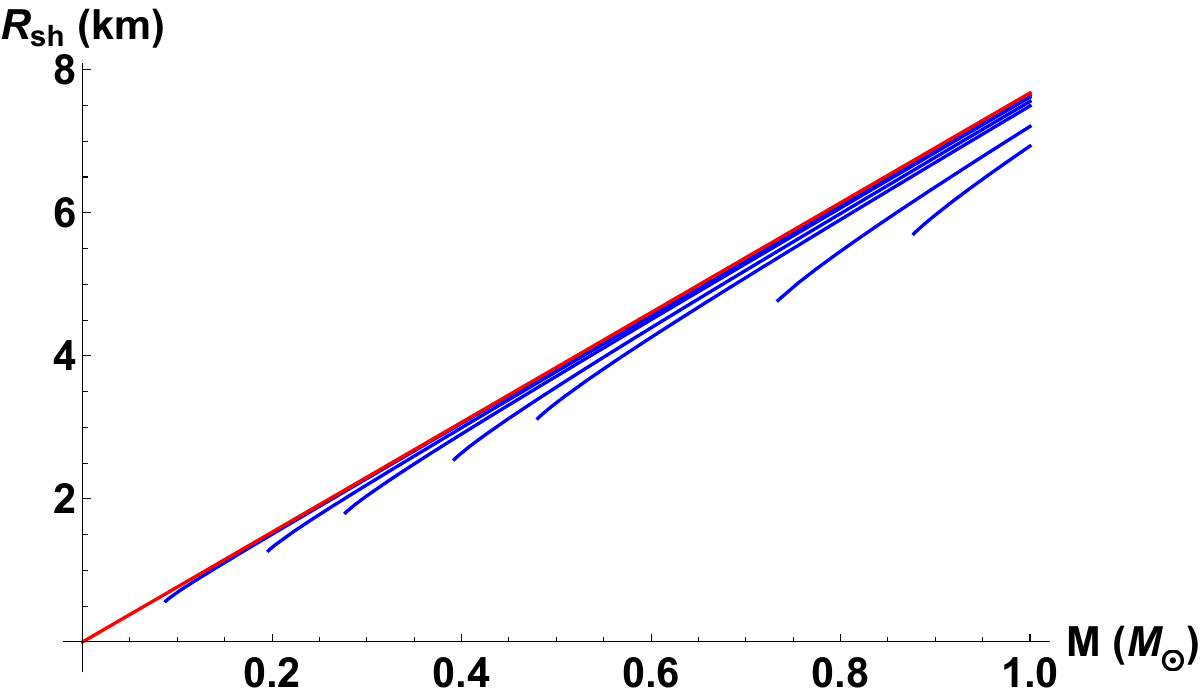}}
\hspace{0.05\textwidth} 
\subcaptionbox[Ratio $D_{sh}/R_{sh}$ when $\Lambda = 0$.]{Ratio $D_{sh}/R_{sh}$ for $\alpha/M_{\odot}^2$ = 0.01, 0.05, 0.1, 0.2, 0.3, 0.7, 1 (in blue from left to right) against the GR results (red) when $\Lambda = 0$. Here we have fixed $\chi = -0.1$ for comparison with \cite{adair2020}, and have also fixed $\theta_0 = \pi/2$. \label{fig:doverr l0}}
{\includegraphics[valign=t,width=0.43\textwidth]{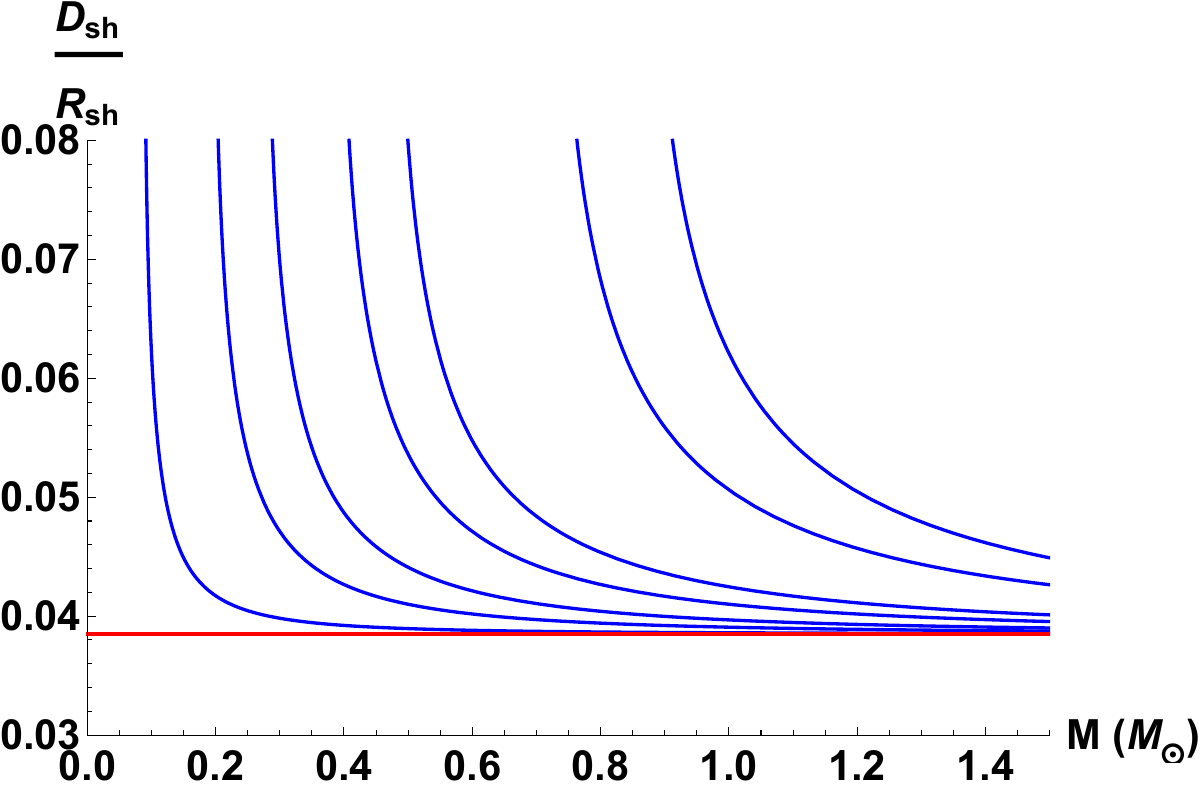}}
\caption[Properties of the black hole shadow when $\Lambda = 0$.]{Properties of the black hole shadow when $\Lambda = 0$.}
\end{figure*}

\begin{figure}
    \centering
    \includegraphics[width=8.6cm]{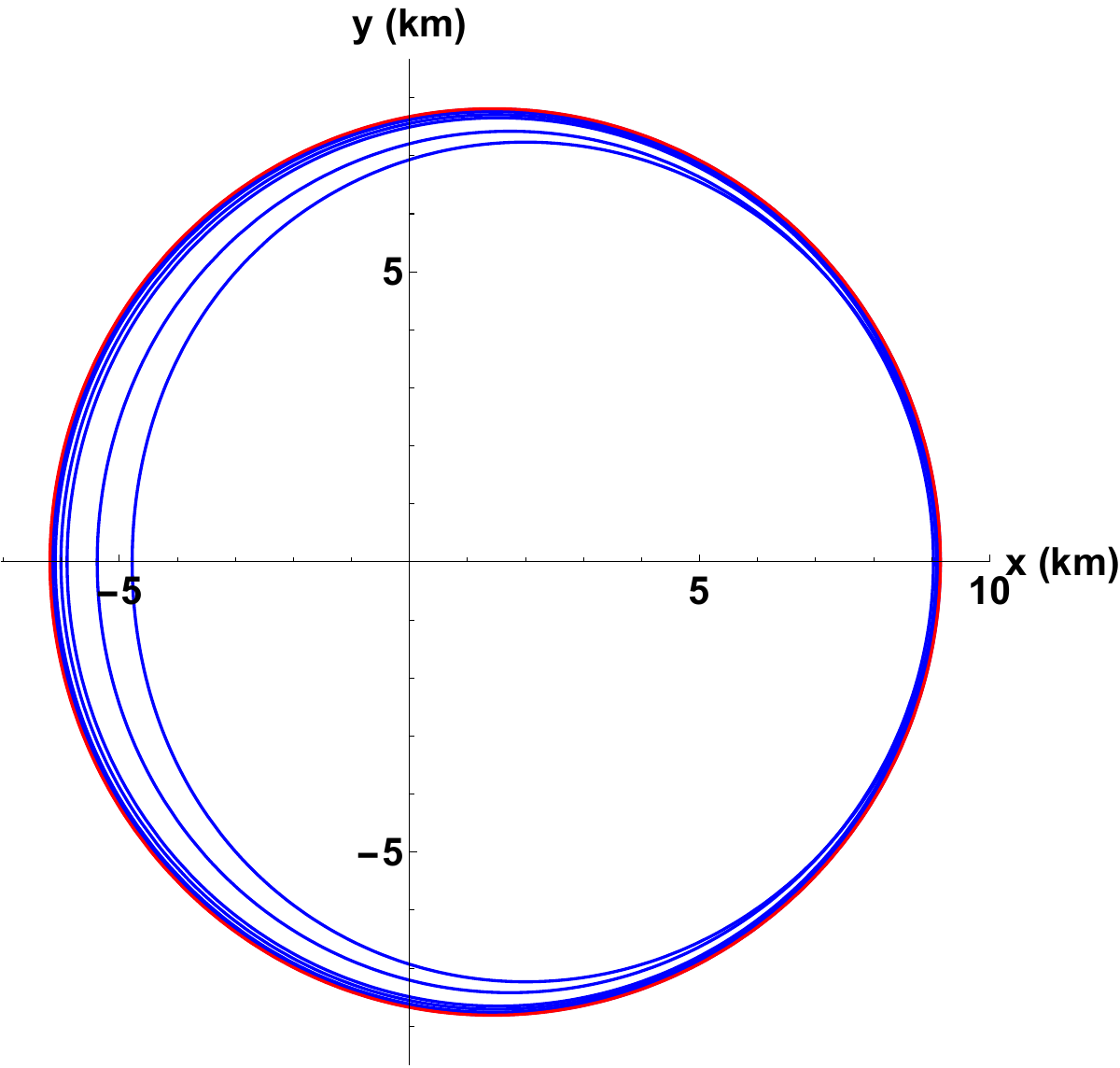}
    \caption[Contour of the black hole shadow when $\Lambda = 0$.]{Contour of the black hole shadow  for $\alpha/M_{\odot}^2$ = 0.01, 0.05, 0.1, 0.2, 0.3, 0.7, 1 (in blue from left to right) against the GR results (red) when $\Lambda = 0$. Here we have fixed $\chi = -0.5$ for comparison with \cite{adair2020}, and have also fixed $\bar{M} = 1$, $\theta_0 = \pi/2$.}
    \label{fig:shadow l0}
\end{figure}

The plots characterizing the properties of the black hole shadow for $\Lambda = 0$ can be seen in figures \ref{fig:rshadow l0} and \ref{fig:doverr l0}. As we have come to expect, we see the greatest disagreement with GR near $M_{min}$, and we see convergence with GR as mass gets large. The contours representing the black hole shadow geometry in this case can be seen in figure \ref{fig:shadow l0}. We notice that as the coupling constant becomes large, the radius of the shadow shrinks and its center is offset as the equations suggest.

\subsubsection{AdS ($\mathbf{\Lambda < 0}$)}

\begin{figure*}
\subcaptionbox[Black hole shadow radius in AdS.]{Black hole shadow radius plotted as a function of mass for $\alpha/\alpha_C$ = 0.002, 0.066, 0.2, 0.5, 0.666, 0.9, 0.98 0.99, 0.997, 0.998, 0.999 (in blue from left to right) against the GR results (red) in an asymptotically anti-de Sitter spacetime. Here we have fixed $\chi = -0.1$ for comparison with \cite{adair2020}, and have also fixed $\theta_0 = \pi/2$. \label{fig:rshadow ads}}[0.43\textwidth]
{\includegraphics[valign=t,width=0.43\textwidth]{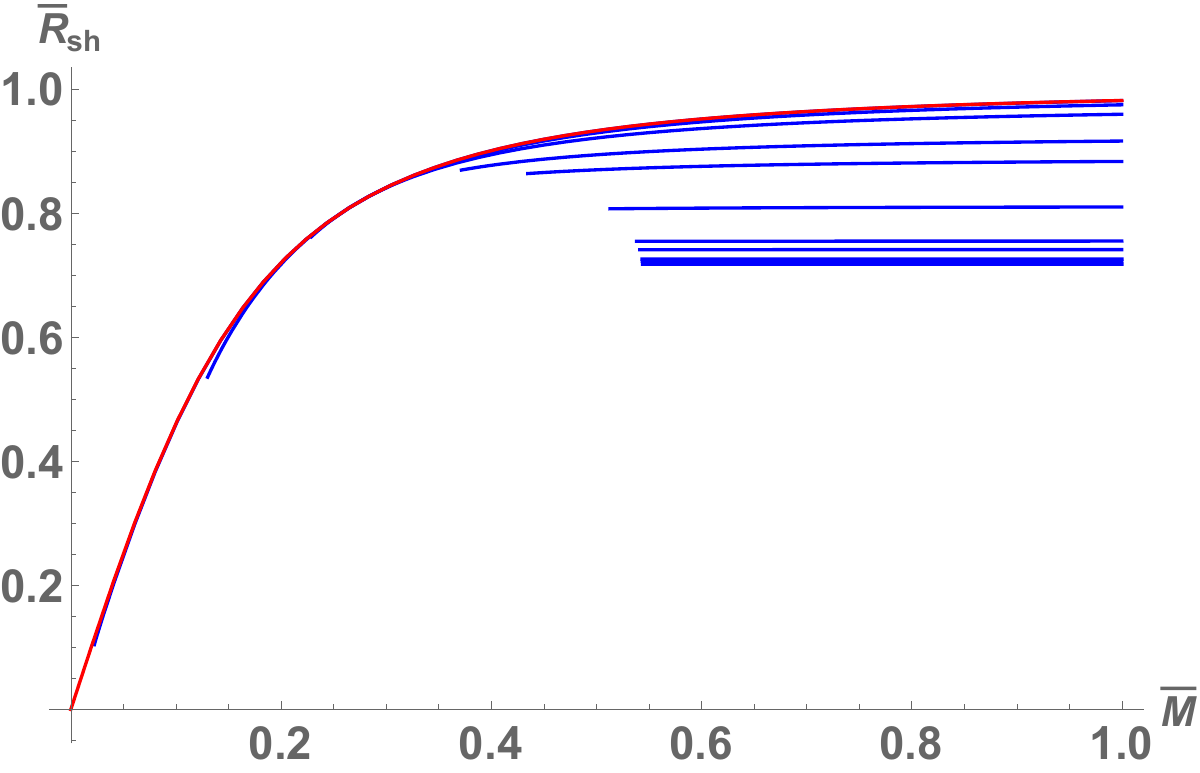}}
\hspace{0.05\textwidth} 
\subcaptionbox[Ratio $D_{sh}/R_{sh}$ in AdS.]{Ratio $D_{sh}/R_{sh}$ for $\alpha/\alpha_C$ = 0.002, 0.066, 0.2, 0.5, 0.666, 0.9, 0.98 0.99, 0.997, 0.998, 0.999 (in blue from left to right) against the GR results (red) in an asymptotically anti-de Sitter spacetime. Here we have fixed $\chi = -0.1$ for comparison with \cite{adair2020}, and have also fixed $\theta_0 = \pi/2$. \label{fig:doverr ads}}[0.43\textwidth]
{\includegraphics[valign=t,width=0.43\textwidth]{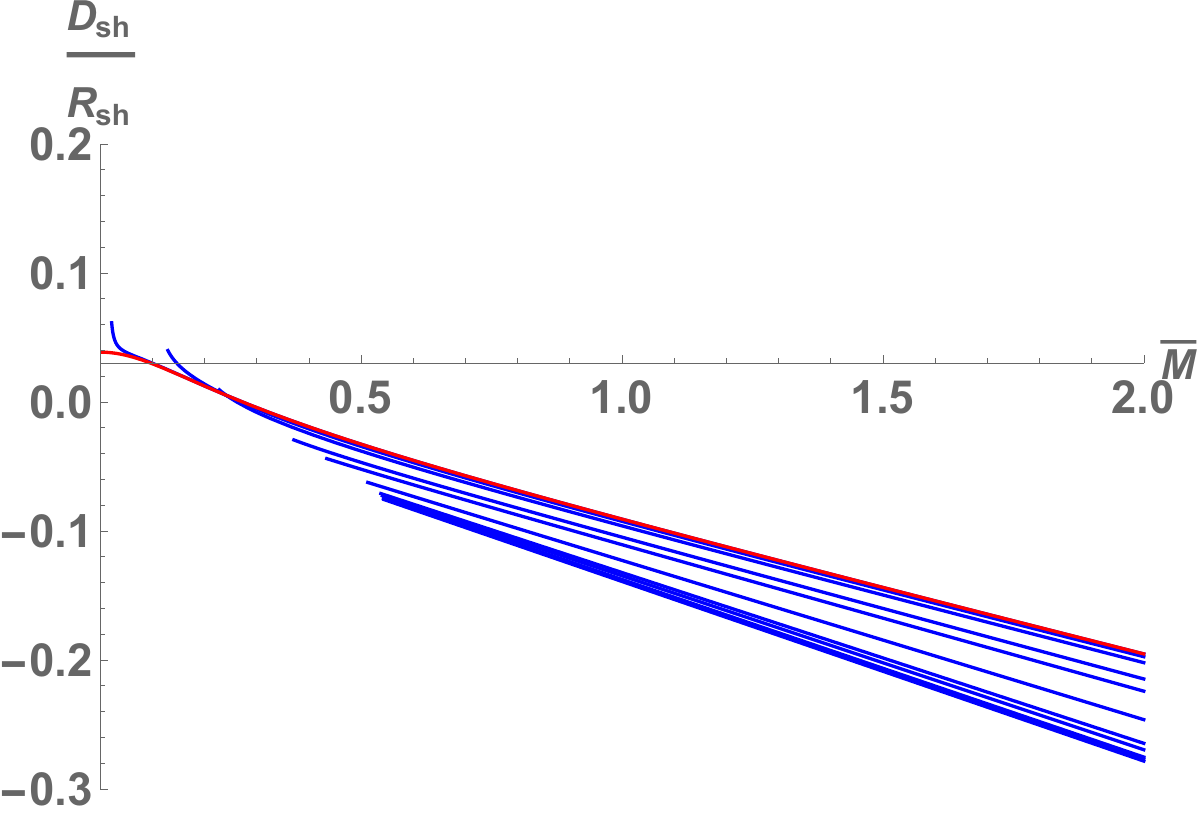}}
\caption[Properties of the black hole shadow in AdS.]{Properties of the black hole shadow when $\Lambda < 0$.}\label{fig:shadowprops ads}
\end{figure*}

\begin{figure}
    \centering
    \includegraphics[width=8.6cm]{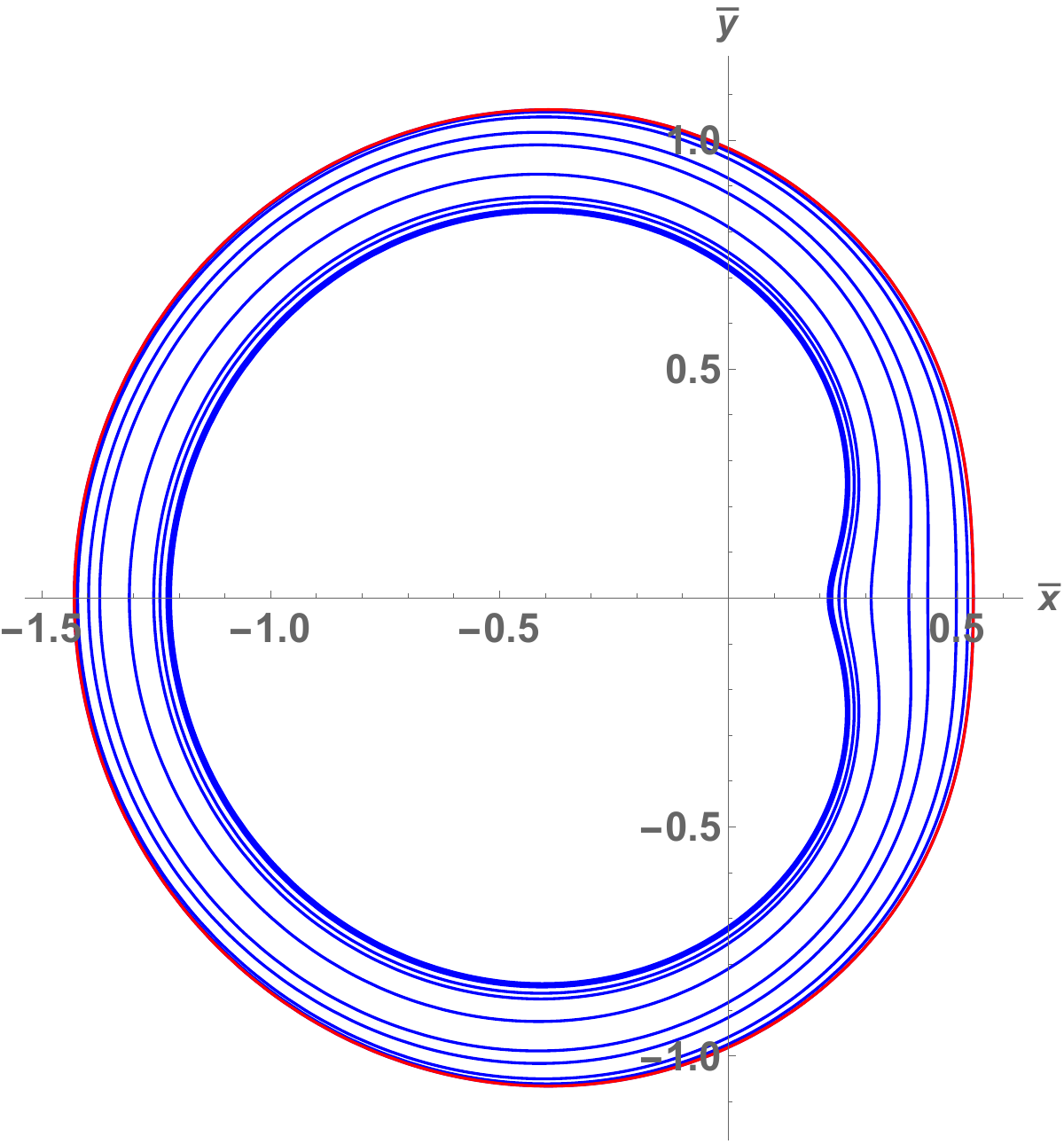}
    \caption[Contour of the black hole shadow in AdS.]{Contour of the black hole shadow  for $\alpha/\alpha_C$ = 0.002, 0.066, 0.2, 0.5, 0.666, 0.9, 0.98 0.99, 0.997, 0.998, 0.999 (in blue from right to left) against the GR results (red) in an asymptotically anti-de Sitter spacetime. Here we have fixed $\chi = -0.5$ for comparison with \cite{adair2020}, and have also fixed $\bar{M} = 1$, $\theta_0 = \pi/2$.}
    \label{fig:shadow ads}
\end{figure}

The black hole shadow results for asymptotically anti-de Sitter spacetimes are plotted in figures \ref{fig:shadowprops ads} and \ref{fig:shadow ads}. Once again we see very tight agreement with GR for small $\alpha$, which in the case of the ratio $D_{sh}/R_{sh}$ crosses over the GR result again in the small mass regime. The condition for the crossing point in figure \ref{fig:doverr ads} turns out to be equivalent to that in equation \eqref{eq:ratioadscross}, implying that the small $\alpha$ solutions cross GR at $\bar{M}=0.234239$. For larger $\alpha$ this crossover behaviour is not manifest. 
The small $\alpha$ shadow radius on the other hand touches the GR result before decreasing again. As alpha approaches criticality, the shadow radius approaches a constant defined only up to the expected minimum mass found in previous sections ($\bar{M}_{min} = 0.544$). 

The effect of increasing the 4DEGB coupling constant in a universe with a negative cosmological constant is to shrink the shadow radius \ref{fig:shadow ads}, with a dimple appearing on the right hand side for large $\alpha$.

\subsubsection{dS ($\mathbf{\Lambda > 0}$)}

\begin{figure*}
\subcaptionbox[Black hole shadow radius in dS.]{Black hole shadow radius plotted as a function of mass for $\alpha$ = 0.0001$\alpha_T$, 0.01$\alpha_T$, 0.2$\alpha_T$, 0.5$\alpha_T$, 0.75$\alpha_T$, 0.9375$\alpha_T$, 0.1$\alpha_C$, 0.5$\alpha_C$, 0.8$\alpha_C$ (in blue from left to right) against the GR results (red) in an asymptotically de Sitter spacetime. Here we have fixed $\chi = -0.1$ for comparison with \cite{adair2020}, and have also fixed $\theta_0 = \pi/2$. \label{fig:rshadow ds}}[0.43\textwidth]
{\includegraphics[valign=t,width=0.43\textwidth]{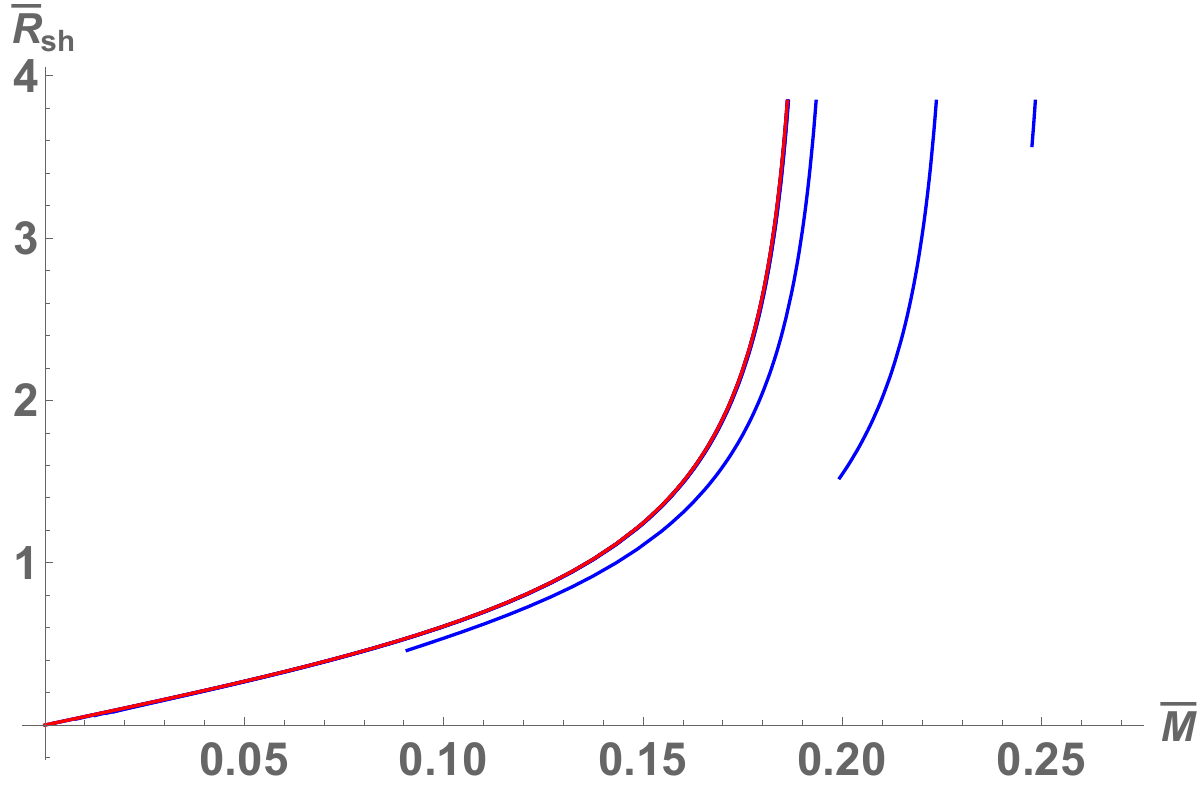}}
\hspace{0.05\textwidth} 
\subcaptionbox[Ratio $D_{sh}/R_{sh}$ in dS.]{Ratio $D_{sh}/R_{sh}$ for $\alpha$ = 0.0001$\alpha_T$, 0.01$\alpha_T$, 0.2$\alpha_T$, 0.5$\alpha_T$, 0.75$\alpha_T$, 0.9375$\alpha_T$, 0.1$\alpha_C$, 0.5$\alpha_C$, 0.8$\alpha_C$ (in blue from left to right) against the GR results (red) in an asymptotically de Sitter spacetime. Here we have fixed $\chi = -0.1$ for comparison with \cite{adair2020}, and have also fixed $\theta_0 = \pi/2$. \label{fig:doverr ds}}[0.43\textwidth]
{\includegraphics[valign=t,width=0.43\textwidth]{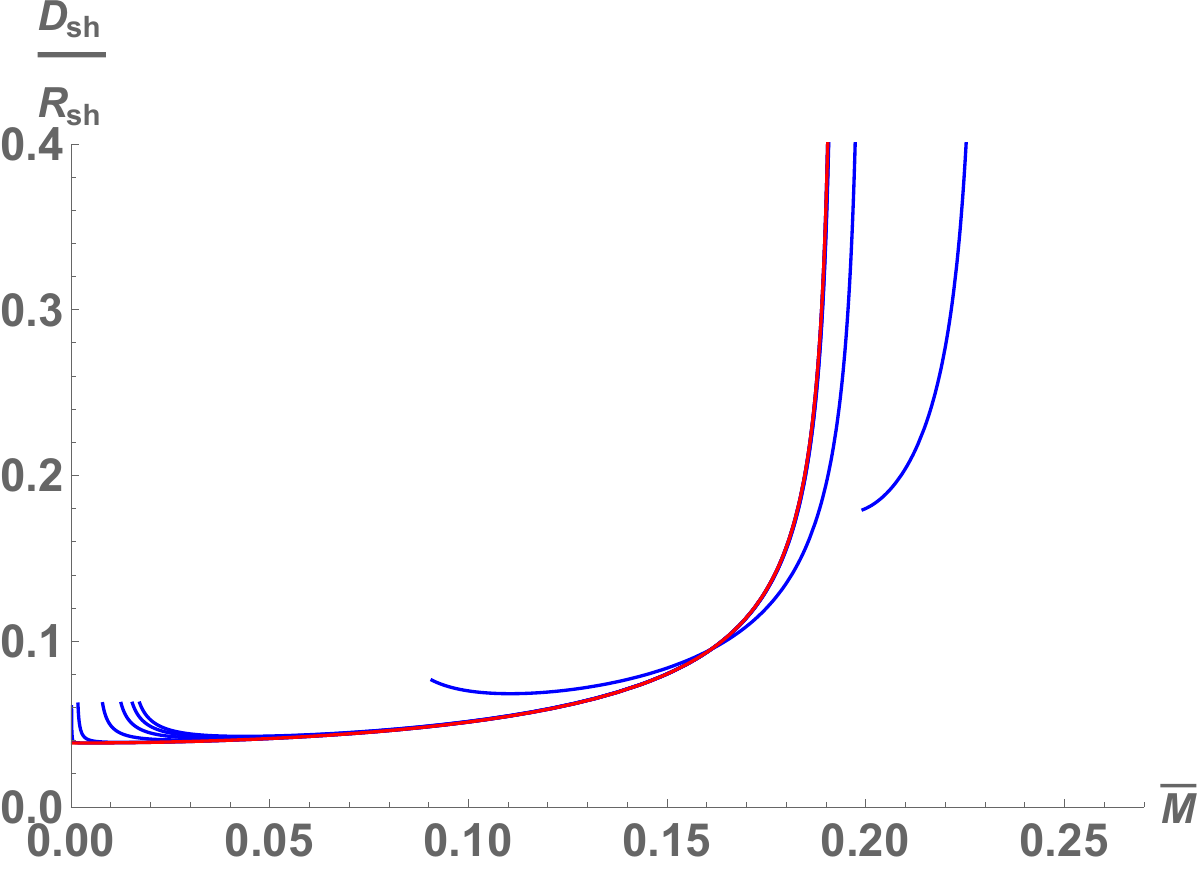}}
\caption[Properties of the black hole shadow in dS.]{Properties of the black hole shadow when $\Lambda > 0$.}\label{fig:shadowprops ds}
\end{figure*}

\begin{figure}
    \centering
    \includegraphics[width=8.6cm]{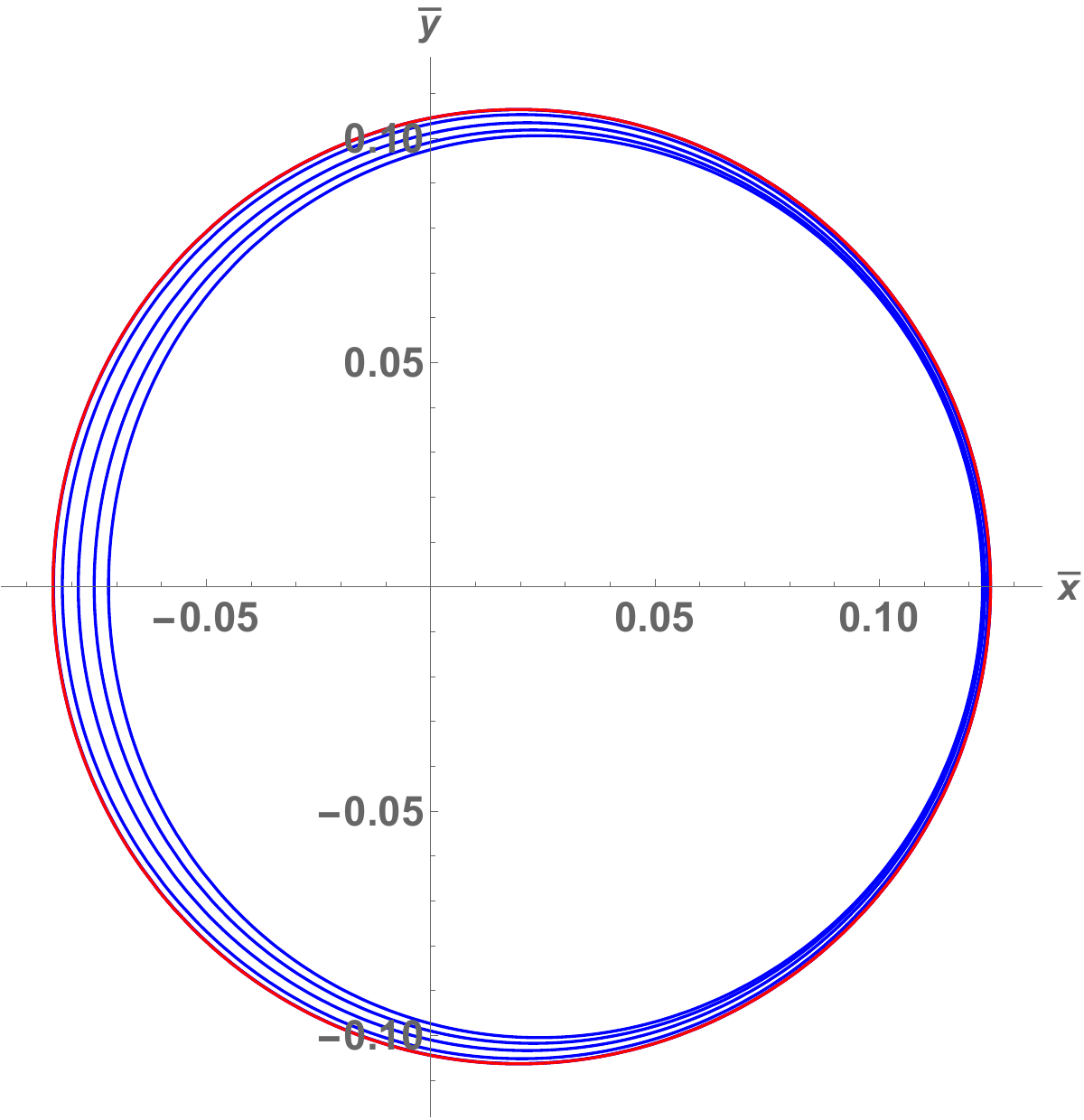}
    \caption[Contour of the black hole shadow in dS.]{Contour of the black hole shadow  for $\alpha$ = 0.0001$\alpha_T$, 0.01$\alpha_T$, 0.2$\alpha_T$, 0.5$\alpha_T$, 0.75$\alpha_T$, 0.9375$\alpha_T$, 0.1$\alpha_C$, 0.5$\alpha_C$, 0.8$\alpha_C$ (in blue from left to right) against the GR results (red) in an asymptotically de Sitter spacetime. Here we have fixed $\chi = -0.5$ for comparison with \cite{adair2020}, and have also fixed $\bar{M} = 0.02$, $\theta_0 = \pi/2$.}
    \label{fig:shadow ds}
\end{figure}

In an asymptotically de Sitter spacetime, as expected, we see strong agreement with the GR curves when $\alpha$ is small, only deviating near $\bar{M}_{\rm{min}}$ for the ratio $\frac{D_{sh}}{R_{sh}}$ (see figure \ref{fig:shadowprops ds}). As $\alpha \to \alpha_C$, we again see the behaviour of the allowed mass region diverging completely from those in GR (ie. masses which correspond to physical black holes in GR represent naked singularities in 4DEGB with large enough $\alpha$). The effect of increased coupling strength on the shadow in asymptotically de Sitter spacetimes can be seen in figure \ref{fig:shadow ds} and is very similar to the analogous case when $\Lambda = 0$.

\section{Summary}

In this paper we investigated slowly rotating solutions to the novel 4D Einstein Gauss-Bonnet theory of gravity for asymptotically flat, de Sitter, and anti-de Sitter spacetimes. These solutions include exact analytic forms for the metric functions $f(r)$ and $p(r)$.  We have examined  black hole horizon structures and angular velocities, radial location/angular momenta of innermost stable circular orbits for massive particles, the location/angular velocities/stabilities associated with the photon spheres, and the geometry of the shadows that characterize black holes in the 4DEGB theory of gravity. In most cases, we found solutions that were similar in form to the corresponding GR results, but with the inclusion of enforced minimum mass points (and in the dS case, maximum mass points). Most results tend to converge with general relativity in the large mass regime, although in AdS we instead sometimes see ``crossover behaviour" where the 4DEGB curves intersect the GR curves before diverging as mass gets larger.  

Analytic properties of the solution were also investigated, and we found that the slowly rotating black hole metric in 4DEGB gravity is regular everywhere except $r=0$ so long as we enforce $\alpha>0$. Solutions are only considered if they conceal this singular point behind a horizon (further investigation into interior solutions is also required). In de Sitter spacetimes we found that the black hole horizon structure is directly analogous to the charged Nariai solutions from the Reissner-Nordström de Sitter metric. We also found that, given a large enough 4DEGB coupling constant, the de Sitter results have no overlap in allowed mass with the GR results. In principle this feature (among others) could be used to set an upper limit on $\alpha$, using empirical evidence of black holes existing within the allowed GR regime. In practice, since asymptotic flatness is a good approximation in our universe, detecting such deviations from GR is a daunting task with how little the ISCO solutions differ in this case.

We also note that  these Lense-Thirring type metrics we have obtained are
linear in $a$, and are the first step   in searching for a full rotating solution in a given gravity theory. Obtaining solutions to   order $a^2$ will be   of great interest in studying tidal deformability, Love numbers, etc., leaving open an interesting avenue for future research. These calculations will be even more involved, requiring an expansion of the scalar and metric functions into spherical harmonic components and then solving the resultant field equations.

A thorough study of black hole shadow geometry suggests that the 4DEGB theory will produce rotating black hole shadows very similar to those generated by an Einsteinian black hole, although slightly smaller in diameter. These similarities between the theories elicit mixed emotions. Since GR is so successful, in principle any appropriate gravity theory should reproduce most or many of the same results. This feature is also what makes experimental verification difficult due to the extreme precision required to measure differences of this order. However, since many of the models outlined here \textit{do} predict unique features for certain small mass black holes, this motivates the search for minimal mass astrophysical black holes with which the 4DEGB theory could be put to the test as a proper observational competitor to GR.

\begin{acknowledgments}
This work was supported in part by the Natural Sciences and Engineering Research Council of Canada.
\end{acknowledgments}


\bibliography{apssamp}

\end{document}